\documentclass[12pt]{article}
\usepackage{a4wide}
\usepackage{longtable,fix-cm,url,units,hyperref}

\usepackage{axodraw2}

\def\num{$\langle$number$\rangle$}
\def\colorname{$\langle$colorname$\rangle$}

\DeclareUrlCommand\file{\urlstyle{tt}}

\def\program#1{\texttt{#1}}

\newcommand\BS{\symbol{`\\}}

\newcommand\name[1]{%
 \bgroup
   \def\{{\symbol{`\{}}%
   \def\}{\symbol{`\}}}%
   \texttt{\BS#1}%
 \egroup
}

\begin{document}

%% ?? To get settings useful to test compatibility of axodraw2 with
%% color.sty, uncomment the following line:
%\SetColor{Blue} \color{green} \pagecolor[cmyk]{0,0.02,0.05,0}
%%   Note green is useful, because it is defined in rgb color model
%% and the apparently equivalent Green, cmyk 1 0 1 0, looks quite
%% different on screen. So we can test the colors are entering graphics
%% correctly. 

%% ?? To test whether offsets work correctly, uncomment the following
%% line:
%\SetOffset(10,20) \SetScaledOffset(10,-20)

\setcounter{page}{0}
\thispagestyle{empty}
%\hfill \begin{minipage}{3.0cm}
%Nikhef 2013-035 \hfill \\
%2016
%\end{minipage}
%\vspace{20mm}

\begin{center}
{\LARGE\bf\sc Axodraw Version 2}
\end{center}
\vspace{5mm}
\begin{center}
{\large John C. Collins$^{\, a}$ and J.A.M. Vermaseren$^{\, b}$} 
\vspace{1cm}\\
{\it $^a$ Department of Physics, Pennsylvania State University, \\
\vspace{0.1cm}
University Park, Pennsylvania 16802, USA} \\
\vspace{0.5cm}
{\it $^b$Nikhef Theory Group \\
\vspace{0.1cm}
Science Park 105, 1098 XG Amsterdam, The Netherlands} \\
\vspace{1.0cm}
(26 May 2016)
\end{center}
\vspace{5mm}

\begin{abstract}
We present version two of the \LaTeX{} graphical style file Axodraw. 
It has a number of new drawing primitives and many extra options, and 
it can now work with \program{pdflatex} to directly produce
output in PDF file format (but with the aid of an auxiliary program).
\end{abstract}

\newpage

\tableofcontents

\newpage

%>>#[ Introduction :
%=========================
\section{Introduction}
\label{sec:intro}

This is the documentation for axodraw2, a \LaTeX{} package for drawing
Feynman graphs (and other simple graphics).  This version is a
substantial update of the original axodraw package \cite{axodraw1},
which was released
in 1994, and which has become rather popular in the preparation of articles in
elementary-particle physics.  One of its advantages is that its
drawing primitives are included in the .tex document, in a
human-writable form. (This also allows convenient production of
axodraw figures by other software, e.g., Jaxodraw
\cite{jaxodraw1,jaxodraw2}.)
This is in distinction to methods that
use a separate program to create graphics files that are read in
during the processing of the \LaTeX{} file.  The objects needed in
Feynman graphs are often difficult to draw at high quality with
conventional computer graphics software.

The original axodraw package has hardly been modified since its
introduction.  The new version addresses several later needs.  A
detailed list of the changes is given in Sec.\ \ref{sec:changes}.

One change arises from the fact that \TeX{} (and hence \LaTeX{})
themselves do not possess sufficiently useful methods of drawing
complicated graphics, so that the drawing of the graphics is actually
done inserting suitable code in the final output file (postscript or
pdf).  The original axodraw worked only with the
\program{latex}-\program{dvips} processing chain to put the diagrams in
the final postscript file.\footnote{A pdf file can be produced from
  the postscript file by a program like \program{ps2pdf}.}  Now we also
have in common use the \program{pdflatex} (and \program{lualatex} and
\program{xelatex}) programs that directly produce pdf.  The new version
of axodraw works with \program{pdflatex}, \program{lualatex}, and
\program{xelatex}, as well as with the \program{latex}-\program{dvips}
method.

Furthermore, more kinds of graphical object and greater flexibility in
their properties have been found useful for Feynman graphs.  The new
version provides a new kind of line, B\'ezier, and is able to make the
various kinds of line doubled.  There is now a very flexible
configuration of arrows.  Many of the changes correspond to
capabilities of JaxoDraw \cite{jaxodraw1,jaxodraw2}, which is a
graphical program for drawing Feynman graphs, and which is able to
write and to import diagrams in the axodraw format.

Finally, substantial improvements have been made in the handling of
colors, with much better compatibility with modern packages used to
set colors in the normal \LaTeX{} part of a document.

Since some of the changes (especially in the internal coding)
introduce potential incompatibilities with the original version of
axodraw, the new version of the style file is given a new name
\file{axodraw2.sty}.  Then the many legacy documents (e.g., on
\url{http://arxiv.org}) that use the old axodraw will continue to use
the old version, and will therefore continue to be compilable without
any need for any possible changes in the source document, and with unchanged
output.  Even so, as regards the coding of diagrams, there are very
few backwardly incompatible changes in axodraw2.

The software is available under the GNU General Public License
\cite{GPL} version 3.

%=========================
\section{Changes}
\label{sec:changes}

\subsection{Changes relative to original, axodraw version 1}
\label{sec:changes.wrt.1}

Relative to the original version of axodraw, the current version,
axodraw2, has the following main changes:
\begin{itemize}

\item A bug that the line bounding an oval did not have a uniform
  width has been corrected.

\item A bug has been corrected that axodraw did not work with the
  revtex4 document class when \verb+\maketitle+ and two-column mode
  were used.

\item Axodraw2 works both when pdf output is produced directly using
  the programs \program{pdflatex}, \program{lualatex}, and
  \program{xelatex}, as well as when a postscript file is produced by
  the latex--dvips method.  The old version only worked when
  postscript output was produced.  However, an auxiliary program is
  needed when using \program{pdflatex}, \program{lualatex}, or
  \program{xelatex}.  See Sec.\ \ref{sec:doc.compile} for how this is
  done.

\item In the original axodraw, a diagram is coded inside a
  \verb+picture+ environment of \LaTeX.  Now, a specialized
  \verb+axopicture+ environment is provided and preferred; it provides
  better behavior, especially when diagrams are to be scaled.

\item In association with this, there are some changes in how scaling
  of diagrams is done.

\item An inconsistency in length units between postscript and \TeX{}
  has been corrected.  All lengths are now specified in terms of
  $\unit[1]{pt} = \unit[1/72.27]{in} = \unit[0.3515]{mm}$.  Previously
  the unit length for graphics was the one defined by postscript to be 
  $\unit[1]{bp} = \unit[1/72]{in} = \unit[0.3528]{mm}$.

\item Substantial improvements have been made in the treatment of
  color.  When named colors are used, axodraw2's use of color is
  generally compatible with that of the modern, \LaTeX-standard
  \file{color.sty} package.  It also provides all the macros that were
  defined in v.\ 1 of axodraw, including those of the \file{colordvi.sty}
  package used by v.\ 1.

\item The various types of line can now be produced as double lines,
  e.g.,
  \begin{axopicture}(35,5)(0,-2)
    \SetWidth{1}
    \Line[double,sep=2.5](0,2)(35,2)
  \end{axopicture}.
  This is commonly used, for example, for notating Wilson lines. 

\item Lines can be made from B\'ezier curves.
  Currently this is only for simple lines, not photon, gluon, or
  zigzag lines.

\item Gluon, photon, and zigzag lines can be dashed.

\item Macros are provided for drawing gluon circles, without the
  endpoint effects given by the corresponding gluon arc macros.

\item The positions and sizes of arrows can be adjusted.  See Sec.\
  \ref{sec:arrows} for all the possibilities.  One example is 
  \begin{axopicture}(30,6)(0,-2)
    \SetWidth{1}
    \Line[arrow,arrowpos=0.8](0,2)(30,2)
  \end{axopicture}

\item Macros for drawing polygons and filled polygons are provided.

\item Macros for drawing rotated boxes are provided.

\item A macro \verb+\ECirc+ is provided for drawing a circle with a
  transparent interior.

\item A macro \verb+\EBoxc+ is provided for drawing a box with a
  specified center.

\item A macro \verb+\AxoGrid+ is provided for drawing a grid.  One
  use is to provide a useful tool in designing pictures.

\item Since there are now many more possibilities to specify the
  properties of a line, optional arguments to the main line drawing
  commands can be used to specify them in a keyword style.  

\item A new macro named \verb+\Arc+ is introduced.  With the aid of
  optional arguments, this unifies the behavior of various arc-drawing
  commands in the original axodraw.

\item For consistency with the \verb+\Gluon+ macro, the
  \verb+\GlueArc+ macro has been renamed to \verb+\GluonArc+, with the old
  macro retained as a synonym.

\item The behavior of arcs is changed to what we think is more natural
  behavior when the specified opening is outside the natural range.

\item What we call macros for drawing objects with postscript text are
  now implemented within \LaTeX{} instead of relying on instructions
  inserted in the postscript code.  Thus all the normal \LaTeX{}
  commands, including mathematics, can now be used in all text
  objects, with proper scaling.  The placement and scaling of text
  objects are more consistent. 

\item Some new named colors are provided:
  \LightYellow{LightYellow}, \LightRed{LightRed},
  \LightBlue{LightBlue}, \LightGray{LightGray},
  \VeryLightBlue{VeryLightBlue}.  
  (LightYellow, LightRed, LightBlue, LightGray, VeryLightBlue.)

\item The macros originally specified as \verb+\B2Text+,
  \verb+\G2Text+, and \verb+\C2Text+ are now named \verb+\BTwoText+,
  \verb+\GTwoText+, and \verb+\CTwoText+.  The intent of the
    original code was to define macros with names \verb+\B2Text+, etc.
    However in normal \TeX, macro names of more than one character
    must only contain letters, unlike typical programming languages
    that also allow digits.  So the rules for \TeX{} macro names mean
    that in defining, for example \verb+\def\B2Text(#1,#2)#3#4{...}+,
    the original version of axodraw actually defined a macro named
    named \verb+\B+, obligatorially followed by \verb+2Text+.  This
    caused a conflict if the user wished to define a macro \verb+\B+.
    If it is desired to retain the old behavior, then the following
    should be placed in the preamble of the .tex file, then the
    axodraw2 package should be invoked in the source document with the
    \texttt{v1compatible} option:
    \begin{verbatim}
          \usepackage[v1compatible]{axodraw2}
    \end{verbatim}

\end{itemize}

\subsection{Changes relative to axodraw4j distributed with JaxoDraw}
\label{sec:changes.wrt.4j}

The JaxoDraw program \cite{jaxodraw2} is distributed with a
version of axodraw called axodraw4j.  As of July 2014, this was
effectively a predecessor of axodraw2, but without the possibility of
working with \program{pdflatex}.  (The suffix ``4j'' is intended to mean ``for
JaxoDraw''.)

The changes in axodraw2 relative to the version of axodraw4j dated
2008/11/19 are the following subset of those listed in Sec.\
\ref{sec:changes.wrt.1}:
\begin{itemize}
\item Correction of the oval-drawing bug.
\item The ability to work with \program{pdflatex}, \program{lualatex},
  and  \program{xelatex}.
\item The improvements in the handling of color.
\item The double and arrow options for B\'ezier lines.
\item The dash option for gluons and photons.
\item Color option for all lines.
\item Correction of inconsistency of length unit between \TeX{} and
  postscript. 
\item Better drawing of double gluons and photons.
\item The gluon circle, polygon, rotated box, \verb+\ECirc+,
  \verb+\EBoxc+, and the \verb+\AxoGrid+ macros
\item A series of ``LongArrow'' macros for drawing lines with the
  arrow at the end.  The same effect could only be achieved in
  axodraw4j with arrowpos=1 option to the basic line-drawing
  commands.
\item A series of macros like \verb+\DashDoubleLine+ to provide access
  to the dashed and double properties in the style of the macros
  provided in v.\ 1 of axodraw.  This is in addition to the optional
  arguments that allow the same effect in axodraw4j and in axodraw2.
\item The \texttt{v1compatible} and other options are provided for the
  package.
\item Better treatment of the scaling of objects.
\item The treatment of ``postscript text objects'' within \LaTeX{}
  itself. 
\end{itemize}

%---------------
\subsection{Backward compatibility, etc}

The official user interface of axodraw2 is backward-compatible with
versions 1 and 4j, with the exception of the issue mentioned above
about the commands that have the signatures \verb+\B2Text+,
\verb+\G2Text+, and \verb+\C2Text+.  There are some minor changes in
the objects that are drawn, mostly concerning the exact dimensions of
default arrows and the scaling of the sizes of text objects.  The
scoping of color changes is significantly different, but improved.

The old axodraw only used the tools available in \LaTeX{} in the early
1990s.  The new version needs a more modern installation.  It has been
extensively tested with TeXLive 2011 and 2016.

We have tested backwards compatibility by compiling the version 1
manual with axodraw2; only a trivially modified preamble was needed.
It also worked to compile Collins's QCD book\cite{qcdbook},
which has a large number
of JaxoDraw figures (processed automatically to pieces of axodraw code
imported into the document); only changes in the preamble were needed.

Axodraw2 uses the following \LaTeX{} packages: \program{keyval},
\program{ifthen}, \program{graphicx}, \program{color}, \program{ifxetex}. 
It defines its own set of 73 named colors --- Sec.\ \ref{sec:colors}
--- which are the same as the 68 defined as dvips-defined names in the
color package, plus 5 more.

In addition axodraw2 provides an \verb+axopicture+ environment
inside of which axodraw2's graphics are coded and drawn.  In the old
axodraw, \LaTeX's \verb+picture+ environment was used instead.  We
recommend the use of \verb+axopicture+ environment in axodraw2, and
that is the only method we document.  However, old diagrams coded with
\verb+picture+ environment continue to work.

%=========================
\section{Installation}
\label{sec:installation}

%---------------
\subsection{Installation from standard \TeX{} distribution}

At the moment of writing, this document, axodraw2 is not part of any
standard \TeX{} distribution.

But it is planned to put it on CTAN such that it should
eventually be part of the standard distributions (TeXLive and
MiKTeX).  After that, axodraw2 will either be installed by default or
can be installed by using the package manager of the \TeX{}
distribution. When available, this will be the easiest method of
installation.

%---------------
\subsection{Manual installation}

%For a manual installation, the minimum that needs to be done is to put
For a manual installation, what needs to be done is to put
the file \file{axodraw2.sty} in a place where it will be found by
the \program{latex} program.  If you wish to use axodraw2 with
\program{pdflatex}, you will also need to compile the \program{axohelp}
program and put it in an appropriate directory.  Documentation can
also be installed if you want.

%--
\subsubsection{Style file texttt{axodraw2.sty}}

If you merely want to try out axodraw2, just put the file
\file{axodraw2.sty} in the same directory as the \file{.tex}
file(s) you are working on. 

Otherwise, put it in an appropriate directory for a \LaTeX{} style
file, and, if necessary, run the texhash program to ensure that the
file is in the \TeX{} system's database of files.  For example,
suppose that you have a TeXLive system installed for all users on a
Unix-like system (e.g., Linux or OS-X), and that TeXLive is installed,
as is usual, under the directory \file{/usr/local/texlive}.  Then an
appropriate place for axodraw2 is in a directory
\file{/usr/local/texlive/texmf-local/tex/latex/axodraw2}.  You will
need to run the \program{texhash} program in this last case.  For such
a system-wide installation, you will probably have to do these
operations as an administrative user (e.g., root), possibly
supplemented by running the relevant commands with the \program{sudo}
program.

%--
\subsubsection{Helper program \program{axohelp}}
\label{sec:axohelp}

If you wish to use axodraw2 with \program{pdflatex}, \program{lualatex},
or \program{xelatex}., then you need to install the \program{axohelp}
program.

On a Unix-like system (e.g., linux or OS-X), you first need to compile
the program by a C compiler.  An appropriate shell command to do this
is
\begin{verbatim}
   cc -o axohelp -O3 axohelp.c -lm
\end{verbatim}
(Note that this is a C compiler, \emph{not} a C++ compiler.)  Most linux
systems have the program \program{cc} already installed.  This also applies to
OS-X at versions below 10.7.  But on OS-X version 10.7 and higher, you
will need to install a compiler, which can be done by installing XCode
and the associated command-line utilities.  If you have the GNU
compilers installed, you might need to use the command \program{gcc}
instead of \program{cc}.

For Microsoft Windows, if you do not have a C compiler available, you
can use the Windows binary \file{axohelp.exe} we have provided.  It
should work with Windows 7 or higher.

In any case once you have the executable (named \program{axohelp} on
unix-like systems, or \program{axohelp.exe} on a Microsoft system), put
it in a directory where it will be found when you run programs from
the command line.

%--
\subsubsection{Testing}

To test whether the installation works, you need a simple test file.
An example is given in Sec.\ \ref{sec:example}, and is provided
with the axodraw2 distribution as \file{example.tex}.

At a command line with the current directory set to the directory
containing the file \file{example.tex}, run the following commands:
\begin{verbatim}
   latex example
   dvips example -o
\end{verbatim}
If all goes well, you will obtain a file \file{example.ps}.  When
you view it, it should contain the diagram shown in Sec.\
\ref{sec:example}.  You can make a pdf file instead by the commands
\begin{verbatim}
   latex example
   dvipdf example
\end{verbatim}
A more extensive test can be made by compiling the manual.

To make a pdf file directly, with \program{pdflatex}, you use the commands
\begin{verbatim}
   pdflatex example
   axohelp example
   pdflatex example
\end{verbatim}
The \program{axohelp} run takes as input a file \file{example.ax1}
produced by the first run of \program{pdflatex} and makes an output
file \file{example.ax2}.  The second run of \program{pdflatex} reads
the \file{example.ax2} file and uses the result to place the axodraw
objects in the \file{example.pdf} file.

%--
\subsubsection{Documentation}

Put the documentation in a place where you can find it.  If you
installed the \file{axodraw2.sty} file in
\file{/usr/local/texlive/texmf-local/tex/latex/axodraw2}, the 
standard place for the documentation would be
\file{usr/local/texlive/texmf-local/doc/latex/axodraw2}.

%=========================
\section{Use}
\label{sec:use}

In this section we show how to use axodraw2, illustrated with an
example.

\subsection{Basic example}
\label{sec:example}

The principles of using axodraw2 are illustrated by the following
complete \LaTeX{} document:
\begin{verbatim}
        \documentclass{article}
        \usepackage{axodraw2}
        \begin{document}
        Example of Feynman graph using axodraw2 macros:
        \begin{center}
          \begin{axopicture}(200,110)
            \SetColor{Red}
            \Arc[arrow](100,50)(40,0,180)
            \Text(100,100){$\alpha P_1 + \beta P_2 + k_\perp$}
            \SetColor{Black}
            \Arc[arrow](100,50)(40,180,360)
            \Gluon(0,50)(60,50){5}{4}
            \Vertex(60,50){2} 
            \Gluon(140,50)(200,50){5}{4}
            \Vertex(140,50){2}
          \end{axopicture}
        \end{center}
        \end{document}
\end{verbatim}
After compilation according to the instructions in Sec.\
\ref{sec:doc.compile}, viewing the resulting file should show the
following Feynman graph:
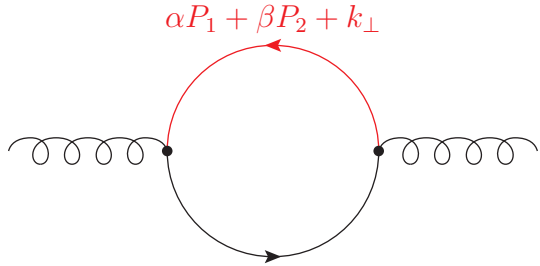
\begin{center}
  \begin{axopicture}(200,110)
    \SetColor{Red}
    \Arc[arrow](100,50)(40,0,180)
    \Text(100,100){$\alpha P_1 + \beta P_2 + k_\perp$}
    \SetColor{Black}
    \Arc[arrow](100,50)(40,180,360)
    \Gluon(0,50)(60,50){5}{4}
    \Vertex(60,50){2} 
    \Gluon(140,50)(200,50){5}{4}
    \Vertex(140,50){2}
  \end{axopicture}
\end{center}
See Sec.\ \ref{sec:examples} for more examples

\emph{Important note about visibility of graphics objects:} If you
view this document on a computer monitor, Feynman graphs drawn with
narrow lines may not fully match what was intended.  This is because
of the way graphics viewers interact with the limited resolution of
computer monitors. To see the example graphs properly, you may need to
use a large enough magnification, or to use an actual print out.

\emph{Note about sending a document to others}: If for example, you
submit an article to arXiv.org, it is likely that their automated
system for processing the file will not run axohelp. So together with
the tex file, you one should also submit the .ax2 file.

%-----------------------
\subsection{Document preparation}
\label{sec:doc.prep}

The general rules for preparation of a document are:
\begin{itemize}

\item Insert the following
   \begin{verbatim}
     \usepackage{axodraw2}
  \end{verbatim}
  in the preamble of the \file{.tex} file.
  There are some options and commands that can be used to change axodraw2's
  behavior from its default.  See Secs.\ \ref{sec:invoke} and
  \ref{sec:settings} for details. 

\item Where you want to insert axodraw2 objects, put them inside an
  axopicture environment, specified in Sec.\ \ref{sec:env},
  \begin{verbatim}
     \begin{axopicture}(x,y)
        ...
     \end{axopicture}
  \end{verbatim}
  Here \texttt{x} and \texttt{y} denote the desired size of the box
  that is to be inserted in the document and that contains the graph.
  An optional offset can be specified (as with \LaTeX's
  \texttt{picture} environment). By default the units are
  $\unit[1]{pt} = \unit[1/72.27]{in} = \unit[0.3515]{mm}$.

\end{itemize}
Full details of all these components are in Sec.\
\ref{sec:reference}. 

The design of graphs can be done manually, and this can be greatly
facilitated with the new \verb:\AxoGrid: command.  A convenient way of
constructing diagrams is to use the graphical program
JaxoDraw~\cite{jaxodraw1,jaxodraw2}, which is what most people
do. This program can export axodraw code.  It also uses axodraw as one
way of making postscript and pdf files.  The original version of
axodraw was used by JaxoDraw until version 1.3. In version 2 of
JaxoDraw, a specially adapted version of \file{axodraw.sty} is used,
named \file{axodraw4j.sty}. The output from version 2 of
JaxoDraw is compatible with axodraw2.

%-----------------------
\subsection{Document compilation}
\label{sec:doc.compile}

\subsubsection{To make a postscript file}
\label{sec:doc.compile.ps}

When a postscript file is needed, you just make the postscript file as
usual.  E.g., when the source file is \file{example.tex}, you run
the following commands:
\begin{verbatim}
   latex example
   dvips example -o
\end{verbatim}
which results in a postscript file \file{example.ps}.  Of course, if
there are cross references to be resolved, you may need multiple runs
of \program{latex}, as usual.  When needed, use of \program{bibtex},
\program{makeindex}, and other similar programs is also as usual.
Instead of \program{latex}, one may also use the \program{dvilualatex}
program, which behaves like \program{latex} except for providing some
extra capabilities that are sometimes useful.

Internally, axodraw uses \TeX's \verb+\special+ mechanism to put
specifications of postscript code into the \file{.dvi} file, and
\program{dvips} puts this code in the postscript file.  This postscript
code performs the geometrical calculations needed to specific
axodraw's objects, and then draws them when the file is displayed or
printed. 

\emph{Important note about configuration of \program{dvips}:} You may
possibly find that when you run \program{dvips} that it spends a lot of
time running \program{mktexpk} to make bitmapped fonts, or that the
postscript
file contains bitmapped type-3 fonts.  This is \emph{not} the default
situation in typical current installations.  But if you do find this
situation, which is highly undesirable in most circumstances, you
should arrange for \program{dvips} to use type 1 fonts.  This can be
done either by appropriately configuring your \TeX{} installation, for
which you will have to locate instructions, or by giving
\program{dvips} its \texttt{-V0} option:
\begin{verbatim}
   dvips -V0 example -o
\end{verbatim}
Once you do this, you should see, from \program{dvips}'s output,
symptoms of its use of type 1 fonts. \emph{Let us re-emphasize that
  you do not have to be concerned with this issue, under
  normal circumstances.  But since things were different within our
  memory, we give some suggestions as to what to do in what are
  currently abnormal circumstances.}

\subsubsection{To make a pdf file via \program{latex}}

There are multiple methods of making pdf files for a latex document;
we will not give all the advantages and disadvantages here.

One way is to convert the postscript file, e.g., by
\begin{verbatim}
   ps2pdf example.ps
\end{verbatim}
You can also produce a pdf file from the dvi file produced by
\program{latex} by the \program{dvipdf} command, e.g,.
\begin{verbatim}
   dvipdf example
\end{verbatim}
\emph{Important note:} The program here is \program{dvipdf} and
\emph{not} the similarly named \program{dvipdfm} or \program{dvipdfmx},
which are incompatible with axodraw.  The reason why \program{dvipdf}
works is that it internally makes a postscript file and then converts
it to pdf.

\subsubsection{To make a pdf file by \program{pdflatex},
  \program{lualatex}, or  \program{xelatex}}

A common and standard way to make a pdf file is the \program{pdflatex}
program, which makes pdf directly.  It has certain advantages, among
which are the possibility of importing a wide variety of graphics file
formats.  (In contrast, the \program{latex} program only handles
encapsulated postscript.)

However, to use axodraw2 with \program{pdflatex}, you need an 
auxiliary program, \program{axohelp}, as in
\begin{verbatim}
   pdflatex example
   axohelp example
   pdflatex example
\end{verbatim}
What happens is that during a run of \program{pdflatex}, axodraw2
%writes a file \file{example.ax1} with specifications of its
writes a file \file{example.ax1} containing specifications of its
graphical objects.  Then running \program{axohelp} reads the
%\file{example.ax1} file, computes the necessary pdf code to draw the
\file{example.ax1} file, computes the necessary pdf code to draw the
objects, and writes the results to \file{example.ax2}.  The next run
of \program{pdflatex} reads \file{example.ax2} and uses it to put the
appropriate code in the output pdf file.

The reason for the extra program is that axodraw needs many
geometrical calculations to place and draw its graphical objects.
\LaTeX{} itself does not provide anything convenient and efficient for
these calculations, while the PDF language does not offer sufficient
computational facilities, unlike the postscript language.

If you modify a document, and recompile with \program{pdflatex}, you
will only need to rerun \program{axohelp} if the modifications
involve axodraw objects.  Axodraw2 will output an appropriate message
when a rerun of \program{axohelp} is needed.

If you wish to use \program{lualatex} or \program{xelatex}, instead of
\program{pdflatex}, then you can simply run the program
\program{lualatex} or \program{xelatex} instead of
\program{pdflatex}. These are equally compatible with axodraw2.

%-----------------------
\subsection{Automation of document compilation}
\label{sec:doc.auto.compile}

It can be useful to automate the multiple steps for compiling a
\LaTeX{} document.  One of us has provided a program \program{latexmk}
to do this --- see \url{http://www.ctan.org/pkg/latexmk/}.  Here we
show how to configure 
\program{latexmk} to run \program{axohelp} as needed when a document is
compiled via the \program{pdflatex} route.

All you need to do is to put the following lines in one of
\program{latexmk}'s initialization files (as specified in its
documentation):
\begin{verbatim}
     add_cus_dep( "ax1", "ax2", 0, "axohelp" );
     sub axohelp { return system "axohelp \"$_[0]\""; }
     $clean_ext .= " %R.ax1 %R.ax2";
\end{verbatim}
The first two lines specify that \program{latexmk} is to make
\file{.ax2} files from \file{.ax1} files by the \program{axohelp}
program, whenever necessary.  (After that \program{latexmk}
automatically also does any further runs of \program{pdflatex} that are
necessary.)  The last line is optional; it adds \file{.ax1} and
\file{.ax2} files to the list of files that will be deleted when
\program{latexmk} is requested to do a clean up of generated,
recreatable files.

\program{Latexmk} is installed by default by the currently common
distributions of \TeX{} software, i.e., TeXLive and MiKTeX.  It has as
an additional requirement a properly installed Perl system.  For the
TeXLive distribution, this requirement is always met.

With the above configuration, you need no change in how you invoke
\program{latexmk} to compile a document, when it uses axodraw2.  For
producing postscript, you can simply use
\begin{verbatim}
     latexmk -ps example
\end{verbatim}
and for producing pdf via \program{pdflatex} you can use
\begin{verbatim}
     latexmk -pdf example
\end{verbatim}
Then \program{latexmk} takes care of whatever runs are needed of all
the relevant programs, now including \program{axohelp}, as well
whatever, possibly multiple, runs are needed for the usual programs
(\program{latex}, \program{pdflatex}, \program{bibtex}, etc).

%>>#] Introduction :
%>>#[ The Commands :

\section{Reference}
\label{sec:reference}

\subsection{Package invocation}
\label{sec:invoke}

To use the axodraw2 package in a \LaTeX{} document, you simply put
\begin{verbatim}
     \usepackage{axodraw2}
\end{verbatim}
in the preamble of the document, as normal.  

The \verb+\usepackage+ command takes optional arguments
(comma-separated list of keywords) in square brackets, e.g.,
\begin{verbatim}
     \usepackage[v1compatible]{axodraw2}
\end{verbatim}
The options supported by axodraw2 are
\begin{itemize}
\item \texttt{v1compatible}: This makes axodraw2's operation more
  compatible with v.\ 1.  It allows the use of \verb+\B2Text+,
  \verb+\G2Text+, and \verb+\C2Text+ as synonyms for the macros named
  \verb+\BTwoText+, \verb+\GTwoText+, and \verb+\CTwoText+.
  (You may wish also to use the \texttt{canvasScaleisUnitLength}
  option, so that the scaling of the units in the \texttt{axopicture}
  environment is the same as it was for the \texttt{picture}
  environment used in v.\ 1.)
\item \texttt{canvasScaleIs1pt}: Unit for canvas dimensions
  in an \texttt{axopicture} environment is fixed at $\unit[1]{pt}$,
\item \texttt{canvasScaleIsObjectScale}: Unit for canvas dimensions
  in an \texttt{axopicture} environment are the same as those set for
  axodraw objects (by the \verb+\SetScale+ macro).  This is the
  default setting, so the option need not be given.
\item \texttt{canvasScaleIsUnitLength}: Unit for canvas dimensions
  in an \texttt{axopicture} environment is the current value of
  \verb+\unitlength+, exactly as for \LaTeX{}'s \texttt{picture}
  environment.  (Thus, this corresponds to the behavior of the
  original axodraw v.\ 1, which simply used the \texttt{picture}
  environment.)
\item \texttt{PStextScalesIndependently}: Axodraw's text objects are
  scaled by the factor set by the \verb+\SetTextScale+ command.
\item \texttt{PStextScalesLikeGraphics}: Axodraw's text objects are
  scaled by the factor set by same factor for its graphics objects,
  i.e., the scale set by the \verb+\SetScale+ command.
\end{itemize}
(N.B. Default scaling factors are initialized to unity.)

\emph{Note:} If you use \program{axodraw}'s commands for placing text
and you use the standard \TeX{} Computer Modern fonts for the
document, then when you compile your document you may get a lot of
warning messages.  These are about fonts not being available in
certain sizes.  To fix this problem invoke the package
\program{fix-cm} in your document's preamble:
\begin{verbatim}
    \usepackage{fix-cm}
\end{verbatim}
It is also possible to use the package \program{lmodern} for the same
purpose.

\subsection{Environment(s)}
\label{sec:env}

The graphical and other objects made by axodraw2 are placed in an
\texttt{axopicture} environment, which is invoked either as
\begin{verbatim}
     \begin{axopicture}(x,y)
        ...
     \end{axopicture}
\end{verbatim}
or
\begin{verbatim}
     \begin{axopicture}(x,y)(xoffset,yoffset)
        ...
     \end{axopicture}
\end{verbatim}
Here, the \dots{} denote sequences of axodraw2 commands, as documented
in later sections, for drawing lines, etc.  The \texttt{axopicture}
environment is just like standard \LaTeX's \texttt{picture}
environment,\footnote{In fact, the \texttt{axopicture} is changed from
  the \texttt{picture} environment only by making some
  axodraw-specific settings. So the \texttt{picture} environment that
  was used in v.\ 1 may also be used with axodraw2; it merely has a
  lack of automation on the setting of the canvas scale relative to
  the object scale, and, in the future, other possible
  initializations.}, except for doing some axodraw-specific
initialization.  It inserts a region of size \texttt{x} by \texttt{y}
(with default units of $\unit[1]{pt} = \unit[1/72.27]{in} =
\unit[0.3515]{mm}$). Here \texttt{x} and \texttt{y} are set to the
numerical values you need.

The positioning of axodraw objects is specified by giving $x$ and $y$
coordinates, e.g., for the ends of lines.  The origin of these
coordinates is, by default, at the lower left corner of the box that
\texttt{axopicture} inserts in your document.  But sometimes,
particularly after editing a graph, you will find this is not
suitable.  To avoid changing a lot of coordinate values to get correct
placement, you can specify an offset by the optional arguments
\texttt{(xoffset,yoffset)} to the \texttt{axopicture} environment,
exactly as for \LaTeX's \texttt{picture} environment.  The offset
\texttt{(xoffset,yoffset)} denotes the position of the bottom left
corner of the box inserted in your document relative to the coordinate
system used for specifying object positions.  Thus
\begin{verbatim}
     \begin{axopicture}(20,20)
        \Line(0,0)(20,20)
     \end{axopicture}
\end{verbatim}
and 
\begin{verbatim}
     \begin{axopicture}(20,20)(-10,20)
        \Line(-10,20)(10,40)
     \end{axopicture}
\end{verbatim}
are exactly equivalent.

Within an \texttt{axopicture} environment, all the commands that can
be used inside an ordinary \texttt{picture} environment can also be
used. 

We can think of the \texttt{axopicture} environment as defining a
drawing canvas for axodraw's graphical and text objects.
There are possibilities for manipulating (separately) the units used
to specify the canvas and the objects.  These can be useful for
scaling a diagram or parts of it from an originally chosen design.
See Secs.\ \ref{sec:units} and \ref{sec:settings} for details.

\subsection{Graphics drawing commands}
\label{sec:commands}

In this section we present commands for drawing graphical objects,
split up by category.  Later, we will give: details of options to the
line-drawing commands, explanations of some details about specifying
gluons and about specifying arrow parameters, and then commands for
textual objects and for adjusting settings (e.g., separation in a
double line).  Mostly, we present the commands by means of examples.
Note that many of the arguments of the commands, notably arguments for
$(x,y)$ coordinate values are delimited by parentheses and commas
instead of the brace delimiters typically used in \LaTeX.

It should also be noted that some commands provide different ways of
performing the same task. For instance
\begin{verbatim}
   \BCirc(50,50){30}
\end{verbatim}
can also be represented by
\begin{verbatim}
   \CCirc(50,50){30}{Black}{White}
\end{verbatim}
when the current color is black. The presence of the BCirc command has been 
maintained both for backward compatibility, and because it represents
a convenient short hand for a common situation. This also holds for similar 
commands involving boxes and triangles. For the new Polygon, FilledPolygon, 
RotatedBox and FilledRotatedBox commands we have selected a more minimal 
scheme.

Similar remarks apply to the new feature of options for line drawing
commands. Originally in v.\ 1, a line with an arrow would be coded as
\begin{verbatim}
   \ArrowLine(30,65)(60,25)
\end{verbatim}
It is now also possible to code using the general \verb+\Line+ macro,
but with a keyword optional argument:
\begin{verbatim}
   \Line[arrow](30,65)(60,25)
\end{verbatim}
One advantage of the option method is a variety of other properties of
an individual line may also be coded, as in
\begin{verbatim}
   \Line[arrow,arrowpos=1](30,65)(60,25)
\end{verbatim}
without the need to use separate global setting for the property, by
the commands listed in Sec.\ \ref{sec:settings}, or by having a
corresponding compulsory argument to the command.
Which way to do things is a matter of user taste in particular
situations.

%--#[ AxoGrid :

\subsubsection{Grid drawing}

\noindent
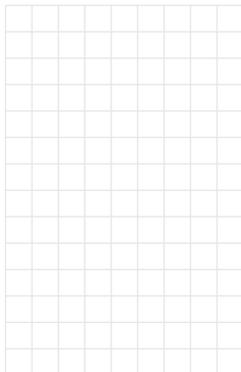
\begin{minipage}{3.83cm}
\begin{axopicture}{(90,140)(-10,0)}
\AxoGrid(0,0)(10,10)(9,14){LightGray}{0.5}
\end{axopicture}
\end{minipage}
\begin{minipage}{11.5cm}
\label{axogrid}
\verb:\AxoGrid(0,0)(10,10)(9,14){LightGray}{0.5}: \hfill \\
This command is used in our examples to allow the reader to compare the 
coordinates in the commands with those of the actual picture. The arguments 
are first the position of the left bottom corner, then two values that tell 
the size of the divisions in the $x$ and $y$ direction. Next there are two 
values that specify how many divisions there should be in the $x$ and $y$ 
direction. Then the color of the lines is given and finally the width of 
the lines. Note that if there are $(n_x,n_y)$ divisions there will be 
$n_x+1$ vertical lines and $n_y+1$ horizontal lines. The temporary use of 
this command can also be convenient when designing pictures manually.
\end{minipage}\vspace{4mm}

%--#] AxoGrid :
%--#[ Line :

\subsubsection{Ordinary straight lines}
\label{sec:Line}

All of the commands in this section can be given optional keyword
arguments, which are defined in Secs.\ \ref{sec:options} and
\ref{sec:arrows}.  These can be used to specify the type of line
(dashed, double), to specify the use of an arrow, and its parameters, and
to specify some of the line's parameters.

The basic line drawing command is \verb+\Line+:\\[3mm]
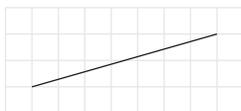
\begin{minipage}{3.83cm}
\begin{axopicture}{(90,40)(-10,0)}
\AxoGrid(0,0)(10,10)(9,4){LightGray}{0.5}
\Line(10,10)(80,30)
\end{axopicture}
\end{minipage}
\begin{minipage}{11.5cm}
\label{line}
\verb:\Line(10,10)(80,30): \hfill \\
In this command we have two coordinates. The (solid) line goes from the 
first to the second.
\end{minipage}\vspace{4mm}

Examples of the use of optional arguments are:\\[3mm]
\begin{minipage}{3.83cm}
\begin{axopicture}{(90,80)(-10,0)}
\AxoGrid(0,0)(10,10)(9,8){LightGray}{0.5}
\Line[color=Magenta,arrow](10,70)(80,70)
\Line[dash](10,50)(80,50)
\Line[arrow,double](10,30)(80,30)
\Line[arrow,dash,double](10,10)(80,10)
\end{axopicture}
\end{minipage}
\begin{minipage}{11.5cm}
\label{line.options}
\begin{verbatim}
\Line[color=Magenta,arrow](10,70)(80,70)
\Line[dash](10,50)(80,50)
\Line[arrow,double](10,30)(80,30)
\Line[arrow,dash,double](10,10)(80,10)
\end{verbatim}
\end{minipage}
\\[4mm]
Details of the specification of arrows, together with alternative
commands for making lines with arrows are given in Sec.\
\ref{sec:arrows}. 

\vspace{4mm}
%--#] Line :
%--#[ DoubleLine :

Alternative commands for dashed and/or double lines are:\\[3mm]
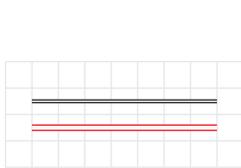
\noindent
\begin{minipage}{3.83cm}
\begin{axopicture}{(90,40)(-10,0)}
\AxoGrid(0,0)(10,10)(9,4){LightGray}{0.5}
\DoubleLine(10,25)(80,25){1}
\DoubleLine[color=Red](10,15)(80,15){2}
\end{axopicture}
\end{minipage}
\begin{minipage}{11.5cm}
\label{doubleline}
\verb:\DoubleLine(10,25)(80,25){1}: \hfill \\
\verb:\DoubleLine[color=Red](10,15)(80,15){2}: \hfill \\
In this command we have two coordinates as in the Line command but two 
lines are drawn. The extra parameter is the separation between the two 
lines. Note however that everything between the lines is blanked out.
\end{minipage}
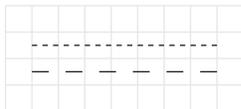
\vspace{4mm}

%--#] DoubleLine :
%--#[ DashLine :

\noindent
\begin{minipage}{3.83cm}
\begin{axopicture}{(90,40)(-10,0)}
\AxoGrid(0,0)(10,10)(9,4){LightGray}{0.5}
\DashLine(10,25)(80,25){2}
\DashLine(10,15)(80,15){6}
\end{axopicture}
\end{minipage}
\begin{minipage}{11.5cm}
\label{dashline}
\verb:\DashLine(10,25)(80,25){2}: \hfill \\
\verb:\DashLine(10,15)(80,15){6}: \hfill \\
In this command we have two coordinates. The dashed line goes from the 
first to the second. The extra parameter is the size of the dashes. The 
space between the dashes is transparent.
\end{minipage}
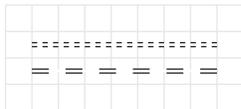
\vspace{4mm}

%--#] DashLine :
%--#[ DashDoubleLine :

\noindent
\begin{minipage}{3.83cm}
\begin{axopicture}{(90,40)(-10,0)}
\AxoGrid(0,0)(10,10)(9,4){LightGray}{0.5}
\DashDoubleLine(10,25)(80,25){1.5}{2}
\DashDoubleLine(10,15)(80,15){1.5}{6}
\end{axopicture}
\end{minipage}
\begin{minipage}{11.5cm}
\label{dashdoubleline}
\verb:\DashDoubleLine(10,25)(80,25){1.5}{2}: \hfill \\
\verb:\DashDoubleLine(10,15)(80,15){1.5}{6}: \hfill \\
In this command we have two coordinates. The dashed lines go from the 
first to the second. The first extra parameter is the separation between 
the lines and the second extra parameter is the size of the dashes.
\end{minipage}\vspace{4mm}

%--#] DashDoubleLine :
%--#[ Arc :

\subsubsection{Arcs}
\label{sec:Arc}

The commands in this section draw circular arcs in types corresponding
to the straight lines of Sec.\ \ref{sec:Line}.  In v.\ 1, some of
these commands had names containing ``Arc'' and some ``CArc''.  Some
kinds had variant names containing ``Arcn'', whose the direction of
drawing was clockwise instead of anticlockwise. In v.\ 2, we have
tried to make the situation more consistent.  First, all the old names
have been retained, for backward compatibility.  Second, a general
purpose command \verb+\Arc+ has been introduced; in a single command,
with the aid of optional arguments, it covers all the variants.  See
Secs.\ \ref{sec:options} and \ref{sec:arrows} for full details.  The
options can be used to specify the type of line (dashed, double,
clockwise or anticlockwise), to specify the use of arrow, and its
parameters, and to specify some of the line's parameters.  The other
commands in this section can also be given optional keyword arguments.

The basic \verb+\Arc+ command has the form\\[3mm]
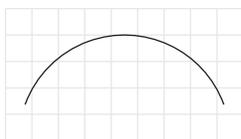
\noindent
\begin{minipage}{3.83cm}
\begin{axopicture}{(90,50)(-10,0)}
\AxoGrid(0,0)(10,10)(9,5){LightGray}{0.5}
\Arc(45,0)(40,20,160)
\end{axopicture}
\end{minipage}
\begin{minipage}{11.5cm}
\label{carc}
\verb:\Arc(45,0)(40,20,160):\hfill \\
In this command we have one coordinate: the center of the circle. Then 
follow the radius of the circle, the start angle and the finishing angle. 
The arc will be drawn counterclockwise.
\end{minipage}\vspace{4mm}

An example of the use of the optional parameters is:\\[3mm]
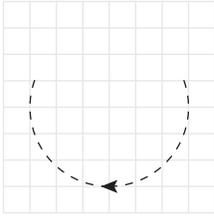
\begin{minipage}{3.83cm}
\begin{axopicture}{(80,80)(-10,0)}
\AxoGrid(0,0)(10,10)(8,8){LightGray}{0.5}
\Arc[arrow,dash,clockwise](40,40)(30,20,160)
\end{axopicture}
\end{minipage}
\begin{minipage}{11.5cm}
\label{carc.opt}
\verb:\Arc[arrow,dash,clockwise](40,40)(30,20,160):
\end{minipage}\vspace{4mm}

Alternative commands for dashed and/or double arcs are as follows.
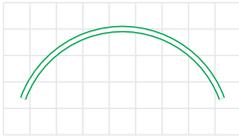
\vspace*{4mm}

%--#] Arc :
%--#[ DoubleArc :

\noindent
\begin{minipage}{3.83cm}
\begin{axopicture}{(90,50)(-10,0)}
\AxoGrid(0,0)(10,10)(9,5){LightGray}{0.5}
\DoubleArc[color=Green](45,0)(40,20,160){2}
\end{axopicture}
\end{minipage}
\begin{minipage}{11.5cm}
\label{doublearc}
\verb:\DoubleArc[color=Green](45,0)(40,20,160){2}:\hfill \\
In this command we have one coordinate: the center of the circle. Then 
follow the radius of the circle, the start angle and the finishing angle. 
The arc will be drawn counterclockwise. The last argument is the line 
separation of the double line.
\end{minipage}
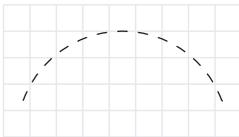
\vspace{4mm}

%--#] DoubleArc :
%--#[ DashArc :

\noindent
\begin{minipage}{3.83cm}
\begin{axopicture}{(90,50)(-10,0)}
\AxoGrid(0,0)(10,10)(9,5){LightGray}{0.5}
\DashArc(45,0)(40,20,160){4}
\end{axopicture}
\end{minipage}
\begin{minipage}{11.5cm}
\label{dasharc}
\verb:\DashArc(45,0)(40,20,160){4}:\hfill \\
In this command we have one coordinate: the center of the circle. Then 
follow the radius of the circle, the start angle and the finishing angle. 
The arc will be drawn counterclockwise. The last argument is the size of 
the dashes.
\end{minipage}
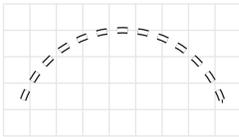
\vspace{4mm}

%--#] DashArc :
%--#[ DashDoubleArc :

\noindent
\begin{minipage}{3.83cm}
\begin{axopicture}{(90,50)(-10,0)}
\AxoGrid(0,0)(10,10)(9,5){LightGray}{0.5}
\DashDoubleArc(45,0)(40,20,160){2}{4}
\end{axopicture}
\end{minipage}
\begin{minipage}{11.5cm}
\label{dashdoublearc}
\verb:\DashDoubleArc(45,0)(40,20,160){2}{4}:\hfill \\
In this command we have one coordinate: the center of the circle. Then 
follow the radius of the circle, the start angle and the finishing angle. 
The arc will be drawn counterclockwise. The last two arguments are the line 
separation of the double line and the size of the dashes.
\end{minipage}\vspace{4mm}

%--#] DashDoubleArc :
%--#[ Bezier :

\subsubsection{B\'ezier lines}
\label{sec:Bezier}

The commands in this section draw B\'ezier curves, specified by 4
points.  The variants are just as for straight lines, Sec.\
\ref{sec:Line}.

All of the commands in this section can be given optional keyword
arguments, which are defined in Sec.\ \ref{sec:options}.  These can be
used to specify the type of line (dashed, double), to specify the use
of an arrow, and its parameters, and to specify some of the line's
parameters.

The basic general purpose command is \verb+\Bezier+:\\[3mm]
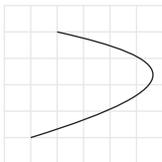
\noindent
\begin{minipage}{3.83cm}
\begin{axopicture}{(60,60)(-25,0)}
\AxoGrid(0,0)(10,10)(6,6){LightGray}{0.5}
\Bezier(10,10)(75,30)(65,40)(20,50)
\end{axopicture}
\end{minipage}
\begin{minipage}{11.5cm}
\label{bezier}
\verb:\Bezier(10,10)(75,30)(65,40)(20,50): \hfill \\
Draws a cubic B\'ezier curve based on the four given points. The first
point is the starting point and the fourth the finishing point. The
second and third points are the two control points.
\end{minipage}\vspace{4mm}

An example of the use of optional arguments is
\\[3mm]
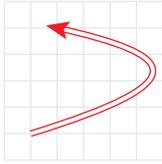
\begin{minipage}{3.83cm}
\begin{axopicture}{(60,60)(-25,0)}
\AxoGrid(0,0)(10,10)(6,6){LightGray}{0.5}
\Bezier[color=Red,arrow,double,arrowpos=1](10,10)%
    (75,30)(65,40)(20,50)
\end{axopicture}
\end{minipage}
\begin{minipage}{11.5cm}
\label{bezier.opt}
\begin{verbatim}
  \Bezier[color=Red,arrow,double,arrowpos=1](10,10)%
    (75,30)(65,40)(20,50)
\end{verbatim}
\end{minipage}\vspace{4mm}

%--#] Bezier :
%--#[ DoubleBezier :
Alternative ways of making dashed and/or double B\'ezier curves
are:\\[3mm]
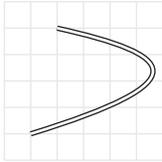
\noindent
\begin{minipage}{3.83cm}
\begin{axopicture}{(60,60)(-25,0)}
\AxoGrid(0,0)(10,10)(6,6){LightGray}{0.5}
\DoubleBezier(10,10)(75,30)(65,40)(20,50){1.5}
\end{axopicture}
\end{minipage}
\begin{minipage}{11.5cm}
\label{doublebezier}
\verb:\DoubleBezier(10,10)(75,30)(65,40)(20,50){1.5}: \hfill \\
Draws a cubic B\'ezier curve based on the four given points. 
The first four arguments are the same as for \verb+\Bezier+.
The final argument is the line separation.
\end{minipage}
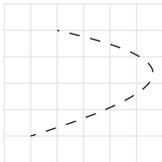
\vspace{4mm}

%--#] DoubleBezier :
%--#[ DashBezier :

\noindent
\begin{minipage}{3.83cm}
\begin{axopicture}{(60,60)(-25,0)}
\AxoGrid(0,0)(10,10)(6,6){LightGray}{0.5}
\DashBezier(10,10)(75,30)(65,40)(20,50){4}
\end{axopicture}
\end{minipage}
\begin{minipage}{11.5cm}
\label{dashbezier}
\verb:\DashBezier(10,10)(75,30)(65,40)(20,50){4}: \hfill \\
Draws a cubic B\'ezier curve based on the four given points. 
The first four arguments are the same as for \verb+\Bezier+.
The final argument is the size of the dashes.
\end{minipage}
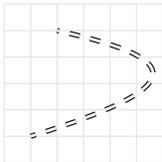
\vspace{4mm}

%--#] DashBezier :
%--#[ DashDoubleBezier :

\noindent
\begin{minipage}{3.83cm}
\begin{axopicture}{(60,60)(-25,0)}
\AxoGrid(0,0)(10,10)(6,6){LightGray}{0.5}
\DashDoubleBezier(10,10)(75,30)(65,40)(20,50){1.5}{4}
\end{axopicture}
\end{minipage}
\begin{minipage}{11.5cm}
\label{dashdoublebezier}
\verb:\DashDoubleBezier(10,10)(75,30)(65,40)(20,50){1.5}{4}:
Draws a cubic B\'ezier curve based on the four given points. 
The first four arguments are the same as for \verb+\Bezier+.
The final two arguments are the line separation and the size of the
dashes.
\end{minipage}\vspace{4mm}

%--#] DashDoubleBezier :
%--#[ Curve :

\subsubsection{Curves}

The commands in this section draw curves through an arbitrary sequence
of points.  They only exist in variants for continuous and dashed
lines.  No optional arguments are allowed.
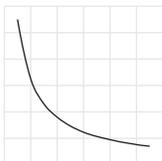
\vspace{4mm}

\noindent
\begin{minipage}{3.83cm}
\begin{axopicture}{(60,60)(-25,0)}
\AxoGrid(0,0)(10,10)(6,6){LightGray}{0.5}
\Curve{(5,55)(10,32.5)(15,23)(20,18)(25,14.65)(30,12.3)(40,9.5)(55,7)}
\end{axopicture}
\end{minipage}
\begin{minipage}{11.5cm}
\label{curve}
\verb:\Curve{(5,55)(10,32.5)(15,23)(20,18): \hfill \\
\verb:       (25,14.65)(30,12.3)(40,9.5)(55,7)}: \hfill \\
Draws a smooth curve through the given points. The $x$ coordinates of the 
points should be in ascending order. The curve is obtained by constructing 
quadratic fits to each triplet of adjacent points and then in each interval 
between two points interpolating between the two relevant parabolas.
\end{minipage}
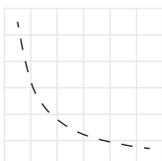
\vspace{4mm}

%--#] Curve :
%--#[ DashCurve :

\noindent
\begin{minipage}{3.83cm}
\begin{axopicture}{(60,60)(-25,0)}
\AxoGrid(0,0)(10,10)(6,6){LightGray}{0.5}
\DashCurve{(5,55)(10,32.5)(15,23)(20,18)(25,14.65)(30,12.3)(40,9.5)(55,7)}{4}
\end{axopicture}
\end{minipage}
\begin{minipage}{11.5cm}
\label{dashcurve}
\verb:\DashCurve{(5,55)(10,32.5)(15,23)(20,18): \hfill \\
\verb:       (25,14.65)(30,12.3)(40,9.5)(55,7)}{4}: \hfill \\
Draws a smooth dashed curve through the given points. The $x$ coordinates of 
the points should be in ascending order. The last argument is the size of 
the dashes.
\end{minipage}\vspace{4mm}

%--#] DashCurve :
%--#[ Gluon :

\subsubsection{Gluon lines}
\label{sec:Gluon}

The basic gluon drawing commands are \verb+\Gluon+, \verb+\GluonArc+,
\verb+\GluonCirc+.  There are also variants for dashed and double
gluons. But arrows aren't possible.

See Sec.\ \ref{sec:gluon.remarks} for additional information on the
shape of gluon lines.

All of the commands in this section can be given optional keyword
arguments, which are defined in Sec.\ \ref{sec:options}.  These can be
used to specify the type of line (dashed, double), and to specify some
of the line's parameters.
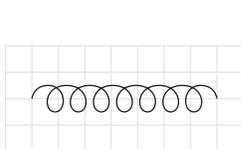
\vspace{3mm}

\noindent
\begin{minipage}{3.83cm}
\begin{axopicture}{(90,40)(-10,0)}
\AxoGrid(0,0)(10,10)(9,4){LightGray}{0.5}
\Gluon(10,20)(80,20){5}{7}
\end{axopicture}
\end{minipage}
\begin{minipage}{11.5cm}
\label{gluon}
\verb:\Gluon(10,20)(80,20){5}{7}: \hfill \\
In this command we have coordinates for the start and end of the line,
the amplitude of the windings and the number of windings.  A negative
value for the amplitude reverses the orientation of the windings ---
see Sec.\ \ref{sec:gluon.remarks} for details.
\end{minipage}
\\[4mm]
Optional arguments can be used, e.g., \hfill \\[3mm]
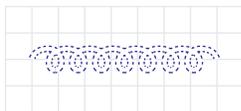
\noindent
\begin{minipage}{3.83cm}
\begin{axopicture}{(90,40)(-10,0)}
\AxoGrid(0,0)(10,10)(9,4){LightGray}{0.5}
\Gluon[color=Blue,dash,dashsize=1,double](10,20)(80,20){4}{7}
\end{axopicture}
\end{minipage}
\begin{minipage}{11.5cm}
\label{gluon.opt}
\verb:\Gluon[color=Blue,dash,double](10,20)(80,20){4}{7}:
\end{minipage}

\vspace{4mm}

%--#] Gluon :
%--#[ DoubleGluon :
\noindent
Examples of the other commands for various types of gluon line are as
follows. They can all take optional arguments.
\\[3mm]
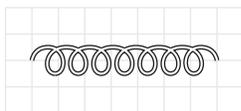
\begin{minipage}{3.83cm}
\begin{axopicture}{(90,40)(-10,0)}
\AxoGrid(0,0)(10,10)(9,4){LightGray}{0.5}
\DoubleGluon(10,20)(80,20){5}{7}{1.3}
\end{axopicture}
\end{minipage}
\begin{minipage}{11.5cm}
\label{doublegluon}
\verb:\DoubleGluon(10,20)(80,20){5}{7}{1.3}:\hfill \\
The first 6 arguments are as in the \verb+\Gluon+ command. The
extra argument is the line separation.
\end{minipage}
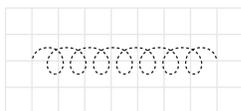
\vspace{4mm}

%--#] DoubleGluon :
%--#[ DashGluon :

\noindent
\begin{minipage}{3.83cm}
\begin{axopicture}{(90,40)(-10,0)}
\AxoGrid(0,0)(10,10)(9,4){LightGray}{0.5}
\DashGluon(10,20)(80,20){5}{7}{1}
\end{axopicture}
\end{minipage}
\begin{minipage}{11.5cm}
\label{dashgluon}
\verb:\DashGluon(10,20)(80,20){5}{7}{1}:\hfill \\
The first 6 arguments are as in the \verb+Gluon+ command. The 
extra argument is the size of the dashes.
\end{minipage}
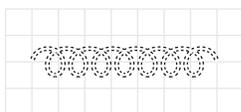
\vspace{4mm}

%--#] DashGluon :
%--#[ DashDoubleGluon :

\noindent
\begin{minipage}{3.83cm}
\begin{axopicture}{(90,40)(-10,0)}
\AxoGrid(0,0)(10,10)(9,4){LightGray}{0.5}
\DashDoubleGluon(10,20)(80,20){5}{7}{1.3}{1}
\end{axopicture}
\end{minipage}
\begin{minipage}{11.5cm}
\label{dashdoublegluon}
\verb:\DashDoubleGluon(10,20)(80,20){5}{7}{1.3}{1}:\hfill \\
The first 7 arguments are as in the \verb+DoubleGluon+
command.
The last two arguments are the line 
separation of the double line and the size of the dashes.
\end{minipage}
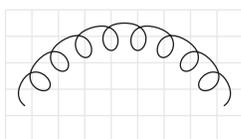
\vspace{8mm}

%--#] DashDoubleGluon :
%--#[ GluonArc :

\noindent
\begin{minipage}{3.83cm}
\begin{axopicture}{(90,50)(-10,0)}
\AxoGrid(0,0)(10,10)(9,5){LightGray}{0.5}
\GluonArc(45,0)(40,20,160){5}{8}
\end{axopicture}
\end{minipage}
\begin{minipage}{11.5cm}
\label{gluonarc}
\verb:\GluonArc(45,0)(40,20,160){5}{8}:\hfill \\
In this command we have one coordinate: the center of the circle. Then 
follow the radius of the circle, the start angle and the finishing angle. 
The arc will be drawn counterclockwise. The final two parameters are the 
amplitude of the windings and the number of windings.
Like the other commands in this section, this command can take
optional arguments, Sec.\ \ref{sec:options}.
\end{minipage}
\vspace{4mm}

%--#] GluonArc :
%--#[ DoubleGluonArc :

\noindent
\begin{minipage}{3.83cm}
\begin{axopicture}{(90,50)(-10,0)}
\AxoGrid(0,0)(10,10)(9,5){LightGray}{0.5}
   \DoubleGluonArc[color=Red](45,0)(40,20,160)%
                             {5}{8}{1.3}
\end{axopicture}
\end{minipage}
\begin{minipage}{11.5cm}
\label{doublegluonarc}
\begin{verbatim}
   \DoubleGluonArc[color=Red](45,0)(40,20,160)%
                             {5}{8}{1.3}
\end{verbatim}
The first 7 arguments are as in the \verb+GluonArc+ command. The extra
argument is the separation in the double line.
\end{minipage}
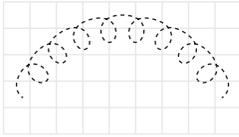
\vspace{4mm}

%--#] DoubleGluonArc :
%--#[ DashGluonArc :

\noindent
\begin{minipage}{3.83cm}
\begin{axopicture}{(90,50)(-10,0)}
\AxoGrid(0,0)(10,10)(9,5){LightGray}{0.5}
\DashGluonArc(45,0)(40,20,160){5}{8}{1.5}
\end{axopicture}
\end{minipage}
\begin{minipage}{11.5cm}
\label{dashgluonarc}
\verb:\DashGluonArc(45,0)(40,20,160){5}{8}{1.5}:\hfill \\
The first 7 arguments are as in the \verb+GluonArc+ command. The extra
argument is the size of the dash segments.
\end{minipage}
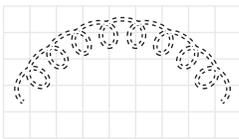
\vspace{4mm}

%--#] DashGluonArc :
%--#[ DashDoubleGluonArc :

\noindent
\begin{minipage}{3.83cm}
\begin{axopicture}{(90,50)(-10,0)}
\AxoGrid(0,0)(10,10)(9,5){LightGray}{0.5}
\DashDoubleGluonArc(45,0)(40,20,160){5}{8}{1.3}{1.5}
\end{axopicture}
\end{minipage}
\begin{minipage}{11.5cm}
\label{dashdoublegluonarc}
\verb:\DashDoubleGluonArc(45,0)(40,20,160){5}{8}{1.3}{1.5}:\hfill \\
The first 7 arguments are as in the \verb+GluonArc+ command. The extra
arguments are the separation of the lines and the size of the dash
segments.
\end{minipage}
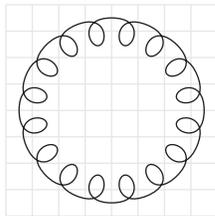
\vspace{10mm}

%--#] DashDoubleGluonArc :
%--#[ GluonCirc :

\noindent
\begin{minipage}{3.83cm}
\begin{axopicture}{(80,80)(-15,0)}
\AxoGrid(0,0)(10,10)(8,8){LightGray}{0.5}
\GluonCirc(40,40)(30,0){5}{16}
\end{axopicture}
\end{minipage}
\begin{minipage}{11.5cm}
\label{gluoncirc}
\verb:\GluonCirc(40,40)(30,0){5}{16}:\hfill \\
The arguments are: Coordinates for the center of the circle, the
radius and a phase, the 
amplitude of the gluon windings and the number of windings.
Like the other commands in this section, this command can take
optional arguments, Sec.\ \ref{sec:options}.  The phase argument
specifies a counterclockwise rotation of the line relative to a
default starting point.
\end{minipage}
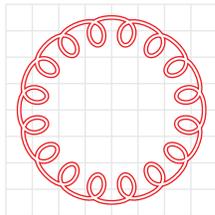
\vspace{4mm}

%--#] GluonCirc :
%--#[ DoubleGluonCirc :

\noindent
\begin{minipage}{3.83cm}
\begin{axopicture}{(80,80)(-15,0)}
\AxoGrid(0,0)(10,10)(8,8){LightGray}{0.5}
\DoubleGluonCirc[color=Red](40,40)(30,0){5}{16}{1.3}
\end{axopicture}
\end{minipage}
\begin{minipage}{11.5cm}
\label{doublegluoncirc}
\verb:\DoubleGluonCirc[color=Red](40,40)(30,0){5}{16}{1.3}:\hfill \\
The first 6 arguments are as for the \verb+GluonCirc+ command.  The
final argument is the line separation.
\end{minipage}
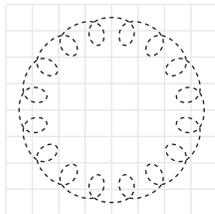
\vspace{4mm}

%--#] DoubleGluonCirc :
%--#[ DashGluonCirc :

\noindent
\begin{minipage}{3.83cm}
\begin{axopicture}{(80,80)(-15,0)}
\AxoGrid(0,0)(10,10)(8,8){LightGray}{0.5}
\DashGluonCirc(40,40)(30,0){5}{16}{1.5}
\end{axopicture}
\end{minipage}
\begin{minipage}{11.5cm}
\label{dashgluoncirc}
\verb:\DashGluonCirc(40,40)(30,0){5}{16}{1.5}:\hfill \\
The first 6 arguments are as for the \verb+GluonCirc+ command.  
The final argument is the size of the dashes.
\end{minipage}
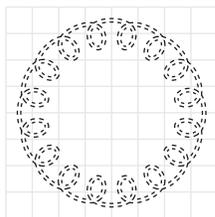
\vspace{4mm}

%--#] DashGluonCirc :
%--#[ DashDoubleGluonCirc :

\noindent
\begin{minipage}{3.83cm}
\begin{axopicture}{(80,80)(-15,0)}
\AxoGrid(0,0)(10,10)(8,8){LightGray}{0.5}
\DashDoubleGluonCirc(40,40)(30,0){5}{16}{1.3}{1.5}
\end{axopicture}
\end{minipage}
\begin{minipage}{11.5cm}
\label{dashdoublegluoncirc}
\verb:\DashDoubleGluonCirc(40,40)(30,0){5}{16}{1.3}{1.5}:\hfill \\
The first 6 arguments are as for the \verb+GluonCirc+ command.  
The final 2 arguments are the line separation and the size of the
dashes.
\end{minipage}\vspace{4mm}

%--#] DashDoubleGluonCirc :
%--#[ Photon :

\subsubsection{Photon lines}
\label{sec:Photon}

The basic drawing commands for drawing photon lines are \verb+\Photon+
and \verb+\PhotonArc+.  There are also variants for dashed and double
photons. But arrows aren't possible.

All of the commands in this section can be given optional keyword
arguments, which are defined in Sec.\ \ref{sec:options}.  These can be
used to specify the type of line (dashed, double), and to specify some
of the line's parameters.
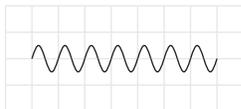
\vspace{3mm}

\noindent
\begin{minipage}{3.83cm}
\begin{axopicture}{(90,40)(-10,0)}
\AxoGrid(0,0)(10,10)(9,4){LightGray}{0.5}
\Photon(10,20)(80,20){5}{7}
\end{axopicture}
\end{minipage}
\begin{minipage}{11.5cm}
\label{photon}
\verb:\Photon(10,20)(80,20){5}{7}: \hfill \\
In this command we have two coordinates, the amplitude of the wiggles and 
the number of wiggles.
A negative value for the amplitude will reverse the orientation of the
wiggles.
The line will be drawn with the number of wiggles rounded to the
nearest half integer.
Like the other commands in this section, this command can take
optional arguments, Sec.\ \ref{sec:options}.
\end{minipage}
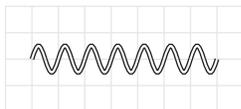
\vspace{4mm}

%--#] Photon :
%--#[ DoublePhoton :

\noindent
\begin{minipage}{3.83cm}
\begin{axopicture}{(90,40)(-10,0)}
\AxoGrid(0,0)(10,10)(9,4){LightGray}{0.5}
\DoublePhoton(10,20)(80,20){5}{7}{1.3}
\end{axopicture}
\end{minipage}
\begin{minipage}{11.5cm}
\label{doublephoton}
\verb:\DoublePhoton(10,20)(80,20){5}{7}{1.3}:\hfill \\
The first 6 arguments are as in the \verb+Photon+ command. The 
extra argument is the line separation.
\end{minipage}
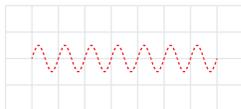
\vspace{4mm}

%--#] DoublePhoton :
%--#[ DashPhoton :

\noindent
\begin{minipage}{3.83cm}
\begin{axopicture}{(90,40)(-10,0)}
\AxoGrid(0,0)(10,10)(9,4){LightGray}{0.5}
\DashPhoton[color=Red](10,20)(80,20){5}{7}{1}
\end{axopicture}
\end{minipage}
\begin{minipage}{11.5cm}
\label{dashphoton}
\verb:\DashPhoton[color=Red](10,20)(80,20){5}{7}{1}:\hfill \\
The first 6 arguments are as in the \verb+Photon+ command. The 
extra argument is the size of the dashes.
\end{minipage}
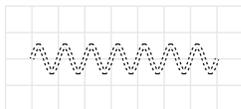
\vspace{4mm}

%--#] DashPhoton :
%--#[ DashDoublePhoton :

\noindent
\begin{minipage}{3.83cm}
\begin{axopicture}{(90,40)(-10,0)}
\AxoGrid(0,0)(10,10)(9,4){LightGray}{0.5}
\DashDoublePhoton(10,20)(80,20){5}{7}{1.3}{1}
\end{axopicture}
\end{minipage}
\begin{minipage}{11.5cm}
\label{dashdoublephoton}
\verb:\DashDoublePhoton(10,20)(80,20){5}{7}{1.3}{1}:\hfill \\
The first 6 arguments are as in the \verb+Photon+ 
command. 
The final 2 arguments are the line separation and the size of the
dashes.
\end{minipage}
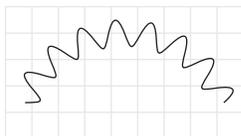
\vspace{10mm}

%--#] DashDoublePhoton :
%--#[ PhotonArc :

\noindent
\begin{minipage}{3.83cm}
\begin{axopicture}{(90,50)(-10,0)}
\AxoGrid(0,0)(10,10)(9,5){LightGray}{0.5}
\PhotonArc(45,0)(40,20,160){5}{8}
\end{axopicture}
\end{minipage}
\begin{minipage}{11.5cm}
\label{photonarc}
\verb:\PhotonArc(45,0)(40,20,160){5}{8}:\hfill \\
In this command we have one coordinate: the center of the circle. Then 
follow the radius of the circle, the start angle and the finishing angle. 
The arc will be drawn counterclockwise. The final two parameters are the 
amplitude of the wiggles and the number of wiggles.
Like the other commands in this section, this command can take
optional arguments, Sec.\ \ref{sec:options}.
\end{minipage}
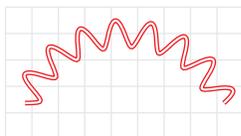
\vspace{4mm}

%--#] PhotonArc :
%--#[ DoublePhotonArc :

\noindent
\begin{minipage}{3.83cm}
\begin{axopicture}{(90,50)(-10,0)}
\AxoGrid(0,0)(10,10)(9,5){LightGray}{0.5}
\DoublePhotonArc[color=Red](45,0)(40,20,160)%
                           {5}{8}{1.3}
\end{axopicture}
\end{minipage}
\begin{minipage}{11.5cm}
\label{doublephotonarc}
\begin{verbatim}
\DoublePhotonArc[color=Red](45,0)(40,20,160)%
                           {5}{8}{1.3}
\end{verbatim}
The first 7 arguments are as in the \verb+PhotonArc+ command. The extra
argument is the separation of the double line.
\end{minipage}
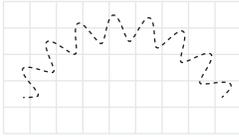
\vspace{4mm}

%--#] DoublePhotonArc :
%--#[ DashPhotonArc :

\noindent
\begin{minipage}{3.83cm}
\begin{axopicture}{(90,50)(-10,0)}
\AxoGrid(0,0)(10,10)(9,5){LightGray}{0.5}
\DashPhotonArc(45,0)(40,20,160){5}{8}{1.5}
\end{axopicture}
\end{minipage}
\begin{minipage}{11.5cm}
\label{dashphotonarc}
\verb:\DashPhotonArc(45,0)(40,20,160){5}{8}{1.5}:\hfill \\
The first 7 arguments are as in the \verb+PhotonArc+ command. The
extra argument is the size of the dash segments.
\end{minipage}
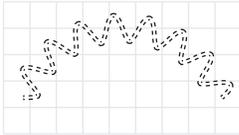
\vspace{4mm}

%--#] DashPhotonArc :
%--#[ DashDoublePhotonArc :

\noindent
\begin{minipage}{3.83cm}
\begin{axopicture}{(90,50)(-10,0)}
\AxoGrid(0,0)(10,10)(9,5){LightGray}{0.5}
\DashDoublePhotonArc(45,0)(40,20,160){5}{8}{1.3}{1.5}
\end{axopicture}
\end{minipage}
\begin{minipage}{11.5cm}
\label{dashdoublephotonarc}
\verb:\DashDoublePhotonArc(45,0)(40,20,160){5}{8}{1.3}{1.5}:\hfill \\
The first 7 arguments are as in the \verb+PhotonArc+ command. The
extra arguments are the separation of the lines and the size of the
dash segments.
\end{minipage}\vspace{4mm}

%--#] DashDoublePhotonArc :
%--#[ ZigZag :

\subsubsection{Zigzag lines}

The basic drawing commands for drawing zigzag lines are \verb+\Zigzag+
and \verb+\ZigzagArc+.  There are also variants for dashed and double
lines. But arrows aren't possible.

All of the commands in this section can be given optional keyword
arguments, which are defined in Sec.\ \ref{sec:options}.  These can be
used to specify the type of line (dashed, double), and to specify some
of the line's parameters.
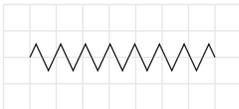
\vspace{4mm}

\noindent
\begin{minipage}{3.83cm}
\begin{axopicture}{(90,40)(-10,0)}
\AxoGrid(0,0)(10,10)(9,4){LightGray}{0.5}
\ZigZag(10,20)(80,20){5}{7.5}
\end{axopicture}
\end{minipage}
\begin{minipage}{11.5cm}
\label{zigzag}
\verb:\ZigZag(10,20)(80,20){5}{7.5}: \hfill \\
In this command we have two coordinates, the amplitude of the sawteeth and 
the number of sawteeth.
A negative value for the amplitude will reverse the orientation of the
sawteeth.
The line will be drawn with the number of sawteeth rounded to the
nearest half integer.
\end{minipage}
\\[3mm]
Like the other commands in this section, this command can take
optional arguments, Sec.\ \ref{sec:options}, e.g.,\\[3mm]
\noindent
\begin{minipage}{3.83cm}
\begin{axopicture}{(90,40)(-10,0)}
\AxoGrid(0,0)(10,10)(9,4){LightGray}{0.5}
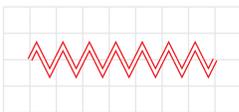
\ZigZag[color=Red,double,sep=1.5](10,20)(80,20){5}{7}
\end{axopicture}
\end{minipage}
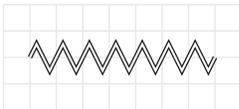
\begin{minipage}{11.5cm}
\label{zigzag.opt}
\verb:\ZigZag[color=Red,double,sep=1.5](10,20)(80,20){5}{7}:
\end{minipage}\vspace{6mm}

%--#] ZigZag :
%--#[ DoubleZigZag :

\noindent
\begin{minipage}{3.83cm}
\begin{axopicture}{(90,40)(-10,0)}
\AxoGrid(0,0)(10,10)(9,4){LightGray}{0.5}
\DoubleZigZag(10,20)(80,20){5}{7}{1.3}
\end{axopicture}
\end{minipage}
\begin{minipage}{11.5cm}
\label{doublezigzag}
\verb:\DoubleZigZag(10,20)(80,20){5}{7}{1.3}:\hfill \\
The first 6 arguments are as in the \verb+ZigZag+ command. The 
extra argument is the line separation.
\end{minipage}
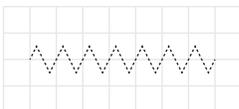
\vspace{4mm}

%--#] DoubleZigZag :
%--#[ DashZigZag :

\noindent
\begin{minipage}{3.83cm}
\begin{axopicture}{(90,40)(-10,0)}
\AxoGrid(0,0)(10,10)(9,4){LightGray}{0.5}
\DashZigZag(10,20)(80,20){5}{7}{1}
\end{axopicture}
\end{minipage}
\begin{minipage}{11.5cm}
\label{dashzigzag}
\verb:\DashZigZag(10,20)(80,20){5}{7}{1}:\hfill \\
The first 6 arguments are as in the \verb+ZigZag+ command. The 
extra argument is the size of the dashes.
\end{minipage}
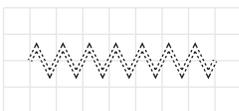
\vspace{4mm}

%--#] DashZigZag :
%--#[ DashDoubleZigZag :

\noindent
\begin{minipage}{3.83cm}
\begin{axopicture}{(90,40)(-10,0)}
\AxoGrid(0,0)(10,10)(9,4){LightGray}{0.5}
\DashDoubleZigZag(10,20)(80,20){5}{7}{1.3}{1}
\end{axopicture}
\end{minipage}
\begin{minipage}{11.5cm}
\label{dashdoublezigzag}
\verb:\DashDoubleZigZag(10,20)(80,20){5}{7}{1.3}{1}:\hfill \\
The first 6 arguments are as in the \verb+ZigZag+ command. 
The extra arguments are the separation of the lines and the size of
the dash segments.
\end{minipage}\vspace{6mm}

%--#] DashDoubleZigZag :
%--#[ ZigZagArc :

\noindent
\begin{minipage}{3.83cm}
\begin{axopicture}{(90,50)(-10,0)}
\AxoGrid(0,0)(10,10)(9,5){LightGray}{0.5}
\ZigZagArc(45,0)(40,20,160){5}{8}
\end{axopicture}
\end{minipage}
\begin{minipage}{11.5cm}
\label{zigzagarc}
\verb:\ZigZagArc(45,0)(40,20,160){5}{8}:\hfill \\
In this command we have one coordinate: the center of the circle. Then
follow the radius of the circle, the start angle and the finishing
angle.  The arc will be drawn counterclockwise. The final two
arguments are the amplitude of the sawteeth and the number of
sawteeth.  Like the other commands in this section, this command can
take optional arguments, Sec.\ \ref{sec:options}.
\end{minipage}\vspace{4mm}

%--#] ZigZagArc :
%--#[ DoubleZigZagArc :

\noindent
\begin{minipage}{3.83cm}
\begin{axopicture}{(90,50)(-10,0)}
\AxoGrid(0,0)(10,10)(9,5){LightGray}{0.5}
\DoubleZigZagArc(45,0)(40,20,160){5}{8}{1.3}
\end{axopicture}
\end{minipage}
\begin{minipage}{11.5cm}
\label{doublezigzagarc}
\verb:\DoubleZigZagArc(45,0)(40,20,160){5}{8}{1.3}:\hfill \\
The first 7 arguments are as for the \verb+ZigZagArc+ command. The
extra argument is the separation in the double line.
\end{minipage}\vspace{4mm}

%--#] DoubleZigZagArc :
%--#[ DashZigZagArc :

\noindent
\begin{minipage}{3.83cm}
\begin{axopicture}{(90,50)(-10,0)}
\AxoGrid(0,0)(10,10)(9,5){LightGray}{0.5}
\DashZigZagArc(45,0)(40,20,160){5}{8}{1.5}
\end{axopicture}
\end{minipage}
\begin{minipage}{11.5cm}
\label{dashzigzagarc}
\verb:\DashZigZagArc(45,0)(40,20,160){5}{8}{1.5}:\hfill \\
The first 7 arguments are as for the \verb+ZigZagArc+ command. The
extra argument is the size of the dash segments.
\end{minipage}\vspace{4mm}

%--#] DashZigZagArc :
%--#[ DashDoubleZigZagArc :

\noindent
\begin{minipage}{3.83cm}
\begin{axopicture}{(90,50)(-10,0)}
\AxoGrid(0,0)(10,10)(9,5){LightGray}{0.5}
\DashDoubleZigZagArc(45,0)(40,20,160){5}{8}{1.3}{1.5}
\end{axopicture}
\end{minipage}
\begin{minipage}{11.5cm}
\label{dashdoublezigzagarc}
\verb:\DashDoubleZigZagArc(45,0)(40,20,160){5}{8}{1.3}{1.5}:\hfill \\
The first 7 arguments are as for the \verb+ZigZagArc+ command. The
final 2 arguments are the separation of the lines and the size of the
dash segments.
\end{minipage}\vspace{4mm}

%--#] DashDoubleZigZagArc :
%--#[ Vertex :

\subsubsection{Vertices,  circles, ovals}
\label{sec:other.graphics}

The commands in this section are for graphical elements other
than those that we conceived of as lines in Feynman graphs.  Many of
these have standard uses as components of Feynman graphs\footnote{Of
  course, none of the commands is restricted to its originally
  envisaged use, or to being used to draw Feynman graphs.  But
  especially the line-drawing commands have been designed from the
  point-of-view of being suitable for the needs of drawing particular
  elements of Feynman graphs.}.  The commands here are mostly shown
in association with other objects, to indicate some of their
properties.
\vspace{4mm}

\noindent
\begin{minipage}{3.83cm}
\begin{axopicture}{(80,50)(-15,0)}
\AxoGrid(0,0)(10,10)(8,5){LightGray}{0.5}
\Line(10,10)(70,10)
\Photon(40,10)(40,40){4}{3}
\Vertex(40,10){1.5}
\end{axopicture}
\end{minipage}
\begin{minipage}{11.5cm}
\label{vertex}
\verb:\Line(10,10)(70,10): \hfill \\
\verb:\Photon(40,10)(40,40){4}{3}: \hfill \\
\verb:\Vertex(40,10){1.5}: \hfill \\
\verb+\Vertex+ gives a vertex, as is often used for connecting lines
in Feynman graphs. It gives a fat dot. The arguments are coordinates
(between parentheses) for its center, and the radius of the dot.
\end{minipage}
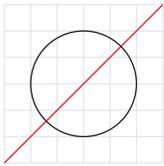
\vspace{4mm}

%--#] Vertex :
%--#[ ECirc :

\noindent
\begin{minipage}{3.83cm}
\begin{axopicture}{(60,60)(-25,0)}
\AxoGrid(0,0)(10,10)(6,6){LightGray}{0.5}
\Red{\Line(0,0)(60,60)}
\ECirc(30,30){20}
\end{axopicture}
\end{minipage}
\begin{minipage}{11.5cm}
\label{ecirc}
\verb:\Red{\Line(0,0)(60,60)}:\\
\verb:\ECirc(30,30){20}:\\
\verb+\ECirc+ draws a circle with its center at the specified
coordinate (first two arguments) and the specified radius (third
argument).  The interior is transparent, so that it does not erase
previously drawn material.
If you need a filled circle, use the \verb+\Vertex+ command (to which
we have defined a synonym \verb+\FCirc+ to match similar commands for
other shapes).
\end{minipage}
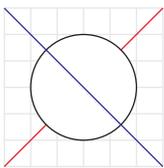
\vspace{4mm}

%--#] ECirc :
%--#[ BCirc :

\noindent
\begin{minipage}{3.83cm}
\begin{axopicture}{(60,60)(-25,0)}
\AxoGrid(0,0)(10,10)(6,6){LightGray}{0.5}
\Red{\Line(0,0)(60,60)}
\BCirc(30,30){20}
\Blue{\Line(60,0)(0,60)}
\end{axopicture}
\end{minipage}
\begin{minipage}{11.5cm}
\label{bcirc}
\verb:\Red{\Line(0,0)(60,60)}:\\
\verb:\BCirc(30,30){20}:\\
\verb:\Blue{\Line(60,0)(0,60)}:\\
\verb+\BCirc+
draws a circle with the center at the specified coordinate (first two 
arguments) and the specified radius (third argument). The interior is
white and opaque, so that it erases previously written objects, but not
subsequently drawn objects.
\end{minipage}
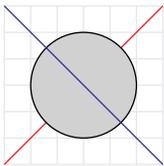
\vspace{4mm}

%--#] BCirc :
%--#[ GCirc :

\noindent
\begin{minipage}{3.83cm}
\begin{axopicture}{(60,60)(-25,0)}
\AxoGrid(0,0)(10,10)(6,6){LightGray}{0.5}
\Red{\Line(0,0)(60,60)}
\GCirc(30,30){20}{0.82}
\Blue{\Line(60,0)(0,60)}
\end{axopicture}
\end{minipage}
\begin{minipage}{11.5cm}
\label{gcirc}
\verb:\Red{\Line(0,0)(60,60)}:\\
\verb:\GCirc(30,30){20}{0.82}:\\
\verb:\Blue{\Line(60,0)(0,60)}:\\
\verb+\GCirc+ draws a circle with the center at the specified
coordinate (first two arguments) and the specified radius (third
argument).  Previously written contents are overwritten and made gray
according to the grayscale specified by the fourth argument (0=black,
1=white).
\end{minipage}
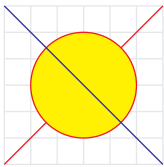
\vspace{4mm}

%--#] GCirc :
%--#[ CCirc :

\noindent
\begin{minipage}{3.83cm}
\begin{axopicture}{(60,60)(-25,0)}
\AxoGrid(0,0)(10,10)(6,6){LightGray}{0.5}
\Red{\Line(0,0)(60,60)}
\CCirc(30,30){20}{Red}{Yellow}
\Blue{\Line(60,0)(0,60)}
\end{axopicture}
\end{minipage}
\begin{minipage}{11.5cm}
\label{ccirc}
\verb:\Red{\Line(0,0)(60,60)}:\\
\verb:\CCirc(30,30){20}{Red}{Yellow}:\\
\verb:\Blue{\Line(60,0)(0,60)}:\\
\verb+\CCirc+ draws a colored circle with the center at the specified
coordinate (first two arguments) and the specified radius (third
argument). The fourth argument is the name of the color for the circle
itself. Its interior is overwritten and colored with the color
specified by name in the fifth argument.
\end{minipage}
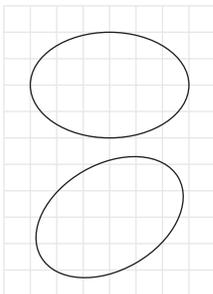
\vspace{4mm}

%--#] CCirc :
%--#[ Oval :

\noindent
\begin{minipage}{3.83cm}
\begin{axopicture}{(80,110)(-15,0)}
\AxoGrid(0,0)(10,10)(8,11){LightGray}{0.5}
\Oval(40,80)(20,30)(0)
\Oval(40,30)(20,30)(30)
\end{axopicture}
\end{minipage}
\begin{minipage}{11.5cm}
\label{oval}
\verb:\Oval(40,80)(20,30)(0):\\
\verb:\Oval(40,30)(20,30)(30):\\
\verb:\Oval: draws an oval.  The first pair of values is the center of
the oval. The next pair forms the half-height and the half-width. The
last argument is a (counterclockwise) rotation angle.  The interior is
transparent, so that it does not erase previously drawn material.
\end{minipage}
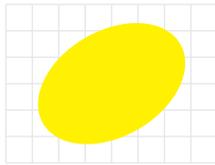
\vspace{4mm}

%--#] Oval :
%--#[ FOval :

\noindent
\begin{minipage}{3.83cm}
\begin{axopicture}{(80,60)(-15,0)}
\AxoGrid(0,0)(10,10)(8,6){LightGray}{0.5}
\SetColor{Yellow}
\FOval(40,30)(20,30)(30)
\end{axopicture}
\end{minipage}
\begin{minipage}{11.5cm}
\label{foval}
\verb:\SetColor{Yellow}:\\
\verb:\FOval(40,80)(20,30)(30):\\
\verb:\FOval: draws an oval filled with the current color overwriting
previously written material. Its arguments are the same as for the
\verb:\Oval: command. 
\end{minipage}
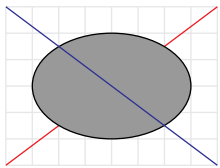
\vspace{4mm}

%--#] FOval :
%--#[ GOval :

\noindent
\begin{minipage}{3.83cm}
\begin{axopicture}{(80,60)(-15,0)}
\AxoGrid(0,0)(10,10)(8,6){LightGray}{0.5}
\Red{\Line(0,0)(80,60)}
\GOval(40,30)(20,30)(0){0.6}
\Blue{\Line(80,0)(0,60)}
\end{axopicture}
\end{minipage}
\begin{minipage}{11.5cm}
\label{goval}
\verb:\Red{\Line(0,0)(80,60)}:\\
\verb:\GOval(40,30)(20,30)(0){0.6}: \\
\verb:\Blue{\Line(80,0)(0,60)}:\\
\verb:\GOval: draws an oval with a gray interior.  
The first 5 arguments are the same as for the \verb:\Oval: command. 
The last argument indicates the
grayscale with which the oval will be filled, overwriting previously
written contents (0=black, 1=white).
\end{minipage}
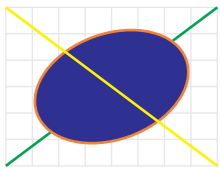
\vspace{4mm}

%--#] GOval :
%--#[ COval :

\noindent
\begin{minipage}{3.83cm}
\begin{axopicture}{(80,60)(-15,0)}
\AxoGrid(0,0)(10,10)(8,6){LightGray}{0.5}
\SetWidth{1}
\Green{\Line(0,0)(80,60)}
\COval(40,30)(20,30)(20){Orange}{Blue}
\Yellow{\Line(80,0)(0,60)}
\end{axopicture}
\end{minipage}
\begin{minipage}{11.5cm}
\label{coval}
\verb:\Green{\Line(0,0)(80,60)}:\\
\verb:\COval(40,30)(20,30)(20){Orange}{Blue}:\\
\verb:\Yellow{\Line(80,0)(0,60)}:\\
\verb:\COval: draws a colored oval.  
The first 5 arguments are the same as for the \verb:\Oval: command. 
The last two arguments are the names of two colors. 
The first is the color of the line that forms the oval and the second is 
the color of the inside.
\end{minipage}\vspace{4mm}

%--#] COval :
%--#[ EBox :

Commands for drawing boxes are in two series.  For the first set, the
box's position is specified by the coordinates of its bottom left
corner and top right corner:\\[4mm]
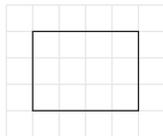
\noindent
\begin{minipage}{3.83cm}
\begin{axopicture}{(60,50)(-25,0)}
\AxoGrid(0,0)(10,10)(6,5){LightGray}{0.5}
\EBox(10,10)(50,40)
\end{axopicture}
\end{minipage}
\begin{minipage}{11.5cm}
\label{ebox}
\verb:\EBox(10,10)(50,40): \hfill \\
Draws a box. The points specified are the bottom left corner and the top 
right corner.
The interior is transparent, so that it does not erase previously
drawn material. 
\end{minipage}
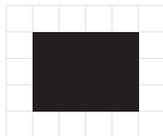
\vspace{4mm}

%--#] EBox :
%--#[ FBox :

\noindent
\begin{minipage}{3.83cm}
\begin{axopicture}{(60,50)(-25,0)}
\AxoGrid(0,0)(10,10)(6,5){LightGray}{0.5}
\FBox(10,10)(50,40)
\end{axopicture}
\end{minipage}
\begin{minipage}{11.5cm}
\label{fbox}
\verb:\FBox(10,10)(50,40): \hfill \\
Draws a box filled with the current color overwriting
previously written material. Its arguments are the same as for the
\verb:\EBox: command. 
\end{minipage}
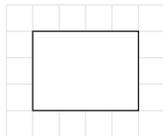
\vspace{4mm}

%--#] FBox :
%--#[ BBox :

\noindent
\begin{minipage}{3.83cm}
\begin{axopicture}{(60,50)(-25,0)}
\AxoGrid(0,0)(10,10)(6,5){LightGray}{0.5}
\BBox(10,10)(50,40)
\end{axopicture}
\end{minipage}
\begin{minipage}{11.5cm}
\label{bbox}
\verb:\BBox(10,10)(50,40): \hfill \\
Draws a blanked-out box. The points specified are the bottom left corner 
and the top right corner.
\end{minipage}
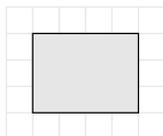
\vspace{4mm}

%--#] BBox :
%--#[ GBox :

\noindent
\begin{minipage}{3.83cm}
\begin{axopicture}{(60,50)(-25,0)}
\AxoGrid(0,0)(10,10)(6,5){LightGray}{0.5}
\GBox(10,10)(50,40){0.9}
\end{axopicture}
\end{minipage}
\begin{minipage}{11.5cm}
\label{gbox}
\verb:\GBox(10,10)(50,40){0.9}: \hfill \\
Draws a box filled with a grayscale given by the fifth argument (black=0, 
white=1). The points specified are the bottom left corner and the top 
right corner.
\end{minipage}
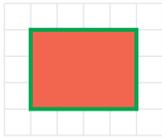
\vspace{4mm}

%--#] GBox :
%--#[ CBox :

\noindent
\begin{minipage}{3.83cm}
\begin{axopicture}{(60,50)(-25,0)}
\AxoGrid(0,0)(10,10)(6,5){LightGray}{0.5}
\SetWidth{1.5}
\CBox(10,10)(50,40){Green}{LightRed}
\end{axopicture}
\end{minipage}
\begin{minipage}{11.5cm}
\label{cbox}
\verb:\CBox(10,10)(50,40){Green}{LightRed}: \hfill \\
Draws a box in the color specified by name in the fifth argument. The
contents are filled with the color specified by name in the sixth
argument. The points specified are the bottom left corner and the top
right corner.
\end{minipage}\vspace{4mm}

%--#] CBox :
%--#[ EBoxc :

For the other series of box-drawing commands, the box's position is
specified by its center, and its width and height.  The command names
end with a ``\texttt{c}'', for ``center'':\\[3mm]
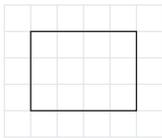
\noindent
\begin{minipage}{3.83cm}
\begin{axopicture}{(60,50)(-25,0)}
\AxoGrid(0,0)(10,10)(6,5){LightGray}{0.5}
\EBoxc(30,25)(40,30)
\end{axopicture}
\end{minipage}
\begin{minipage}{11.5cm}
\label{eboxc}
\label{boxc}
\verb:\EBoxc(30,25)(40,30): \hfill \\
Draws a box. The first two numbers give the center of the box. The next two 
numbers are the width and the height of the box. Instead of \verb:\EBoxc: 
one may also use \verb:\Boxc:.

There is also the similar command \verb:\FBoxc: that draws a filled box.
\end{minipage}
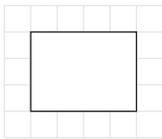
\vspace{4mm}

%--#] EBoxc :
%--#[ BBoxc :

\noindent
\begin{minipage}{3.83cm}
\begin{axopicture}{(60,50)(-25,0)}
\AxoGrid(0,0)(10,10)(6,5){LightGray}{0.5}
\BBoxc(30,25)(40,30)
\end{axopicture}
\end{minipage}
\begin{minipage}{11.5cm}
\label{bboxc}
\verb:\BBoxc(30,25)(40,30): \hfill \\
Draws a box of which the contents are blanked out. The arguments are
the same as for the \verb+\EBoxc+ command.
\end{minipage}
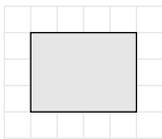
\vspace{4mm}

%--#] BBoxc :
%--#[ GBoxc :

\noindent
\begin{minipage}{3.83cm}
\begin{axopicture}{(60,50)(-25,0)}
\AxoGrid(0,0)(10,10)(6,5){LightGray}{0.5}
\GBoxc(30,25)(40,30){0.9}
\end{axopicture}
\end{minipage}
\begin{minipage}{11.5cm}
\label{gboxc}
\verb:\GBoxc(30,25)(40,30){0.9}: \hfill \\
Draws a box filled with a grayscale given by the fifth argument (black=0, 
white=1).
The first 4 arguments are the same as for the \verb+\EBoxc+ command.
\end{minipage}
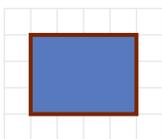
\vspace{4mm}

%--#] GBoxc :
%--#[ CBoxc :

\noindent
\begin{minipage}{3.83cm}
\begin{axopicture}{(60,50)(-25,0)}
\AxoGrid(0,0)(10,10)(6,5){LightGray}{0.5}
\SetWidth{1.5}
\CBoxc(30,25)(40,30){Brown}{LightBlue}
\end{axopicture}
\end{minipage}
\begin{minipage}{11.5cm}
\label{cboxc}
\verb:\CBoxc(30,25)(40,30){Brown}{LightBlue}: \hfill \\
Draws a box in the color specified by name in the fifth argument. The
contents are filled with the color specified by name in the sixth
argument.
The first 4 arguments are the same as for the \verb+\EBoxc+ command.
\end{minipage}
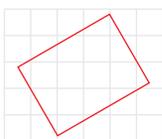
\vspace{4mm}

%--#] BBoxc :
%--#] CBoxc :
%--#[ RotatedBox :

\noindent
\begin{minipage}{3.83cm}
\begin{axopicture}{(60,50)(-25,0)}
\AxoGrid(0,0)(10,10)(6,5){LightGray}{0.5}
\RotatedBox(30,25)(40,30){30}{Red}
\end{axopicture}
\end{minipage}
\begin{minipage}{11.5cm}
\label{rotatedbox}
\verb:\RotatedBox(30,25)(40,30){30}{Red}: \hfill \\
Draws a rotated box. The first two numbers give the center of the
box. The next two numbers are the width and the height of the box. The
fifth argument is the counterclockwise rotation angle and the sixth
argument is the color of the box.  The interior of the box is
transparent.
\end{minipage}
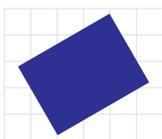
\vspace{4mm}

%--#] RotatedBox :
%--#[ FilledRotatedBox :

\noindent
\begin{minipage}{3.83cm}
\begin{axopicture}{(60,50)(-25,0)}
\AxoGrid(0,0)(10,10)(6,5){LightGray}{0.5}
\FilledRotatedBox(30,25)(40,30){30}{Blue}
\end{axopicture}
\end{minipage}
\begin{minipage}{11.5cm}
\label{filledrotatedbox}
\verb:\FilledRotatedBox(30,25)(40,30){30}{Blue}: \hfill \\
Draws a rotated box.
The first 4 arguments are the same as for the \verb+\RotatedBox+ command.
The
fifth argument is the counterclockwise rotation angle and the sixth
argument is the color of the inside of the box. If a differently
colored outline is needed, it should be written with the
\verb+RotatedBox+ command.
\end{minipage}
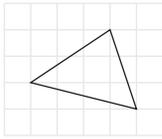
\vspace{4mm}

%--#] FilledRotatedBox :
%--#[ ETri :

\noindent
\begin{minipage}{3.83cm}
\begin{axopicture}{(60,50)(-25,0)}
\AxoGrid(0,0)(10,10)(6,5){LightGray}{0.5}
\ETri(10,20)(50,10)(40,40)
\end{axopicture}
\end{minipage}
\begin{minipage}{11.5cm}
\label{etri}
\verb:\ETri(10,20)(50,10)(40,40): \hfill \\
Draws a triangle. The three points specified are the corners of the 
triangle.
The interior is transparent.

There is also the similar command \verb:\FTri: that draws a filled triangle.
\end{minipage}
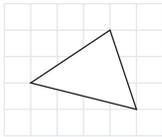
\vspace{4mm}

%--#] ETri :
%--#[ BTri :

\noindent
\begin{minipage}{3.83cm}
\begin{axopicture}{(60,50)(-25,0)}
\AxoGrid(0,0)(10,10)(6,5){LightGray}{0.5}
\BTri(10,20)(50,10)(40,40)
\end{axopicture}
\end{minipage}
\begin{minipage}{11.5cm}
\label{btri}
\verb:\BTri(10,20)(50,10)(40,40): \hfill \\
Draws a blanked-out triangle. The three points specified are the corners of 
the triangle.
\end{minipage}
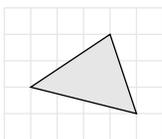
\vspace{4mm}

%--#] BTri :
%--#[ GTri :

\noindent
\begin{minipage}{3.83cm}
\begin{axopicture}{(60,50)(-25,0)}
\AxoGrid(0,0)(10,10)(6,5){LightGray}{0.5}
\GTri(10,20)(50,10)(40,40){0.9}
\end{axopicture}
\end{minipage}
\begin{minipage}{11.5cm}
\label{gtri}
\verb:\GTri(10,20)(50,10)(40,40){0.9}: \hfill \\
Draws a triangle of which the content are filled with the grayscale 
specified by the seventh argument (black=0, white=1). The three points 
specified are the corners of the triangle.
\end{minipage}
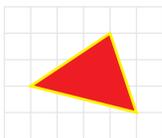
\vspace{4mm}

%--#] GTri :
%--#[ CTri :

\noindent
\begin{minipage}{3.83cm}
\begin{axopicture}{(60,50)(-25,0)}
\AxoGrid(0,0)(10,10)(6,5){LightGray}{0.5}
\SetWidth{1}
\CTri(10,20)(50,10)(40,40){Red}{Yellow}
\end{axopicture}
\end{minipage}
\begin{minipage}{11.5cm}
\label{ctri}
\verb:\CTri(10,20)(50,10)(40,40){Red}{Yellow}: \hfill \\
Draws a triangle in the color named in the seventh argument. The
contents are filled with the color named in the eightth argument. The
three points specified are the corners of the triangle.
\end{minipage}
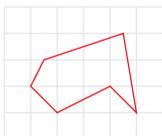
\vspace{4mm}

%--#] CTri :
%--#[ Polygon :

\noindent
\begin{minipage}{3.83cm}
\begin{axopicture}{(60,50)(-25,0)}
\AxoGrid(0,0)(10,10)(6,5){LightGray}{0.5}
\Polygon{(10,20)(20,10)(40,20)(50,10)(45,40)(15,30)}{Red}
\end{axopicture}
\end{minipage}
\begin{minipage}{11.5cm}
\label{polygon}
\verb:\Polygon{(10,20)(20,10)(40,20)(50,10): \hfill \\
  \verb:         (45,40)(15,30)}{Red}: \hfill \\
Draws a polygon. The first argument is a sequence of two dimensional
points which form the corners of the polygon. The second argument is
the name of the color of the polygon. The interior is transparent.
\end{minipage}
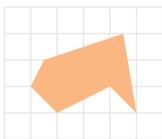
\vspace{4mm}

%--#] Polygon :
%--#[ FilledPolygon :

\noindent
\begin{minipage}{3.83cm}
\begin{axopicture}{(60,50)(-25,0)}
\AxoGrid(0,0)(10,10)(6,5){LightGray}{0.5}
\FilledPolygon{(10,20)(20,10)(40,20)(50,10)(45,40)(15,30)}{Apricot}
\end{axopicture}
\end{minipage}
\begin{minipage}{11.5cm}
\label{filledpolygon}
\verb:\FilledPolygon{(10,20)(20,10)(40,20)(50,10): \hfill \\
  \verb:               (45,40)(15,30)}{Apricot}: \hfill \\
Draws a polygon. The first argument is a sequence of two dimensional
points which form the corners of the polygon. The second argument is
the name of the color of the interior.
\end{minipage}\vspace{4mm}

%--#] FilledPolygon :
%--#[ LinAxis :

\noindent
\begin{minipage}{3.83cm}
\begin{axopicture}{(100,50)(-5,0)}
\AxoGrid(0,0)(10,10)(10,5){LightGray}{0.5}
\LinAxis(10,30)(90,30)(4,5,5,0,1)
\LinAxis(10,10)(90,10)(4,5,5,2,1)
\end{axopicture}
\end{minipage}
\begin{minipage}{11.5cm}
\label{linaxis}
\verb:\LinAxis(10,30)(90,30)(4,5,5,0,1):\\
\verb:\LinAxis(10,10)(100,10)(4,5,5,2,1): \\
\verb+\LinAxis+($x_1$,$y_1$)($x_2$,$y_2$)($N_D$,$d$,hashsize,offset,width)
    draws a line to be used as an axis in a graph. Along the axis 
    are hash marks. Going from the first coordinate to the second, the 
    hash marks are on the left side if `hashsize', which is the size of the 
    hash marks, is positive and on the right side if it is negative. 
    $N_D$ is the number of `decades', indicated by fat hash marks, and 
    $d$ is the (integer) number of subdivisions inside each decade. The offset 
    parameter tells to which subdivision the first coordinate 
    corresponds. When it is zero, this coordinate corresponds to a fat 
    mark of a decade. Because axes have their own width, this is 
    indicated with the last parameter.
%Draws a line with subdivisions that can be used as the axis on a histogram 
%or other figure. The first four arguments are the endpoints of the axis. 
%Then we have the number of decades, the number of divisions inside each 
%decade, the size of the hash marks, the offset in divisions at which we 
%start and the linewidth. The hashmarks will be on the left side when going 
%from point 1 to point 2.
\end{minipage}
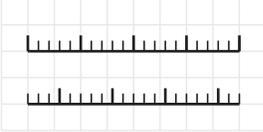
\vspace{4mm}

%--#] LinAxis :
%--#[ LogAxis :

\noindent
\begin{minipage}{3.83cm}
\begin{axopicture}{(100,40)(-5,0)}
\AxoGrid(0,0)(10,10)(10,4){LightGray}{0.5}
\LogAxis(0,30)(100,30)(4,3,0,1)
\LogAxis(0,10)(100,10)(4,3,3,1)
\end{axopicture}
\end{minipage}
\begin{minipage}{11.5cm}
\label{logaxis}
\verb:\LogAxis(0,30)(100,30)(4,3,0,1): \hfill \\
\verb:\LogAxis(0,10)(100,10)(4,3,3,1): \hfill \\
\verb+\LogAxis+($x_1$,$y_1$)($x_2$,$y_2$)($N_L$,hashsize
    ,offset,width) \hfill \\
    This draws a line to be used as a logarithmic axis in a graph. Along 
    the axis are hash marks. Going from the first coordinate to the second, 
    the hash marks are on the left side if `hashsize', which is the size of 
    the hash marks, is positive and on the right side if it is negative. 
    $N_L$ is the number of orders of magnitude, indicated by fat hash 
    marks. The offset parameter tells to which integer subdivision the 
    first coordinate corresponds. When it is zero, this coordinate 
    corresponds to a fat mark, which is identical to when the value would 
    have been 1. Because axes have their own width, this is indicated with 
    the last parameter.
%Draws a line with subdivisions that can be used as the axis on a histogram 
%or other figure. The first four arguments are the endpoints of the axis. 
%Then we have the number of orders of magnitude, 
%the size of the hash marks, the offset inside a logarithm at which we 
%start and the linewidth. The hashmarks will be on the left side when going 
%from point 1 to point 2.
\end{minipage}\vspace{4mm}

%--#] LogAxis :
%>>#] The Commands :
%>>#[ Text :

\subsection{Text}
\label{sec:text}

%--#[ Implementation :

Axodraw2 provides several commands for inserting text into diagrams.
Some are for plain text, with a chosen placement and angle.  Some
allow placement of text inside boxes.  There are two sets of commands.
Some we call \TeX-text commands; these use the standard \LaTeX{} fonts
as used in the rest of the document.  The others we call
postscript-text commands; these use a user-specified standard
postscript font or, if the user wishes, the usual document font, at a
user-chosen size.

[\emph{Side issue:} In version 1 of axodraw, the difference between
the classes of text command was caused by a serious implementation
difficulty.  With the then-available \LaTeX{} technology, certain
graphic effects, could not be achieved within \LaTeX, at least not
easily.  So direct programming in postscript was resorted to, with the
result that normal \LaTeX{} commands, including mathematics, were not
available in the postscript-text commands.  With the greatly improved
methods now available, this has all changed, and the restrictions have
gone.  But since the commands and their basic behavior is already
defined, we have retained the distinction between \TeX{}-text commands
and postscript-text commands.]

In the original version of Axodraw the commands for two lines inside a
box were \verb:B2Text:, \verb:G2Text: and \verb:C2Text:. This causes
some problems explained in Sec.\ \ref{sec:changes.wrt.1}.  If you need to
retain compatibility with v.\ 1 on this issue, e.g., with old files or
old diagrams or for personal preference, you can use the
\texttt{v1compatible} option when loading axodraw2 --- see Sec.\
\ref{sec:invoke}.

\vspace{4mm}

%--#] Implementation :
%--#[ Text :

\subsubsection{\TeX-type text}

Illustrated by examples, the commands to insert text are as follows:

\medskip

\noindent
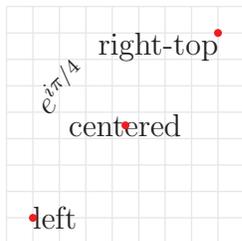
\begin{minipage}{3.83cm}
\begin{axopicture}{(90,90)(-10,0)}
\AxoGrid(0,0)(10,10)(9,9){LightGray}{0.5}
\Text(10,10)[l]{left}
\Text(45,45){centered}
\Text(80,80)[rt]{right-top}
\Text(20,60)(45){$e^{i\pi/4}$}
\SetColor{Red}
\Vertex(10,10){1.5}
\Vertex(45,45){1.5}
\Vertex(80,80){1.5}
\end{axopicture}
\end{minipage}
\begin{minipage}{11.5cm}
\label{text}
\verb:\Text(10,10)[l]{left}: \hfill \\
\verb:\Text(45,45){centered}: \hfill \\
\verb:\Text(80,80)[rt]{right-top}: \hfill \\
\verb:\Text(20,60)(45){$e^{i\pi/4}$}: \hfill \\
\verb:\SetColor{Red}: \hfill \\
\verb:\Vertex(10,10){1.5}: \hfill \\
\verb:\Vertex(45,45){1.5}: \hfill \\
\verb:\Vertex(80,80){1.5}: \hfill \\
\verb+\Text+ writes text in the current \LaTeX{} font.  The most
general form is \verb+\Text(x,y)(theta)[pos]{text}+; but either or
both of the theta and pos arguments (and their delimiters) can be omitted.
It puts the text
at focal point $(x,y)$, with a rotation by anticlockwise angle theta.
The default angle is zero, and the default position is to 
center the text horizontally and vertically at the focal point.  The
position letters are any relevant combination of `l', `r', `t', and
`b', as in the various 
\TeX/\LaTeX{} box commands to indicate left, right, top or bottom 
adjustment with respect to the focal point. No indication means
centered.
\end{minipage}\vspace{4mm}

%--#] Text :
%--#[ rText :

\noindent
\begin{minipage}{3.83cm}
\begin{axopicture}{(90,90)(-10,0)}
\AxoGrid(0,0)(10,10)(9,9){LightGray}{0.5}
\rText(10,10)[l][l]{left-left}
\rText(45,45)[][u]{upside}
\rText(80,10)[r][r]{right-right}
\rText(20,60)[][r]{$e^{i\pi}$}
\SetColor{Red}
\Vertex(10,10){1.5}
\Vertex(45,45){1.5}
\Vertex(80,10){1.5}
\end{axopicture}
\end{minipage}
\begin{minipage}{11.5cm}
\label{rtext}
\verb:\rText(10,10)[l][l]{left-left}: \hfill \\
\verb:\rText(45,45)[][u]{upside}: \hfill \\
\verb:\rText(80,10)[r][r]{right-right}: \hfill \\
\verb:\rText(20,60)[][r]{$e^{i\pi}$}: \hfill \\
\verb:\SetColor{Red}: \hfill \\
\verb:\Vertex(10,10){1.5}: \hfill \\
\verb:\Vertex(45,45){1.5}: \hfill \\
\verb:\Vertex(80,10){1.5}: \hfill \\
The \verb:\rText: command gives a subset of the functionality of the
\verb+\Text+ command.  It is used for backward compatibility with
Axodraw v.\ 1. The general form of the command is
\verb:\rText(x,y)[mode][rotation]{text}:. 
Unlike the case with the \verb:\Text: command and typical standard
\LaTeX{} commands, if the option letters are omitted, the square
brackets must be retained.  
The coordinates $(x,y)$ are
the focal point of the text.  The third argument is \verb+l+,
\verb+r+, or empty to indicate the justification of the text.  The
fourth argument is \verb+l+, \verb+r+, \verb+u+, or empty to indicate
respectively whether the text is rotated left (anticlockwise) by 90
degrees, is rotated right (clockwise) by 90 degrees, is upside-down,
or is not rotated.  The fifth argument is the text.  
This command is retained only for backward compatibility;
for new diagrams it is probably better to use the the \verb:\Text:.
\end{minipage}\vspace{4mm}

%--#] rText :
%--#[ SetPFont :

\subsubsection{Postscript-type text}
\label{sec:PSText}

The remaining text-drawing commands can use postscript fonts with an
adjustable size.

To set the font for later text-drawing commands in this class, the
\verb:\SetPFont: command sets the `Postscript'
font, e.g.,
\begin{verbatim}
   \SetPFont{Helvetica}{20}
\end{verbatim}
(This font is initialized by axodraw2 to Times-Roman at 10pt.)
The font set in this way is used in the \verb:PText:, \verb:BText:,
\verb:GText:, \verb:CText:, \verb:BTwoText:, \verb:GTwoText: and
\verb:CTwoText: commands. The fonts that can be used are the 35 fonts
that are made available by Adobe and that are normally available in
all postscript interpreters, including printers.  The fonts, together
with the names used to specify them in the normal font-setting
commands of \TeX{} and \LaTeX{}, are shown in Table \ref{tab:Pfont}.

\begin{table}
\begin{tabular}{|l|l|l|l|}
\hline
Font name                   & \LaTeX{} & Font name                & \LaTeX{} \\
\hline
AvantGarde-Book             & pagk  & Helvetica-Narrow            & phvrrn\\
AvantGarde-BookOblique      & pagko & Helvetica-NarrowOblique     & phvron\\
AvantGarde-Demi             & pagd  & NewCenturySchlbk-Bold       & pncb  \\
AvantGarde-DemiOblique      & pagdo & NewCenturySchlbk-BoldItalic & pncbi \\
Bookman-Demi                & pbkd  & NewCenturySchlbk-Italic     & pncri \\
Bookman-DemiItalic          & pbkdi & NewCenturySchlbk-Roman      & pncr  \\
Bookman-Light               & pbkl  & Palatino-Bold               & pplb  \\
Bookman-LightItalic         & pbkli & Palatino-BoldItalic         & pplbi \\
Courier-Bold                & pcrb  & Palatino-Italic             & pplri \\
Courier-BoldOblique         & pcrbo & Palatino-Roman              & pplr  \\
Courier                     & pcrr  & Symbol                      & psyr  \\
Courier-Oblique             & pcrro & Times-Bold                  & ptmb  \\
Helvetica-Bold              & phvb  & Times-BoldItalic            & ptmbi \\
Helvetica-BoldOblique       & phvbo & Times-Italic                & ptmri \\
Helvetica-NarrowBold        & phvbrn& Times-Roman                 & ptmr  \\
Helvetica-NarrowBoldOblique & phvbon& ZapfChancery-MediumItalic   & pzcmi \\
Helvetica                   & phvr  & ZapfDingbats                & pzdr  \\
Helvetica-Oblique           & phvro &                             &       \\
\hline
\end{tabular}
\caption{Available postscript fonts and their corresponding names in
  \LaTeX.}
\label{tab:Pfont}
\end{table}
If you prefer to use the normal document font (which would normally be
Computer Modern in the common document classes), you simply leave the
fontname empty, e.g,.
\begin{verbatim}
   \SetPFont{}{20}
\end{verbatim}
As for the second, fontsize argument, leaving it empty uses the size
that \LaTeX{} is using at the moment the text-drawing command starts,
e.g.,
\begin{verbatim}
   \SetPFont{Helvetica-Bold}{}
\end{verbatim}
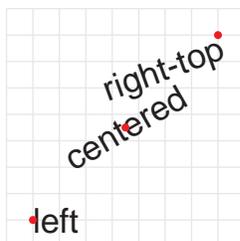
\vspace{3mm}

%--#] SetPFont :
%--#[ PText :

\noindent
\begin{minipage}{3.83cm}
\begin{axopicture}{(90,90)(-10,0)}
\AxoGrid(0,0)(10,10)(9,9){LightGray}{0.5}
\SetPFont{Helvetica}{13}
\PText(10,10)(0)[l]{left}
\PText(45,45)(30)[]{centered}
\PText(80,80)(20)[rt]{right-top}
%\PText(20,60)(140)[]{$e^{i\pi}$}
\SetColor{Red}
\Vertex(10,10){1.5}
\Vertex(45,45){1.5}
\Vertex(80,80){1.5}
\end{axopicture}
\end{minipage}
\begin{minipage}{11.5cm}
\label{ptext}
\verb:\SetPFont{Helvetica}{13}: \hfill \\
\verb:\PText(10,10)(0)[l]{left}: \hfill \\
\verb:\PText(45,45)(30)[]{centered}: \hfill \\
\verb:\PText(80,80)(20)[rt]{right-top}: \hfill \\
%\verb:\PText(20,60)(90)[]{$e^{i\pi}$}: \hfill \\
\verb:\SetColor{Red}: \hfill \\
\verb:\Vertex(10,10){1.5}: \hfill \\
\verb:\Vertex(45,45){1.5}: \hfill \\
\verb:\Vertex(80,80){1.5}: \hfill \\
The \verb:\PText: command writes %text 
in Axodraw's current Postscript font. 
The first two arguments give the focal point, the third argument is a 
rotation angle and the fourth argument is as in the various \TeX/\LaTeX{} 
box commands to indicate left, right, top or bottom adjustment with respect 
to the focal point. No indication means centered.

Note that use of normal \LaTeX{} font setting commands or of math-mode
will not normally have the desired effect.
\end{minipage}
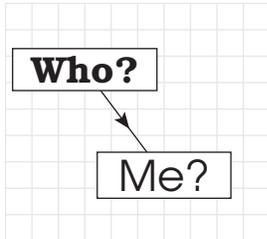
\vspace{4mm}

%--#] PText :
%--#[ BText :

\noindent
\begin{minipage}{4.53cm}
\begin{axopicture}{(110,110)(-10,0)}
\AxoGrid(0,0)(10,10)(10,9){LightGray}{0.5}
\ArrowLine(30,65)(60,25)
\SetPFont{Bookman-Demi}{14}
\BText(30,65){Who?}
\SetPFont{AvantGarde-Book}{16}
\BText(60,25){Me?}
\end{axopicture}
\end{minipage}
\begin{minipage}{10.8cm}
\label{btext}
\verb:\ArrowLine(30,65)(60,25): \hfill \\
\verb:\SetPFont{Bookman-Demi}{14}: \hfill \\
\verb:\BText(30,65){Who?}: \hfill \\
\verb:\SetPFont{AvantGarde-Book}{16}: \hfill \\
\verb:\BText(60,25){Me?}: \hfill \\
The \verb:\BText: command writes a centered box with text in it. It uses 
Axodraw's current Postscript font.
\end{minipage}
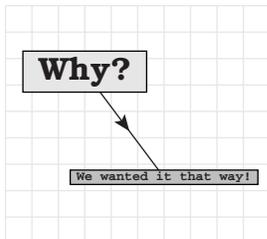
\vspace{4mm}

%--#] BText :
%--#[ GText :

\noindent
\begin{minipage}{4.53cm}
\begin{axopicture}{(110,110)(-10,0)}
\AxoGrid(0,0)(10,10)(10,9){LightGray}{0.5}
\ArrowLine(30,65)(60,25)
\SetPFont{Bookman-Demi}{12}
\GText(30,65){0.9}{Why?}
\SetPFont{Courier-Bold}{5}
\GText(60,25){0.75}{We wanted it that way!}
\end{axopicture}
\end{minipage}
\begin{minipage}{10.8cm}
\label{gtext}
\verb:\ArrowLine(30,65)(60,25): \hfill \\
\verb:\SetPFont{Bookman-Demi}{12}: \hfill \\
\verb:\GText(30,65){0.9}{Why?}: \hfill \\
\verb:\SetPFont{Courier-Bold}{5}: \hfill \\
\verb:\GText(60,25){0.75}{We wanted it that way!}: \hfill \\
The \verb:\GText: command writes a centered box with text in it. It uses 
Axodraw's current Postscript font. The third argument is the grayscale
with which
the box will be filled. 0 is black and 1 is white.
\end{minipage}
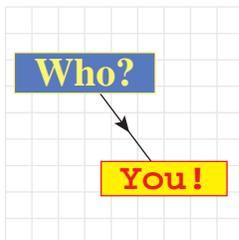
\vspace{4mm}

%--#] GText :
%--#[ CText :
 
\noindent
\begin{minipage}{4.53cm}
\begin{axopicture}{(110,110)(-10,0)}
\AxoGrid(0,0)(10,10)(9,9){LightGray}{0.5}
\ArrowLine(30,65)(60,25)
\SetPFont{Times-Bold}{15}
\CText(30,65){LightYellow}{LightBlue}{Who?}
\SetPFont{Courier-Bold}{14}
\CText(60,25){Red}{Yellow}{You!}
\end{axopicture}
\end{minipage}
\begin{minipage}{10.8cm}
\label{ctext}
\verb:\ArrowLine(30,65)(60,25): \hfill \\
\verb:\SetPFont{Times-Bold}{15}: \hfill \\
\verb:\CText(30,65){LightYellow}{LightBlue}{Who?}: \hfill \\
\verb:\SetPFont{Courier-Bold}{14}: \hfill \\
\verb:\CText(60,25){Red}{Yellow}{You!}: \hfill \\
The \verb:\CText: command writes a centered box with text in it. It uses 
Axodraw's current Postscript font. The third argument is the color of
the box and
the text. The fourth argument is the color with which the box will be 
filled.
\end{minipage}
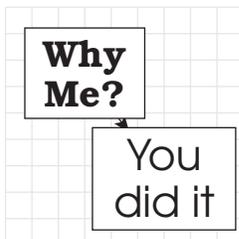
\vspace{4mm}

%--#] CText :
%--#[ BTwoText :
\noindent
\begin{minipage}{4.53cm}
\begin{axopicture}{(110,110)(-10,0)}
\AxoGrid(0,0)(10,10)(9,9){LightGray}{0.5}
\ArrowLine(30,65)(60,25)
\SetPFont{Bookman-Demi}{14}
\BTwoText(30,65){Why}{Me?}
\SetPFont{AvantGarde-Book}{16}
\BTwoText(60,25){You}{did it}
\end{axopicture}
\end{minipage}
\begin{minipage}{10.8cm}
\label{btwotext}
\verb:\ArrowLine(30,65)(60,25): \hfill \\
\verb:\SetPFont{Bookman-Demi}{14}: \hfill \\
\verb:\BTwoText(30,65){Why}{Me?}: \hfill \\
\verb:\SetPFont{AvantGarde-Book}{16}: \hfill \\
\verb:\BTwoText(60,25){You}{did it}: \hfill \\
The \verb:\BTwoText: command writes a centered box with two lines of text in 
it. It uses Axodraw's current Postscript font.
\end{minipage}
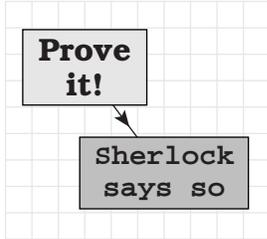
\vspace{4mm}

%--#] BTwoText :
%--#[ GTwoText :

\noindent
\begin{minipage}{4.53cm}
\begin{axopicture}{(110,110)(-10,0)}
\AxoGrid(0,0)(10,10)(10,9){LightGray}{0.5}
\ArrowLine(30,65)(60,25)
\SetPFont{Bookman-Demi}{12}
\GTwoText(30,65){0.9}{Prove}{it!}
\SetPFont{Courier-Bold}{11}
\GTwoText(60,25){0.75}{Sherlock}{says so}
\end{axopicture}
\end{minipage}
\begin{minipage}{10.8cm}
\label{gtwotext}
\verb:\ArrowLine(30,65)(60,25): \hfill \\
\verb:\SetPFont{Bookman-Demi}{12}: \hfill \\
\verb:\GTwoText(30,65){0.9}{Prove}{it!}: \hfill \\
\verb:\SetPFont{Courier-Bold}{11}: \hfill \\
\verb:\GTwoText(60,25){0.75}{Sherlock}{says so}: \hfill \\
The \verb:\GTwoText: command writes a centered box with two lines of text in 
it. It uses Axodraw's current Postscript font. The third argument is the 
grayscale with which the box will be filled. 0 is black and 1 is white.
\end{minipage}
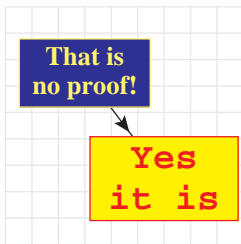
\vspace{4mm}

%--#] GTwoText :
%--#[ CTwoText :
 
\noindent
\begin{minipage}{4.53cm}
\begin{axopicture}{(110,110)(-10,0)}
\AxoGrid(0,0)(10,10)(9,9){LightGray}{0.5}
\ArrowLine(30,65)(60,25)
\SetPFont{Times-Bold}{10}
\CTwoText(30,65){LightYellow}{Blue}{That is}{no proof!}
\SetPFont{Courier-Bold}{14}
\CTwoText(60,25){Red}{Yellow}{Yes}{it is}
\end{axopicture}
\end{minipage}
\begin{minipage}{10.8cm}
\label{ctwotext}
\verb:\ArrowLine(30,65)(60,25): \hfill \\
\verb:\SetPFont{Times-Bold}{10}: \hfill \\
\verb:\CTwoText(30,65){LightYellow}{Blue}: \\
    \verb:{That is}{no proof!}: \hfill \\
\verb:\SetPFont{Courier-Bold}{14}: \hfill \\
\verb:\CTwoText(60,25){Red}{Yellow}{Yes}{it is}: \hfill \\
The \verb:\CTwoText: command writes a centered box with two lines of text in 
it. It uses Axodraw's current Postscript font. The third argument is
the color of both
the box and the text. The fourth argument is the color with which the box 
will be filled.
\end{minipage}\vspace{4mm}
 
%--#] CTwoText :
%--#[ Features :

Note that because you can now use \LaTeX{} commands for the text
arguments of the commands described in this section, the effects of
the \verb+\BTwoText+, \verb+\GTwoText+, and \verb+\CTwoText+ can be
achieved also by the use of regular \verb:\BText: etc commands.
Mathematics can also be used.  (None of these was possible in v.\ 1 of
axodraw.)  Here are some examples: 
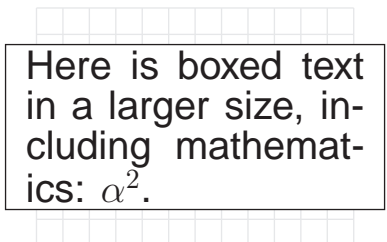
\vspace{4mm}

\noindent
\begin{minipage}{5.5cm}
\begin{axopicture}{(150,90)(-10,0)}
\AxoGrid(0,0)(10,10)(12,9){LightGray}{0.5}
\SetPFont{Helvetica}{15}
\BText(60,45){%
    \begin{minipage}{4.5cm}
      Here is boxed text in a larger size, including
      mathematics: $\alpha^2$.
    \end{minipage}%
}
\end{axopicture}
\end{minipage}
\begin{minipage}{8.5cm}
\label{btext2}
\begin{verbatim}
\SetPFont{Helvetica}{15}
\BText(70,45){%
    \begin{minipage}{4.5cm}
      Here is boxed text in a
      larger size, including
      mathematics: $\alpha^2$.
    \end{minipage}%
}
\end{verbatim}
This example shows that the \verb:\BText: command can also be used
with minipages and other \LaTeX{} methods to make more complicated
boxed texts.
\end{minipage}
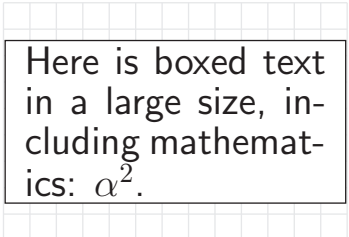
\vspace{4mm}

\noindent
\begin{minipage}{5.5cm}
\begin{axopicture}{(150,90)(-10,0)}
\AxoGrid(0,0)(10,10)(13,9){LightGray}{0.5}
\SetPFont{}{15}
\BText(65,45){%
    \begin{minipage}{4cm}
      \sffamily Here is boxed text in a
      large size, including
      mathematics: $\alpha^2$.
    \end{minipage}%
}
\end{axopicture}
\end{minipage}
\begin{minipage}{8.5cm}
\label{btext2.mod}
\begin{verbatim}
\SetPFont{}{15}
\BText(65,45){%
    \begin{minipage}{4cm}
      \sffamily Here is boxed text in a 
      large size, including
      mathematics: $\alpha^2$.
    \end{minipage}%
}
\end{verbatim}
But if you use mathematics, the text may be more elegant if you use
the document font, which has matching fonts for text and mathematics.
Use of a sans-serif font (by \verb:\sffamily:) may be better in a diagram.
\end{minipage}
\vspace{4mm}

%--#] Features :
%>>#] Text :
%>>#[ Options :

\subsection{Options}
\label{sec:options}

Almost all of axodraw2's line-drawing commands take optional
arguments.  The form here is familiar from many standard \LaTeX{}
commands.  The optional arguments are placed in square brackets after
the command name, and are made of a comma-separated list of items of
the form: \texttt{keyword} or \texttt{keyword=value}.  The required
arguments are placed afterwards.

Optional arguments can be used to set particular characteristics of a
line, e.g., whether it is dashed or has an arrow.  They can also be
used to set some of the line's parameters, to be used instead of
default values.  (The default values can be adjusted by commands
listed in Sec.\ \ref{sec:settings}.  Those commands are useful for
adjusting parameters that apply to multiple lines, while the optional
arguments are useful for setting parameters for individual lines.)

The original axodraw only had different command names to determine
whether lines were dashed, or had arrows, etc.  The new version
retains these commands,
but now the basic commands
(\verb:\Line:, \verb:\Arc:, \verb:\Gluon:, etc) can also be treated as
generic commands, with the different varieties (dashed, double, and/or
with an arrow) being set by options.  

The same set of options are available for all types of line.  However,
not all apply or are implemented for particular types of line.  Thus,
\texttt{clockwise} is irrelevant for a straight line, while
\texttt{arrow} is not implemented for gluons, photons and zigzag
lines.  Warnings are given for unimplemented features, while
inapplicable arguments are ignored.

The full set of options.
\begin{center}
\begin{tabular}{ll}
 color=\colorname    & Set the line in this color. \\
 colour=\colorname   & Same as color=\colorname. \\
 dash                & Use a dashed line. \\
 dsize=\num          & Set the dash size (when a line is dashed). \\
 dashsize=\num       & Same as dsize=\num. \\
 double              & Use a double line. \\
 sep=\num            & Sets the separation for a double line. \\
 linesep=\num        & Same as sep=\num. \\
 width=\num          & Sets line width for this line only.\\[2mm]
 clock               & For arcs, makes the arc run clockwise. \\
 clockwise           & For arcs, makes the arc run clockwise. \\[2mm]
 arrow               & Use an arrow.\\
 flip                & If there is an arrow, its direction is flipped. \\

 arrowpos=\num     & The number should be between zero and one and\\
                   & indicates where along the line the arrow should be. \\
                   & 1 is at the end. 0.5 is halfway (the initial default).\\ 
 arrowaspect=\num  & See Sec.\ \ref{sec:arrows}. \\
 arrowlength=\num  & See Sec.\ \ref{sec:arrows}. \\
 arrowheight=\num  & See Sec.\ \ref{sec:arrows}. \\
 arrowinset=\num   & See Sec.\ \ref{sec:arrows}. \\
 arrowscale=\num   & See Sec.\ \ref{sec:arrows}. \\
 arrowstroke=\num  & See Sec.\ \ref{sec:arrows}. \\
 arrowwidth=\num   & See Sec.\ \ref{sec:arrows}. \\
 inset=\num        & Same as arrowinset.\\
\end{tabular}
\end{center}
The options without an extra argument, e.g., \texttt{arrow}, are
actually of a boolean type.  That is, they can also be used with a
suffix ``\texttt{=true}'' or ``\texttt{=false}'', e.g.,
\texttt{arrow=true} or \texttt{arrow=false}.

If an option is not provided, its default value is used. Defaults are
no dashes, no double lines, anticlockwise arcs, no arrow and if an
arrow is asked for, its position is halfway along the line. Other
arrow settings are explained in Sec.\ \ref{sec:arrows}.  There are
also default values for dash size (3) and the separation of double
lines (2).

The full set of the generic line commands with their syntax is
\begin{center}
  \begin{tabular}{l}
     \verb+\Line[options](x1,y1)(x2,y2)+  \\
     \verb+\Arc[options](x,y)(r,theta1,theta2)+  \\
     \verb+\Bezier[options](x1,y1)(x2,y2)(x3,y3)(x4,y4)+  \\
     \verb+\Gluon[options](x1,y1)(x2,y2){amplitude}{windings}+  \\
     \verb+\GluonArc[options](x,y)(r,theta1,theta2){amplitude}{windings}+  \\
     \verb+\GluonCirc[options](x,y)(r,phase){amplitude}{windings}+   \\
     \verb+\Photon[options](x1,y1)(x2,y2){amplitude}{windings}+  \\
     \verb+\PhotonArc[options](x,y)(r,theta1,theta2){amplitude}{windings}+  \\
     \verb+\ZigZag[options](x1,y1)(x2,y2){amplitude}{windings}+  \\
     \verb+\ZigZagArc[options](x,y)(r,theta1,theta2){amplitude}{windings}+  \\
  \end{tabular}
\end{center}
The applicability of the options is as follows
\begin{center}
  \begin{tabular}{lcc}
                        & Arrow, etc & Clockwise \\
     \verb+\Line+       &    Y       &     N     \\
     \verb+\Arc+        &    Y       &     Y     \\
     \verb+\Bezier+     &    Y       &     N     \\
     \verb+\Gluon+      &    N       &     N     \\
     \verb+\GluonArc+   &    N       &     Y     \\
     \verb+\GluonCirc+  &    N       &     N     \\
     \verb+\Photon+     &    N       &     N     \\
     \verb+\PhotonArc+  &    N       &     Y     \\
     \verb+\ZigZag+     &    N       &     N     \\
     \verb+\ZigZagArc+  &    N       &     Y     \\
  \end{tabular}
\end{center}
The arrow options include those for setting the arrow dimensions.
Options not indicated in the last table apply to all cases.

%{\sc The next options still have to be implemented, but it seems the most 
%sensible thing to do.}\vspace{3mm}
%
%The third family is the one of the shapes:
%
%\begin{center}
%\begin{minipage}{14cm}
%\begin{verbatim}
%\Box[options](x1,y1)(x2,y2)
%\Tri[options](x1,y1)(x2,y2)(x3,y3)
%\Polygon[options]{(x1,y1)(x2,y2)...(xn,yn)}
%\Circ[options](x1,y1){radius}
%\Oval[options](x1,y1)(height,width)(rotation)
%\end{verbatim}
%\end{minipage}
%\end{center}
%
%\noindent The options here are:
%\begin{center}
%\begin{tabular}{ll}
% centered            & For boxes: x1,y1 is the center. x2,y2 is width,
%                       height \\
% blanked             & Inside is blanked out. \\
% inside              & (Over)write only the inside. \\
% color,line=$<$color$>$  & Main color. \\
% filled,fill=$<$color$>$ & When both the outline and the inside are written. \\
% gray,grayscale=\num & Inside is in gray. Filled overwrites this. \\
% rotation=\num       & Only for centered boxes: rotation angle.
%\end{tabular}
%\end{center}
%The options gray and filled imply blanked. Hence it is not needed to use 
%blanked when either of those options is used. The default values are that 
%none of these options are used.

Some examples are:
\begin{verbatim}
   \Line[double,sep=1.5,dash,dsize=4](10,10)(70,30)
   \Line[double,sep=1.5,arrow,arrowpos=0.6](10,10)(70,30)
\end{verbatim}
 
The options can also be used on the more explicit commands as extra 
options. Hence it is possible to use
\begin{verbatim}
   \DoubleLine[dash,dsize=4](10,10)(70,30){1.5}
\end{verbatim}
instead of the first line in the previous example.

One may notice that some of the options are not accessible with the more 
explicit commands. For example, it is possible to put arrows on B\'ezier 
curves only by using the option `arrow' for the B\'ezier command.

%>>#] Options :
%>>#[ Remarks about Gluons :
%
\subsection{Remarks about Gluons}
\label{sec:gluon.remarks}

There are 12 commands that concern gluons. This allows much freedom in 
developing one's own style. Gluons can be drawn as single solid lines, as 
double lines, as dashed lines and as dashed double lines.

Gluons have an amplitude and a number of windings. By varying these 
quantities one may obtain completely different gluons as in:

\noindent
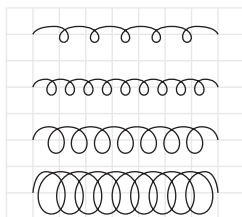
\begin{minipage}{3.83cm}
\begin{axopicture}{(90,90)(-10,0)}
\AxoGrid(0,0)(10,10)(9,8){LightGray}{0.5}
\Gluon(10,70)(80,70){3}{5}
\Gluon(10,50)(80,50){3}{9}
\Gluon(10,30)(80,30){5}{7}
\Gluon(10,10)(80,10){8}{9}
\end{axopicture}
\end{minipage}
\begin{minipage}{11.5cm}
\label{gluons}
\verb:\Gluon(10,70)(80,70){3}{5}: \hfill \\
\verb:\Gluon(10,50)(80,50){3}{9}: \hfill \\
\verb:\Gluon(10,30)(80,30){5}{7}: \hfill \\
\verb:\Gluon(10,10)(80,10){8}{9}:
\end{minipage}\vspace{4mm}

One may change the orientation of the windings by reversing the
direction in which the gluon is drawn and/or changing the sign of the
amplitude:

\noindent
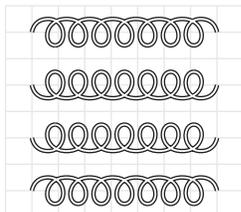
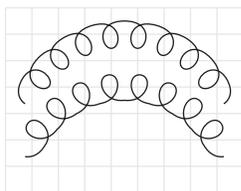
\begin{minipage}{3.83cm}
\begin{axopicture}{(90,90)(-10,0)}
\AxoGrid(0,0)(10,10)(9,8){LightGray}{0.5}
\DoubleGluon(10,70)(80,70){5}{7}{1.2}
\DoubleGluon(80,50)(10,50){5}{7}{1.2}
\DoubleGluon(10,30)(80,30){-5}{7}{1.2}
\DoubleGluon(80,10)(10,10){-5}{7}{1.2}
\end{axopicture}
\end{minipage}
\begin{minipage}{11.5cm}
\label{gluonss}
\verb:\DoubleGluon(10,70)(80,70){5}{7}{1.2}: \hfill \\
\verb:\DoubleGluon(80,50)(10,50){5}{7}{1.2}: \hfill \\
\verb:\DoubleGluon(10,30)(80,30){-5}{7}{1.2}: \hfill \\
\verb:\DoubleGluon(80,10)(10,10){-5}{7}{1.2}:
\end{minipage}\vspace{4mm}

\noindent
\begin{minipage}{3.83cm}
\begin{axopicture}{(90,70)(-10,0)}
\AxoGrid(0,0)(10,10)(9,7){LightGray}{0.5}
\GluonArc(45,20)(40,20,160){5}{8}
\GluonArc(45,0)(40,20,160){-5}{8}
\end{axopicture}
\end{minipage}
\begin{minipage}{11.5cm}
\label{gluonarcA}
\verb:\GluonArc(45,20)(40,20,160){5}{8}:\hfill \\
\verb:\GluonArc(45,0)(40,20,160){-5}{8}:\hfill \\
Here one can see that the sign of the amplitude gives a completely 
different aspect to a gluon on an arc segment.
\end{minipage}\vspace{4mm}

There are two ways of drawing a gluon circle. One is with the command 
GluonCirc and the other is an arc of 360 degrees with the GluonArc command. 
The second way has a natural attachment point, because the GluonArc 
command makes gluons with a begin- and endpoint. 
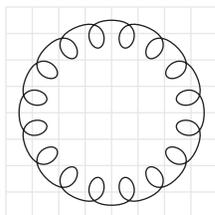
\vspace{4mm}

\noindent
\begin{minipage}{3.83cm}
\begin{axopicture}{(80,80)(-15,0)}
\AxoGrid(0,0)(10,10)(8,8){LightGray}{0.5}
\GluonCirc(40,40)(30,0){5}{16}
\end{axopicture}
\end{minipage}
\begin{minipage}{11.5cm}
%\label{gluoncirc}
\verb:\GluonCirc(40,40)(30,0){5}{16}:\hfill \\
This is the `complete circle'. If one likes to attach one or more lines to 
it one should take into account that the best places for this are at a 
distance radius+amplitude from the center of the circle. One can rotate the 
circle by using the phase argument.
\end{minipage}
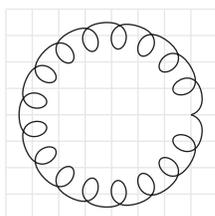
\vspace{4mm}

\noindent
\begin{minipage}{3.83cm}
\begin{axopicture}{(80,80)(-15,0)}
\AxoGrid(0,0)(10,10)(8,8){LightGray}{0.5}
\GluonArc(40,40)(30,0,360){5}{16}
\end{axopicture}
\end{minipage}
\begin{minipage}{11.5cm}
\label{gluonarc360}
\verb:\GluonArc(40,40)(30,0,360){5}{16}:\hfill \\
In the 360 degree arc there is a natural point of attachment. Of course 
there is only one such point. If one needs more than one such point one 
should use more than one arc segment.
\end{minipage}\vspace{4mm}

Some examples are:

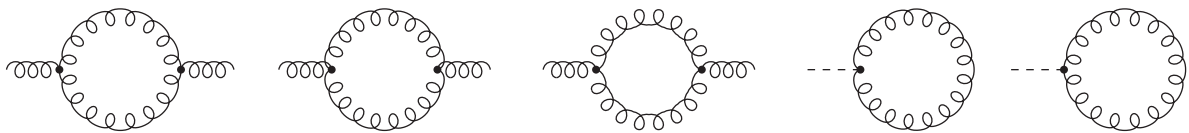
\begin{center} \begin{axopicture}{(460,60)(0,0)}
\Gluon(7,30)(27,30){3}{3}
\GluonCirc(50,30)(20,0){3}{16}
\Gluon(73,30)(93,30){3}{3}
\Vertex(27,30){1.5}
\Vertex(73,30){1.5}
\Gluon(110,30)(130,30){3}{3}
\GluonArc(150,30)(20,0,180){3}{8}
\GluonArc(150,30)(20,180,360){3}{8}
\Gluon(170,30)(190,30){3}{3}
\Vertex(130,30){1.5}
\Vertex(170,30){1.5}
\Gluon(210,30)(230,30){3}{3}
\GluonArc(250,30)(20,0,180){-3}{8}
\GluonArc(250,30)(20,180,360){-3}{8}
\Gluon(270,30)(290,30){3}{3}
\Vertex(230,30){1.5}
\Vertex(270,30){1.5}
\DashLine(310,30)(330,30){3}
\GluonArc(350,30)(20,-180,180){3}{16}
\Vertex(330,30){1.5}
\DashLine(387,30)(407,30){3}
\GluonCirc(430,30)(20,0){3}{16}
\Vertex(407,30){1.5}
\end{axopicture} \end{center}
This picture was generated with the code:
\begin{verbatim}
\begin{center} \begin{axopicture}{(460,60)(0,0)}
   \Gluon(7,30)(27,30){3}{3}
   \GluonCirc(50,30)(20,0){3}{16}
   \Gluon(73,30)(93,30){3}{3}
   \Vertex(27,30){1.5}
   \Vertex(73,30){1.5}
   \Gluon(110,30)(130,30){3}{3}
   \GluonArc(150,30)(20,0,180){3}{8}
   \GluonArc(150,30)(20,180,360){3}{8}
   \Gluon(170,30)(190,30){3}{3}
   \Vertex(130,30){1.5}
   \Vertex(170,30){1.5}
   \Gluon(210,30)(230,30){3}{3}
   \GluonArc(250,30)(20,0,180){-3}{8}
   \GluonArc(250,30)(20,180,360){-3}{8}
   \Gluon(270,30)(290,30){3}{3}
   \Vertex(230,30){1.5}
   \Vertex(270,30){1.5}
   \DashLine(310,30)(330,30){3}
   \GluonArc(350,30)(20,-180,180){3}{16}
   \Vertex(330,30){1.5}
   \DashLine(387,30)(407,30){3}
   \GluonCirc(430,30)(20,0){3}{16}
   \Vertex(407,30){1.5}
\end{axopicture} \end{center}
\end{verbatim}

%>>#] Remarks about Gluons :
%>>#[ Arrows :

\subsection{Remarks about arrows}
\label{sec:arrows}

%--#[ General :

The old Axodraw arrows were rather primitive little triangles. The JaxoDraw 
program has introduced fancier arrows which the user can also customize. 
There are parameters connected to this as shown in the figure:
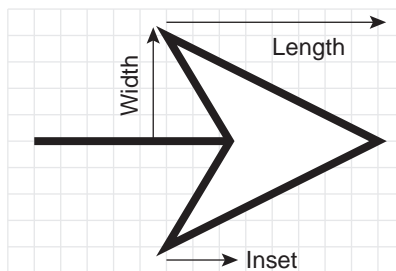
\begin{center}
\begin{axopicture}{(150,100)(0,0)}
\AxoGrid(0,0)(10,10)(15,10){LightGray}{0.5}
\SetWidth{3}
%\Line(10,50)(130,50)
%\FilledPolygon{(140,50)(90,90)(105,50)(90,10)}{White}
%\Polygon{(140,50)(90,90)(105,50)(90,10)}{Black}
%\SetWidth{0.5}
%\LongArrow(85,50)(85,90)
%\LongArrow(90,5)(105,5)
%\LongArrow(90,95)(140,95)
%\SetPFont{Helvetica}{9}
%\PText(110,85)(0)[l]{Length}
%\PText(76,71)(90)[c]{Width}
%\PText(110,5)(0)[l]{Inset}
\Line[arrow,arrowinset=0.3,arrowaspect=1,arrowwidth=40,arrowpos=1,
       arrowstroke=3](10,50)(100,50)
\SetWidth{0.5}
\LongArrow(55,50)(55,90)
\LongArrow(60,5)(84,5)
\LongArrow(60,95)(140,95)
\SetPFont{Helvetica}{9}
\PText(100,85)(0)[l]{Length}
\PText(46,71)(90)[c]{Width}
\PText(90,5)(0)[l]{Inset}
\end{axopicture}\vspace{2mm} \\
\verb:\Line[arrow,arrowinset=0.3,arrowaspect=1,arrowwidth=40,arrowpos=1,:\\
\verb:arrowstroke=3](10,50)(100,50):
\end{center}
The full set of parameters is:
\begin{description}
\item[aspect]   A multiplicative parameter when the length is calculated 
from the width. The normal formula is: 
$\mbox{length}=2\times \mbox{width}\times \mbox{aspect}$.
\item[inset] The fraction of the length that is taken inward.
\item[length] The full length of the arrowhead.
\item[position] The position of the arrow in the line as a fraction of the 
length of the line.
\item[scale]    A scale parameter for the complete arrowhead.
\item[stroke]   The width of the line that makes up the arrowhead. If the 
value is not set (default value is zero) the arrow is filled and overwrites 
whatever was there. In the case of a stroke value the contents are 
overwritten in the background color.
\item[width] The half width of the arrowhead.
\end{description}
The parameters can be set in two ways. One is with one of the commands
\begin{center}
\begin{tabular}{ll}
\verb:\SetArrowScale{number}: & Initial value is 1. \\
\verb:\SetArrowInset{number}: & Initial value is 0.2 \\
\verb:\SetArrowAspect{number}: & Initial value is 1.25 \\
\verb:\SetArrowPosition{number}: & Initial value is 0.5 \\
\verb:\SetArrowStroke{number}: & Initial value is 0 \\
\end{tabular} \vspace{2mm} \\
\end{center}
(A complete list of commands for setting defaults is in
Sec.\ \ref{sec:settings}.)
These commands determine settings that will hold for all following
commands, up to the end of whatever \LaTeX{} or \TeX{} grouping the
default setting is given in.  E.g., setting a default value inside an
\texttt{axopicture} environment sets it until the end of the
environment only.  (Thus the settings obey the normal rules of
\LaTeX{} for scoping.)

The other way is to use one or more of these parameters as options in a 
command that uses an arrow. The general use of options is in Sec.\
\ref{sec:options}. The options that are available are 
\begin{center}
\begin{tabular}{ll}
   arrow              & initial default=false \\
   arrowscale=\num    & initial default=1 \\
   arrowwidth=\num    & initial default=0 \\
   arrowlength=\num   & initial default=0 \\
   arrowpos=\num      & initial default=0.5 \\
   arrowinset=\num    & initial default=0.2 \\
   arrowstroke=\num   & initial default=0 \\
   arrowaspect=\num   & initial default=1.25 \\
   flip               & initial default=false
\end{tabular}
\end{center}
The arrow option tells the program to draw an arrow. Without it no
arrow will be drawn. The flip option indicates that the direction of
the arrow should be reversed from the `natural' direction. 

When
neither the width nor the length are specified, but instead both are
given as zero, they are computed from the line width (and the line
separation when there is a double line). The formula is:
\begin{eqnarray}
   \mbox{Arrowwidth} & = & 
   1.2 \times \left( \mbox{linewidth} 
              + 0.7 \times \mbox{separation}
              + 1
         \right) 
     \times \mbox{arrowscale},
\\
\label{arrowlength}
   \mbox{Length} & = &
   2 \times \mbox{arrowwidth} \times  \mbox{arrowaspect}. 
\end{eqnarray}
%If however $\mbox{linewidth} + \frac{1}{4} \times \mbox{separation} <
%0.5$ the formula for the arrow width becomes $\mbox{arrowwidth} = 2.5
%\times \mbox{arrowscale}$.
If, however, $1.2 \times(\mbox{linewidth}+0.7\times\mbox{separation}+1)$ is less
than 2.5, the formula for the arrow width becomes
$\mbox{arrowwidth}=2.5\times\mbox{arrowscale}$.

If only one of the arrowwidth or the arrowlength parameters is zero,
it is computed from the other non-zero parameter using formula
(\ref{arrowlength}). When both are non-zero, those are the values that
are used.

The position of the arrowhead is a bit tricky. The arrowpos parameter is a 
fraction of the length of the line and indicates the position of the center 
of the arrowhead. This means that when arrowpos is one, the arrowhead 
sticks out beyond the end of the line by half the arrowlength. When for 
instance the line width is 0.5, the default length of the arrowhead 
defaults to 6.25. Hence if one would like to compensate for this one should 
make the line 3.125 points shorter. Usually 3 pt will be sufficient.

Because of backward compatibility axodraw2 has many individual commands for 
lines with arrows. We present them here, together with some `options' 
varieties.
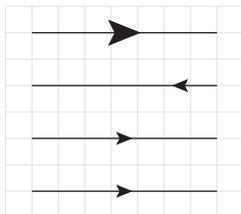
\vspace{4mm}

%--#] General :
%--#[ ArrowLine :

\noindent
\begin{minipage}{3.83cm}
\begin{axopicture}{(90,80)(-10,0)}
\AxoGrid(0,0)(10,10)(9,8){LightGray}{0.5}
\Line[arrow,arrowscale=2](10,70)(80,70)
\Line[arrow,arrowpos=0.8,flip](10,50)(80,50)
\Line[arrow](10,30)(80,30)
\ArrowLine(10,10)(80,10)
\end{axopicture}
\end{minipage}
\begin{minipage}{11.5cm}
\label{arrowline}
\verb:\Line[arrow,arrowscale=2](10,70)(80,70): \hfill \\
\verb:\Line[arrow,arrowpos=0.8,flip](10,50)(80,50): \hfill \\
\verb:\Line[arrow](10,30)(80,30): \hfill \\
\verb:\ArrowLine(10,10)(80,10): \hfill \\
The default position for the arrow is halfway (arrowpos=0.5). With the line 
command and the options we can put the arrow in any position.
\end{minipage}
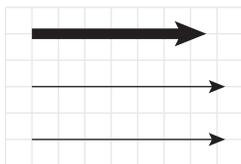
\vspace{4mm}

%--#] ArrowLine :
%--#[ LongArrow :

\noindent
\begin{minipage}{3.83cm}
\begin{axopicture}{(90,60)(-10,0)}
\AxoGrid(0,0)(10,10)(9,6){LightGray}{0.5}
\Line[arrow,arrowpos=1](10,30)(80,30)
\LongArrow(10,10)(80,10)
\SetWidth{4}
\LongArrow[arrowscale=0.8](10,50)(70,50)
\end{axopicture}
\end{minipage}
\begin{minipage}{11.5cm}
\label{longarrow}
\verb:\Line[arrow,arrowpos=1](10,30)(80,30): \hfill \\
\verb:\LongArrow(10,10)(80,10): \hfill \\
\verb:\SetWidth{4}: \hfill \\
\verb:\LongArrow[arrowscale=0.8](10,50)(70,50): \hfill \\
The \verb:\LongArrow: command just places the arrowhead at the end of the 
line. The size of the arrowhead is a function of the linewidth.
\end{minipage}
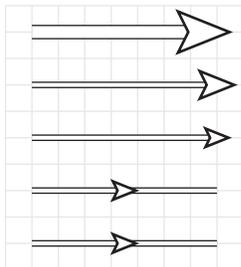
\vspace{4mm}

%--#] LongArrow :
%--#[ ArrowDoubleLine :

\noindent
\begin{minipage}{3.83cm}
\begin{axopicture}{(90,100)(-10,0)}
\AxoGrid(0,0)(10,10)(9,10){LightGray}{0.5}
\SetArrowStroke{1}
\Line[arrow,arrowpos=1,double,sep=5,arrowscale=1.3](10,90)(75,90)
\Line[arrow,arrowpos=1,double,sep=2,arrowscale=1.5](10,70)(80,70)
\Line[arrow,arrowpos=1,double,sep=2](10,50)(80,50)
\Line[arrow,double,sep=2](10,30)(80,30)
\ArrowDoubleLine(10,10)(80,10){2}
\end{axopicture}
\end{minipage}
\begin{minipage}{11.5cm}
\label{arrowdoubleline}
\verb:\SetArrowStroke{1}: \hfill \\
\verb:\Line[arrow,arrowpos=1,double,sep=5,arrowscale=1.3]: \hfill \\
     \verb:    (10,90)(75,90): \hfill \\
\verb:\Line[arrow,arrowpos=1,double,sep=2,arrowscale=1.5]: \hfill \\
     \verb:    (10,70)(80,70): \hfill \\
\verb:\Line[arrow,arrowpos=1,double,sep=2](10,50)(80,50): \hfill \\
\verb:\Line[arrow,double,sep=2](10,30)(80,30): \hfill \\
\verb:\ArrowDoubleLine(10,10)(80,10){2}: \hfill \\
As one can see, the arrows also work with double lines.
\end{minipage}\vspace{4mm}

%--#] ArrowDoubleLine :
%--#[ ArrowDashLine :

\noindent
\begin{minipage}{3.83cm}
\begin{axopicture}{(90,80)(-10,0)}
\AxoGrid(0,0)(10,10)(9,8){LightGray}{0.5}
\Line[arrow,arrowpos=0.3,dash,dsize=3,arrowscale=1.5](10,70)(80,70)
\DashArrowLine(10,50)(80,50){3}
\Line[arrow,dash,dsize=3](10,30)(80,30)
\ArrowDashLine(10,10)(80,10){3}
\end{axopicture}
\end{minipage}
\begin{minipage}{11.5cm}
\label{arrowdashline}
\verb:\Line[arrow,arrowpos=0.3,dash,dsize=3,arrowscale=1.5]: \\
       \verb:(10,70)(80,70): \\
\verb:\DashArrowLine(10,50)(80,50){3}: \\
\verb:\Line[arrow,dash,dsize=3](10,30)(80,30): \\
\verb:\ArrowDashLine(10,10)(80,10){3}: \\
We have not taken provisions for the dashes to be centered in the 
arrowhead, because at times that is nearly impossible. The commands 
\verb:\ArrowDashLine: and \verb:\DashArrowLine: are identical.
\end{minipage}\vspace{4mm}

%--#] ArrowDashLine :
%--#[ ArrowDashDoubleLine :

\noindent
\begin{minipage}{3.83cm}
\begin{axopicture}{(90,80)(-10,0)}
\AxoGrid(0,0)(10,10)(9,8){LightGray}{0.5}
\SetArrowStroke{0.5}
\Line[arrow,arrowpos=1,dash,dsize=3,double,sep=1.5,arrowscale=1.5](10,70)(80,70)
\DashArrowDoubleLine(10,50)(80,50){1.5}{3}
\Line[arrow,dash,dsize=3,double,sep=1.5](10,30)(80,30)
\ArrowDashDoubleLine(10,10)(80,10){1.5}{3}
\end{axopicture}
\end{minipage}
\begin{minipage}{11.5cm}
\label{arrowdashdoubleline}
\verb:\SetArrowStroke{0.5}: \\
\verb:\Line[arrow,arrowpos=1,dash,dsize=3,double: \\
       \verb:,sep=1.5,arrowscale=1.5](10,70)(80,70): \\
\verb:\DashArrowDoubleLine(10,50)(80,50){1.5}{3}: \\
\verb:\Line[arrow,dash,dsize=3](10,30)(80,30): \\
\verb:\ArrowDashDoubleLine(10,10)(80,10){1.5}{3}: \\
The \verb:\ArrowDashDoubleLine: and \verb:\DashArrowDoubleLine: 
commands are identical.
\end{minipage}\vspace{4mm}

%--#] ArrowDashDoubleLine :
%--#[ LongArrowDashLine :

\noindent
\begin{minipage}{3.83cm}
\begin{axopicture}{(90,80)(-10,0)}
\AxoGrid(0,0)(10,10)(9,8){LightGray}{0.5}
\Line[arrow,arrowpos=0,dash,dsize=3,arrowscale=1.5,flip](10,70)(80,70)
\DashLongArrowLine(10,50)(80,50){3}
\Line[arrow,arrowpos=1,dash,dsize=3](10,30)(80,30)
\LongArrowDashLine(10,10)(80,10){3}
\end{axopicture}
\end{minipage}
\begin{minipage}{11.5cm}
\label{longarrowdashline}
\verb:\Line[arrow,arrowpos=0,dash,dsize=3,arrowscale=1.5: \\
       \verb:,flip](10,70)(80,70): \\
\verb:\DashLongArrowLine(10,50)(80,50){3}: \\
\verb:\Line[arrow,arrowpos=1,dash,dsize=3](10,30)(80,30): \\
\verb:\LongArrowDashLine(10,10)(80,10){3}: \\
The commands 
\verb:\LongArrowDashLine:, \verb:\DashLongArrowLine:, 
\verb:\LongArrowDash: and \verb:\DashLongArrow: are identical.
\end{minipage}
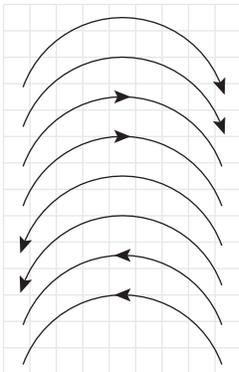
\vspace{4mm}

%--#] LongArrowDashLine :
%--#[ ArrowArc :

\noindent
\begin{minipage}{3.83cm}
\begin{axopicture}{(90,140)(-10,0)}
\AxoGrid(0,0)(10,10)(9,14){LightGray}{0.5}
\Arc[arrow,arrowpos=1,clock](45,95)(40,160,20)
\LongArrowArcn(45,80)(40,160,20)
\Arc[arrow,arrowpos=0.5,clock](45,65)(40,160,20)
\ArrowArcn(45,50)(40,160,20)
\Arc[arrow,arrowpos=1](45,35)(40,20,160)
\LongArrowArc(45,20)(40,20,160)
\Arc[arrow,arrowpos=0.5](45,5)(40,20,160)
\ArrowArc(45,-10)(40,20,160)
\end{axopicture}
\end{minipage}
\begin{minipage}{11.5cm}
\label{arrowarc}
\verb:\Arc[arrow,arrowpos=0,flip](45,95)(40,20,160): \\
\verb:\LongArrowArcn(45,80)(40,20,160): \\
\verb:\Arc[arrow,arrowpos=0.5](45,65)(40,20,160): \\
\verb:\ArrowArcn(45,50)(40,20,160): \\
\verb:\Arc[arrow,arrowpos=1](45,35)(40,20,160): \\
\verb:\LongArrowArc(45,20)(40,20,160): \\
\verb:\Arc[arrow,arrowpos=0.5](45,5)(40,20,160): \\
\verb:\ArrowArc(45,-10)(40,20,160): \\
The \verb:Arc: and the \verb:CArc: commands are identical.
\end{minipage}
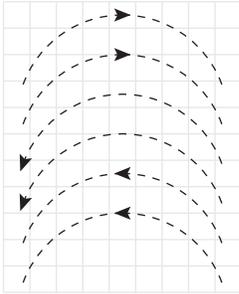
\vspace{4mm}

%--#] ArrowArc :
%--#[ ArrowDashArc :

\noindent
\begin{minipage}{3.83cm}
\begin{axopicture}{(90,110)(-10,0)}
\AxoGrid(0,0)(10,10)(9,11){LightGray}{0.5}
\Arc[arrow,dash,dsize=3,arrowpos=0.5,clock](45,65)(40,160,20)
\ArrowDashArcn(45,50)(40,160,20){3}
\Arc[arrow,dash,dsize=3,arrowpos=1](45,35)(40,20,160)
\LongArrowDashArc(45,20)(40,20,160){3}
\Arc[arrow,dash,dsize=3,arrowpos=0.5](45,5)(40,20,160)
\ArrowDashArc(45,-10)(40,20,160){3}
\end{axopicture}
\end{minipage}
\begin{minipage}{11.5cm}
\label{arrowdasharc}
\verb:\Arc[arrow,dash,dsize=3,arrowpos=0.5]: \\
       \verb:(45,65)(40,20,160): \\
\verb:\ArrowDashArcn(45,50)(40,20,160){3}: \\
\verb:\Arc[arrow,dash,dsize=3,arrowpos=1]: \\
       \verb:(45,35)(40,20,160): \\
\verb:\LongArrowDashArc(45,20)(40,20,160){3}: \\
\verb:\Arc[arrow,dash,dsize=3,arrowpos=0.5]: \\
       \verb:(45,5)(40,20,160): \\
\verb:\ArrowDashArc(45,-10)(40,20,160){3}: \\
The \verb:DashArrowArc: and the \verb:ArrowDashArc: commands are identical.
So are the commands \verb:DashArrowArcn: and \verb:ArrowDashArcn:.
\end{minipage}
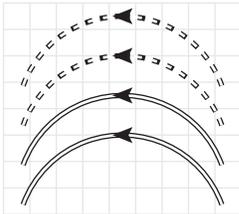
\vspace{4mm}

%--#] ArrowDashArc :
%--#[ ArrowDashDoubleArc :

\noindent
\begin{minipage}{3.83cm}
\begin{axopicture}{(90,80)(-10,0)}
\AxoGrid(0,0)(10,10)(9,8){LightGray}{0.5}
\Arc[arrow,dash,dsize=3,double,sep=1.5,arrowpos=0.5](45,35)(40,20,160)
\ArrowDashDoubleArc(45,20)(40,20,160){1.5}{3}
\Arc[arrow,double,sep=1.5,arrowpos=0.5](45,5)(40,20,160)
\ArrowDoubleArc(45,-10)(40,20,160){1.5}
\end{axopicture}
\end{minipage}
\begin{minipage}{11.5cm}
\label{arrowdashdoublearc}
\verb:\Arc[arrow,dash,dsize=3,double,sep=1.5: \\
      \verb:,arrowpos=0.5](45,35)(40,20,160): \\
\verb:\ArrowDashDoubleArc(45,20)(40,160,20){1.5}{3}: \\
\verb:\Arc[arrow,double,sep=1.5,arrowpos=0.5]: \\
      \verb:(45,5)(40,20,160): \\
\verb:\ArrowDoubleArc(45,-10)(40,20,160){1.5}: \\
Other commands involving Long do not exist. The options can take care of 
their functionality.
\end{minipage}\vspace{4mm}

%--#] ArrowDashDoubleArc :
%--#[ Bezier :
 
Computing the position of the arrow in a B\'ezier curve is a bit complicated. 
Let us recall the definition of a cubic B\'ezier curve:
\begin{eqnarray}
 x & = & x_0 (1-t)^3 + 3 x_1 t (1-t)^2 + 3 x_2 t^2 (1-t) + x_3 t^3 
            \nonumber \\
 y & = & y_0 (1-t)^3 + 3 y_1 t (1-t)^2 + 3 y_2 t^2 (1-t) + y_3 t^3
\end{eqnarray}
Computing the length of the curve is done with the integral
\begin{eqnarray}
   L & = & \int_0^1 dt
   \sqrt{ \left( \frac{dx}{dt} \right)^2 + \left( \frac{dy}{dt} \right )^2 },
\end{eqnarray}
which is an integral over the square root of a quartic polynomial. This we 
do with a 16 point Gaussian quadrature and it gives us more than enough 
accuracy\footnote{We need to compute the length of the B\'ezier curve also 
when we want to put a dash pattern on it. The exact dash size is determined 
such that an integer number of patterns fits in the line.}. Let us assume 
now that we want the arrow at 0.6 of the length. To find the exact fraction 
of the length involves finding the upper limit of the integral for which 
the length is $0.6 L$. This requires an iteration procedure till we have a 
reasonable accuracy for the position $(x,y)$. After that we have to calculate 
the derivative in this point as well.

Because the B\'ezier curves are new commands in axodraw2 there is no need for 
backwards compatibility in the use of arrows. Hence all arrow commands are 
done by means of the options. Some examples are:
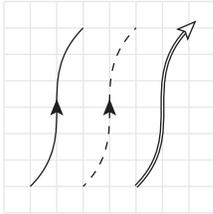
\vspace{4mm}

\noindent
\begin{minipage}{3.83cm}
\begin{axopicture}{(80,80)(-15,0)}
\AxoGrid(0,0)(10,10)(8,8){LightGray}{0.5}
\Bezier[arrow](10,10)(30,30)(10,50)(30,70)
\Bezier[arrow,dash,dsize=3](30,10)(50,30)(30,50)(50,70)
\Bezier[arrow,arrowpos=1,double,sep=1,arrowstroke=0.5](50,10)(70,30)(50,50)(70,70)
\end{axopicture}
\end{minipage}
\begin{minipage}{11.5cm}
\label{arrowbezier}
\verb:\Bezier[arrow](10,10)(30,30)(10,50)(30,70): \\
\verb:\Bezier[arrow,dash,dsize=3](30,10)(50,30): \\
       \verb:(30,50)(50,70): \\
\verb:\Bezier[arrow,arrowpos=1,double,sep=1,arrowstroke: \\
       \verb:=0.5](50,10)(70,30)(50,50)(70,70):
\end{minipage}\vspace{4mm}

%--#] Bezier :
%>>#] Arrows :
%>>#[ Settings :

\subsection{Units and scaling}
\label{sec:units}

When you have constructed a diagram, you may need to change its scale,
to make it larger or smaller.  Axodraw2 provides ways of doing this,
for scaling diagrams without recoding all the individual coordinates.
However the requirements for the nature of the scaling change between
different cases. For example, suppose a diagram is designed for use in
a journal article and you wish to use it in the slides for a seminar.
Then you will want to enlarge both the geometric size of the diagram's
objects and the text labels it contains.  But if you wish to use a
scaled diagram in another place in a journal article, you will wish to
scale its lines etc, but will probably not wish to scale the text (to
preserve its legibility).

Axodraw2 therefore provides tools for the different situations, so we
will now explain what to do.  The commands to achieve this all appear
in the list of parameter-setting commands in Sec.\ \ref{sec:settings}.

\subsubsection{Scaling for slides}

Suppose the original diagram is
\begin{center}
\begin{minipage}{10cm}
\begin{verbatim}
   \SetPFont{Helvetica-Oblique}{12}
   Document text.  Then diagram:
   \begin{axopicture}(60,43)
      \Arc[arrow](30,0)(30,0,180)
      \Text(30,33)[b]{$\alpha P_1$}
      \CText(30,10){Red}{Yellow}{Arc}
   \end{axopicture}
\end{verbatim}
\end{minipage}
\end{center}
to give
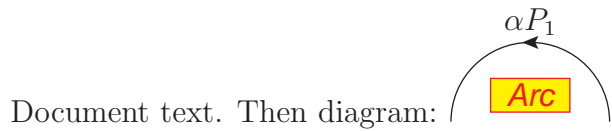
\begin{center}
   \SetPFont{Helvetica-Oblique}{12}
   Document text.  Then diagram:
   \begin{axopicture}(60,43)
      \Arc[arrow](30,0)(30,0,180)
      \Text(30,33)[b]{$\alpha P_1$}
      \CText(30,10){Red}{Yellow}{Arc}
   \end{axopicture}
\end{center}
Then you could double the scale of the diagram by
\begin{center}
\begin{minipage}{10cm}
\begin{verbatim}
   \SetScale{2}
   \fontsize{24}{26}\selectfont
   \SetPFont{Helvetica-Oblique}{12}
   Document text.  Then diagram:
   \begin{axopicture}(60,43)
      \Arc[arrow](30,0)(30,0,180)
      \Text(30,33)[b]{$\alpha P_1$}
      \CText(30,10){Red}{Yellow}{Arc}
   \end{axopicture}
\end{verbatim}
\end{minipage}
\end{center}
to get
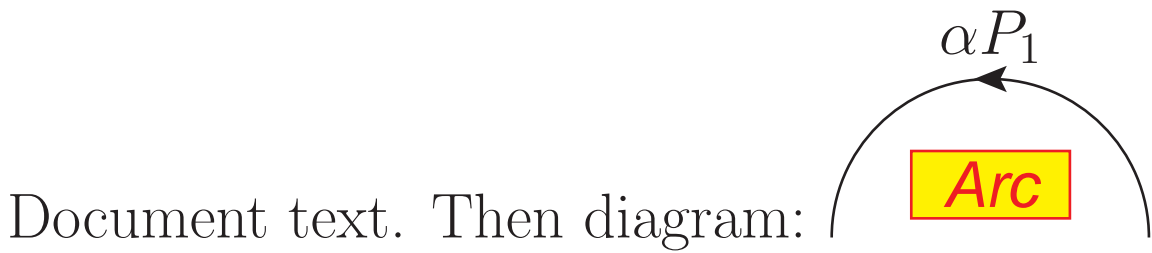
\begin{center}
   \SetScale{2}
   \fontsize{24}{26}\selectfont
   \SetPFont{Helvetica-Oblique}{12}
   Document text.  Then diagram:
   \begin{axopicture}(60,43)
      \Arc[arrow](30,0)(30,0,180)
      \Text(30,33)[b]{$\alpha P_1$}
      \CText(30,10){Red}{Yellow}{Arc}
   \end{axopicture}
\end{center}
We have changed the size of the document font, as would be appropriate
for a make slides for a presentation; this we did by the
\verb+\fontsize+ command.  The arc and the space inserted
in the document for the diagram have scaled up.  The label inserted by
the \verb:\Text: command has changed to match the document font.  The
postscript text in the \verb:\CText: was specified to be at
$\unit[12]{pt}$, but is now scaled up also.  

The above behavior is what axodraw2 does by default, and is what v.\ 1
did.

\subsubsection{Scaling within article}

If you wanted to make an enlarged figure in a journal article, you
would not change the document font.  But the obvious modification to
the previous example is
\begin{center}
\begin{minipage}{10cm}
\begin{verbatim}
   \SetScale{2}
   \SetPFont{Helvetica-Oblique}{12}
   Document text.  Then diagram:
   \begin{axopicture}(60,43)
      \Arc[arrow](30,0)(30,0,180)
      \Text(30,33)[b]{$\alpha P_1$}
      \CText(30,10){Red}{Yellow}{Arc}
   \end{axopicture}
\end{verbatim}
\end{minipage}
\end{center}
which gives
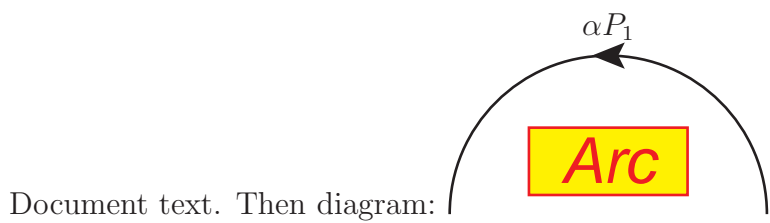
\begin{center}
   \SetScale{2}
   \SetPFont{Helvetica-Oblique}{12}
   Document text.  Then diagram:
   \begin{axopicture}(60,43)
      \Arc[arrow](30,0)(30,0,180)
      \Text(30,33)[b]{$\alpha P_1$}
      \CText(30,10){Red}{Yellow}{Arc}
   \end{axopicture}
\end{center}
The label $\alpha P_1$ is now not enlarged, since it copies the
behavior of the document font.  But the postscript text is enlarged,
which is probably undesirable.  If you were scaling down the diagram
instead of scaling it up, the situation would be worse, because the
postscript font would be difficult to read.

So in this situation, of scaling the diagram while keeping the
document font intact, you probably also want to leave unchanged the
size of the postscript font.  You can achieve this by the
\verb:\PSTextScalesLikeGraphicsfalse: command:
\begin{center}
\begin{minipage}{10cm}
\begin{verbatim}
   \SetScale{2}
   \PSTextScalesLikeGraphicsfalse
   \SetPFont{Helvetica-Oblique}{12}
   Document text.  Then diagram:
   \begin{axopicture}(60,43)
      \Arc[arrow](30,0)(30,0,180)
      \Text(30,33)[b]{$\alpha P_1$}
      \CText(30,10){Red}{Yellow}{Arc}
   \end{axopicture}
\end{verbatim}
\end{minipage}
\end{center}
\begin{center}
   \SetScale{2}
   \PSTextScalesLikeGraphicsfalse
   \SetPFont{Helvetica-Oblique}{12}
   Document text.  Then diagram:
   \begin{axopicture}(60,43)
      \Arc[arrow](30,0)(30,0,180)
      \Text(30,33)[b]{$\alpha P_1$}
      \CText(30,10){Red}{Yellow}{Arc}
   \end{axopicture}
\end{center}

To achieve this on a document-wide basis, which is probably what you
want, you can use the \texttt{PStextScalesIndependently} option when you
load axodraw2 --- see Sec.\ \ref{sec:invoke}.

Nevertheless, if you turn off the default scaling of postscript text,
%you may still want to scale text.  To do this you can use the
you may still want to scale text.  For this you can use the
\verb:\SetTextScale: command, as in \verb:\SetTextScale{1.2}:.  This
only has an effect when you have turned off the scaling of postscript
text with graphics objects; but then it applies to \TeX{} text
inserted by axodraw2's \verb:\Text: and \verb:\rText: commands, as
well text inserted by axodraw2's ``postscript-text'' commands.

If you are confused by the above, we recommend experimentation to
understand how to achieve the effects that you specifically need.  We
could have made the set of commands and options simpler, but only at
the expense of not being able to meet the demands of the different
plausible situations that we could imagine and have to deal with
ourselves.

\subsubsection{Canvas and object scales}

When you use \verb:\SetScale: outside an \verb:axopicture:
environment, as above, the scaling applies to both the axodraw2
objects and the space inserted for the \texttt{axopicture} environment
in the document, as is natural.  But you may find you need to scale a
subset of objects inside the diagram, e.g.,
\begin{center}
  \begin{minipage}{10cm}
    \verb:\begin{axopicture}:(\dots)\\
    \hspace*{1cm} (First block)\\
    \verb:\SetScale{0.5}:\\
    \hspace*{1cm} (Second block)\\
    \verb:\end{axopicture}:
  \end{minipage}
\end{center}
In this case, the units for specifying the objects in the second block
are different from those for specifying the \verb:axopicture:
environment's size (as well as the first block of objects).  We thus
distinguish object units from canvas units, where ``canvas'' refers to
the \verb:axopicture: environment as a whole.  

Another complication is that the \LaTeX{} \verb+picture+ environment
has is own \verb:\unitlength: parameter.  In v.\ 1 of axodraw, the
canvas scale was determined by \LaTeX's \verb:\unitlength:.  But there
was an independent unit for the object scale; this was the one
determined by axodraw's \verb:\SetScale: command. Also, not all
objects used the object scale.  The situation therefore got quite
confusing. In v.\ 1, if, as is often natural, you wished to scale the
canvas as well as the objects, you would have needed to set \LaTeX's
\verb:\unitlength: parameter as well as using axodraw's
\verb:\SetScale: command.

So now we have arranged things so that the canvas and object scales
are tied by default, provided that you use axodraw2's \verb:\SetScale:
command, and that axodraw diagrams are inside \verb+axopicture+
environments (in contrast to the \verb+picture+ environment used in
the original axodraw).
However, it may be necessary to keep backward compatibility in some
cases, and we weren't certain that the new behavior is exactly what is
always desired.  So in axodraw2, we have provided three choices, given
by the \texttt{canvasScaleIs1pt}, \texttt{canvasScaleIsObjectScale},
and \texttt{canvasScaleIsUnitLength} options when loading axodraw2 ---
see Sec.\ \ref{sec:invoke}.  Naturally,
\texttt{canvasScaleIsObjectScale} is the default.  If you wish to
change the setting mid-document, there are corresponding commands ---
Sec.\ \ref{sec:settings}.

\subsection{Settings}
\label{sec:settings}

Axodraw2 has a number of parameters that can be set by the user.  The
parameters include defaults for line types, dimensions, etc.  The
parameters can be set either inside the axopicture environment or
outside.  If they are set outside they modify the default value for
subsequent pictures. If set inside they only affect the current
picture.  (In general, the parameters obey the usual rules for the
scope of \LaTeX{} variables.)  In many cases, the parameters provide
default values for a command to draw an object and can be overridden
for a single object by using an optional parameter in invoking the
command for the object.

The unit for lengths is the current object scale, as set by the
\verb+\SetScale+ command.

\break 

The parameter-setting commands are:
\begin{center}
\def\arraystretch{1.4}
%
% See preamble for definition of \name
\def\descr#1#2{%
  % #1 = command-syntax, #2 = description
  \name{#1} & #2\\
  \hline
}
\def\descrL#1#2{%
  % #1 = command-syntax, #2 = description
  % Set #1 on separate line
  \multicolumn{2}{|l|}{\name{#1}}   \\
                & #2\\
  \hline
}
\def\category#1{%
  % #1 = name of category
  \multicolumn{2}{l}{#1:}
  \\
  \hline
}
\catcode`\#=13
\def#{\#}
\begin{longtable}{|p{5cm}|p{10.2cm}|}
\hline
\endfirsthead
   Command & Commentary 
\\
\hline
%====================
\category{Lines}
\descr{SetDashSize\{\#1\}}{
  This sets the default size for the size of the dashes of dashed
  lines. Its initial value is 3.
}
\descr{SetLineSep\{\#1\}}{
  This sets the default separation of double lines. Its initial value
  is 2.
}
\descr{SetWidth\{\#1\}}{
  This sets the default width of lines.  Its initial value is 0.5.
}
%====================
\category{Arrows}
\descr{SetArrowAspect\{\#1\}}{
   See Sec.\ \ref{sec:arrows}.
}
\descr{SetArrowInset\{\#1\}}{
   See Sec.\ \ref{sec:arrows}.
}
\descr{SetArrowPosition\{\#1\}}{
  Determines where the 
arrowhead is on a line. The position is the fraction of the length of the 
line. 
}
\descr{SetArrowScale\{\#1\}}{
   A scale parameter for the 
   entire head of an arrow.
}
\descr{SetArrowStroke\{\#1\}}{
    This parameter determines the linewidth of the arrowhead if it is just 
    outlined. Its initial value is zero (filled arrowhead).
}
%====================
\category{Scaling}
\descr{canvasScaleOnept}{
      Sets canvas scale to $\unit[1]{pt}$. 
}
\descr{canvasScaleObjectScale}{
      Sets canvas scale to equal the value set by \name{SetScale} in
      units of points.  This is the initial default of axodraw2,
      unless overridden. 
}
\descr{canvasScaleUnitLength}{
      The canvas scale is the same as \LaTeX's length parameter
      \name{unitlength}. 
}
\descr{SetScale\{\#1\}}{
    This sets a scale factor. 
    This factor applies a magnification factor to all 
    axodraw2 graphics objects.  When the setting that
    postscript-text-scales-like-graphics is set (as is true by
    default), it also applies to axodraw2's ``postscript-text''
    writing commands (\name{PText}, \name{BText}, etc), but not to
    its \TeX{}-text commands (\name{Text} etc).  The initial scale
    factor is unity.
}
\descr{SetTextScale\{\#1\}}{
    This factor applies a magnification factor to all 
    axodraw2 text objects, but \emph{only when} the setting that
    postscript-text-scales-like-graphics is turned off.
}
\descrL{PSTextScalesLikeGraphicsfalse}{
      Text drawn by all of Axodraws's text commands scales with the
      factor set by \name{SetTextScale}.
      See Sec.\ \ref{sec:text}. 
}
\descrL{PSTextScalesLikeGraphicstrue}{
      (Default setting.)  Text drawn by Axodraw's postscript-text
      commands scales with the same factor as graphics objects, as set
      by \name{SetScale}.   Text drawn by Axodraw's \TeX{}-text
      commands is unscaled.
      See Sec.\ \ref{sec:text}.
}
%
%====================
\category{Offsets}
\descr{SetOffset(\#1,\#2)}{
  Sets an offset value 
  for all commands of 
  axodraw2. Its value is not affected by the scale variable.
}
\descr{SetScaledOffset(\#1,\#2)}{
  Sets an offset for 
  all commands of axodraw2. This 
  offset is affected by the scale factor.
}
%
%====================
\category{Color}
\descr{SetColor\{\#1\}}{
    Sets the named color,
    for both axodraw2 objects and regular text.  See Sec.\
    \ref{sec:colors} for details on using color with axodraw2.
}
\descr{textRed}{
    Alternative command for setting named a color
    for both axodraw2 objects and regular text.  See Sec.\
    \ref{sec:colors} for details on using color with axodraw2.
    There is one such command for each axodraw2 named color.
}
%====================
\category{Font}
\descr{SetPFont\{\#1\}\{\#2\}}{
    Sets the Postscript 
    font, and its size in units of points.  See Sec.\ \ref{sec:PSText}
    for the commands that use this font, for a table of the names of
    the fonts.  An empty first argument, instead of a font name,  (as in
    \name{SetPFont\{\}\{20\}} indicates that the normal document font is
    to be used at the indicated size.  An empty second argument,
    instead of the font size, (as in \name{SetPFont\{Helvetica\}\{\}} or
    \name{SetPFont\{\}\{\}}) indicates that the font size is to be
    \LaTeX's document font size at the time the text-making command is
    executed. 
}
\end{longtable}
\end{center}

%>>#] Settings :
%>>#[ Colors :

\subsection{Colors}
\label{sec:colors}

\TeX{} and \LaTeX{} by themselves do not provide any means to set
colors in a document.  Instead, one must use a suitable package to
achieve the effect; the current standard one is \file{color.sty}.
Such a package performs its work by passing graphics commands to the
viewable output file.  Since axodraw also works in a similar fashion,
there is a potentiality for conflicts.  

Axodraw version 1, released in 1994, used the package
\file{colordvi.sty} for applying color to normal textual material,
and its own separate methods for applying color to its graphical
objects.  They both defined the same convenient set of named colors
that could be used, but they had to be set separately for text and
graphics\footnote{The named colors corresponded to ones defined by the
  \program{dvips} program.}. The \file{colordvi.sty} package also had
an important disadvantage that its color settings did not respect
\TeX{} grouping and \LaTeX{} environments, so that a color setting
made for text in an environment continued to apply after the end of
the environment.

Since then, the available tools, notably in the powerful
\file{color.sty}, have greatly improved.  But this has introduced
both real and potential incompatibilities with the older methods.
Note that \file{color.sty} is currently the most standard way for
implementing color, and is a required part of \LaTeX{} distributions,
as part of the graphics bundle.

In the new version of axodraw, we have arranged to have compatibility
with \file{color.sty}, while allowing as much backward compatibility
as we could with the user interface from v.\ 1.  We fully rely on
\file{color.sty} for setting color\footnote{Except for certain hard
  wired settings in double lines and stroked arrows.}.  But to keep
the best of the old methods, we have defined all the named colors that
were defined in the old version, together with a few extra ones. We
have also defined color-setting commands in the style of
\file{colordvi.sty}, but they now apply uniformly to both text and
axodraw graphical objects, and they respect \TeX{} and \LaTeX{}
grouping and environments.

This results in some changes in behavior in certain situations.  We
think the new behavior is more natural from the user's point of view;
but it is a change.  

There are two classes of graphics-drawing command in axodraw.  One
class has no explicit color argument, and uses the currently set
color; the line-drawing commands are typical of these.  Other commands
have explicit color arguments, and these arguments are named colors.
The named colors are a union of those axodraw defines, with those
defined by \file{color.sty} together with any further ones defined
by the user.

\subsubsection{How to use colors}

Axodraw works with named colors --- see Sec.\ \ref{sec:defined.colors}
--- which are a standard set of 68 originally defined by the \program{dvips}
program and the \file{colordvi.sty}, plus 5 extra colors defined in
axodraw2.  (In addition there are several named colors that are
normally defined by default by \file{color.sty}, and that can also
be used.)

To use them we have several possibilities to specify colors.  Which to
use is mostly a matter of user preference or convenience.
\begin{itemize}

\item The axodraw command \verb+\SetColor{colorname}+: sets the color
  to be the named color for everything until the end of the current
  environment (or \TeX{} group, as relevant.)  The initial default
  color is Black, of course.  An example:
  \begin{center}
    \begin{minipage}{4cm}
        \SetColor{Red}
        Now red is used:\\
        \begin{axopicture}(0,40)
            \Line(0,10)(40,30)
        \end{axopicture}
    \end{minipage}
    \begin{minipage}{7cm}
    \label{SetColor}
    \begin{verbatim}
    \SetColor{Red}
    Now red is used:\\
    \begin{axopicture}(0,40)
        \Line(0,10)(40,30)
    \end{axopicture}
    \end{verbatim}
    \end{minipage}
  \end{center}

\item Completely equivalently, one can use the command
  \verb+\color{colorname}+ defined by the standard \file{color.sty}
  package, with any of its options, e.g., \verb+\color{Red}+ or
  \verb+\color[rgb]{1,0,0}+.  In fact \verb+\SetColor+ is now a
  synonym for \verb+\color+, retained for backward compatibility.

\item The named colors defined by axodraw2 are listed in Sec.\
  \ref{sec:defined.colors}.  Extra ones can be defined by axodraw2's
  \verb+\newcolor+ command.

\item For each of the named colors defined by axodraw2 (and others
  defined by the use of the \verb+\newcolor+ command), there is a
  macro whose name is ``text'' followed by the color name, e.g.,
  \verb+\textMagenta+.  This behaves just like the corresponding call
  to \verb+\SetColor+ or \verb+\color+.  Thus we have
  \begin{center}
    \begin{minipage}{4cm}
        \textMagenta
        Now magenta is used: \hfill \\
        \begin{axopicture}(0,40)
            \Line(0,10)(40,30)
        \end{axopicture}
    \end{minipage}
    \begin{minipage}{7cm}
    \label{textName}
    \begin{verbatim}
    \textMagenta
    Now magenta is used:\\
    \begin{axopicture}(0,40)
        \Line(0,10)(40,30)
    \end{axopicture}
    \end{verbatim}
    \end{minipage}
  \end{center}
  These macros correspond to macros defined by the venerable
  \file{colordvi.sty} package, but now have what is normally an advantage
  that their scope is delimited by the enclosing environment.  
  \begin{center}
    \begin{minipage}{5cm}
        Normal text, then
        \begin{center}
          \Large \bf \color{Blue}
          Large, bold blue\\
          \begin{axopicture}(40,20)
              \Gluon(0,10)(40,10){4}{4}
          \end{axopicture}\\
        \end{center}
        And normal text afterward.
    \end{minipage}
    \begin{minipage}{7.7cm}
    \label{scope}
    \begin{verbatim}
    Normal text, then
    \begin{center}
        \Large \bf \color{Blue}
        Large, bold blue
        \begin{axopicture}(40,20)
          \Gluon(0,10)(40,10){4}{4}
        \end{axopicture}\\
    \end{center}
    And normal text afterward.
    \end{verbatim}
    \end{minipage}
  \end{center}

\item A delimited section of text can be set in a color by using a
  macro named by the color (e.g., $\verb+\Red+$):
  \begin{center}
    \begin{minipage}{6cm}
        In the middle of black text,
        \textcolor{Red}{red text and
          \begin{axopicture}(30,10)
          \Gluon(0,5)(30,5){3}{4}
          \end{axopicture}\ 
          gluon%
        }.
        Then continue \dots
    \end{minipage}
    \begin{minipage}{7.3cm}
    \label{Red}
    \begin{verbatim}
    In the middle of black text,
    \Red{red text and
      \begin{axopicture}(30,10)
          \Gluon(0,5)(30,5){3}{4}
      \end{axopicture}\ 
      gluon%
    }.
    Then continue \dots
    \end{verbatim}
    \end{minipage}
  \end{center}
  These macros correspond to macros defined by the \file{colordvi.sty}
  package, but they now apply to axodraw objects as well.

\item The same effect, for named colors, can be achieved by
  \file{color.sty}'s \verb+\textcolor+ macro.  Thus
  \verb+\textcolor{Red}{...}+ is equivalent to \verb+\Red{...}+.

\end{itemize}

It is also possible to define new named colors, in the CMYK
system. This means that each color is defined by four numbers. New
colors can be introduced with the \verb:\newcolor{#1}{#2}: command as
in \verb:\newcolor{LightRed}{0 0.75 0.7 0}:. This use of this command
defines a named color for use in axodraw, with corresponding macros
\verb:\LightRed: and \verb:\textLightRed{#1}:, and also makes the name
known to \file{color.sty}.  (Use of \file{color.sty}'s
\verb:\definecolor: macro is not supported here: it will affect only
normal \LaTeX{} text, but not axodraw objects, and it will fail to
define the extra macros.)

We define the CMYK values for the named colors in the
\file{axodraw2.sty} file.  These override the definitions provided
by \file{color.sty} (in its file dvipsnam.def), which are the same
(at least currently).

There can be differences in how colors render on different devices.
In principle, there should be compensations made by the driver to
compensate for individual device properties. Our experience is however
that such compensations are not always implemented well enough. Most
notorious are differences between the shades of green on the screen,
on projectors, and on output from a printer.  These colors are usually
much too light on a projector and one way to correct this is to
redefine those colors when the output is prepared for a projector,
e.g., by use of axodraw's \verb:\newcolor{#1}{#2}: macro.  An example
is illustrated by
\begin{center}
  \color{green} 
  \begin{axopicture}(100,20)
  \Text(25,15){color.sty's green}
  \Line[width=2](0,0)(50,0)
  \end{axopicture}
  \color{Green} 
  \begin{axopicture}(100,20)
  \Text(25,15){axodraw's Green}
  \Line[width=2](0,0)(50,0)
  \end{axopicture}
\end{center}
coded by
\begin{verbatim}
  \color{green} 
  \begin{axopicture}(100,20)
  \Text(25,15){color.sty's green}
  \Line[width=2](0,0)(50,0)
  \end{axopicture}
%
  \color{Green} 
  \begin{axopicture}(100,20)
  \Text(25,15){axodraw's Green}
  \Line[width=2](0,0)(50,0)
  \end{axopicture}
\end{verbatim}
On a typical screen or projector, we find that the two greens are
quite distinct, the ``green'' being much lighter than the
``Green''\footnote{The ``green'' is defined in the RGB scheme from the
  values $(0,1,0)$, while ``Green'' is defined in the CMYK scheme from
  the values $(1,0,1,0)$.}.  But on the paper output from our
printers, they give close results.

\subsubsection{Defined named colors}
\label{sec:defined.colors}

The first set of predefined colors are those defined by dvips (and
defined in \file{colordvi.sty}, or in \file{color.sty} with the
use of both of its usenames and dvipsnames options).  They are
\begin{quote}
\sloppy
\GreenYellow{GreenYellow},
\Yellow{Yellow},
\Goldenrod{Goldenrod},
\Dandelion{Dandelion},
\Apricot{Apricot},
\Peach{Peach},
\Melon{Melon},
\YellowOrange{YellowOrange},
\Orange{Orange},
\BurntOrange{BurntOrange},
\Bittersweet{Bittersweet},
\RedOrange{RedOrange},
\Mahogany{Mahogany},
\Maroon{Maroon},
\BrickRed{BrickRed},
\Red{Red},
\OrangeRed{OrangeRed},
\RubineRed{RubineRed},
\WildStrawberry{WildStrawberry},
\Salmon{Salmon},
\CarnationPink{CarnationPink},
\Magenta{Magenta},
\VioletRed{VioletRed},
\Rhodamine{Rhodamine},
\Mulberry{Mulberry},
\RedViolet{RedViolet},
\Fuchsia{Fuchsia},
\Lavender{Lavender},
\Thistle{Thistle},
\Orchid{Orchid},
\DarkOrchid{DarkOrchid},
\Purple{Purple},
\Plum{Plum},
\Violet{Violet},
\RoyalPurple{RoyalPurple},
\BlueViolet{BlueViolet},
\Periwinkle{Periwinkle},
\CadetBlue{CadetBlue},
\CornflowerBlue{CornflowerBlue},
\MidnightBlue{MidnightBlue},
\NavyBlue{NavyBlue},
\RoyalBlue{RoyalBlue},
\Blue{Blue},
\Cerulean{Cerulean},
\Cyan{Cyan},
\ProcessBlue{ProcessBlue},
\SkyBlue{SkyBlue},
\Turquoise{Turquoise},
\TealBlue{TealBlue},
\Aquamarine{Aquamarine},
\BlueGreen{BlueGreen},
\Emerald{Emerald},
\JungleGreen{JungleGreen},
\SeaGreen{SeaGreen},
\Green{Green},
\ForestGreen{ForestGreen},
\PineGreen{PineGreen},
\LimeGreen{LimeGreen},
\YellowGreen{YellowGreen},
\SpringGreen{SpringGreen},
\OliveGreen{OliveGreen},
\RawSienna{RawSienna},
\Sepia{Sepia},
\Brown{Brown},
\Tan{Tan},
\Gray{Gray},
\Black{Black},
White.
\end{quote}
In addition \file{axodraw2.sty} defines the following extra colors:
\begin{quote}
\LightYellow{LightYellow},
\LightRed{LightRed},
\LightBlue{LightBlue}, 
\LightGray{LightGray},
\VeryLightBlue{VeryLightBlue}.
\end{quote}

Note that \file{color.sty} by default also defines a set of other
named colors: black, white, red, green, blue, cyan, magenta, and
yellow (with purely lower-case names).  Depending on properties of
your screen, projector or printer, these may or may not agree with the
similarly named axodraw colors (which have capitalized names).  These
names can also be used in the \verb+\SetColor+ and \verb+\color+
commands and for color names to those axodraw commands that take named
colors for arguments.

%\subsection{Background issues on color}
%\label{sec:color.issues}

%>>#] Colors :
%>>#[ Some examples :

\section{Some examples}
\label{sec:examples}

\subsection{A Feynman diagram}

When computing the singlet part of structure functions in polarized Deep 
Inelastic Scattering one approach is to use spin two currents to determine 
all anomalous dimensions. At the three loop level this can give diagrams 
like the following:
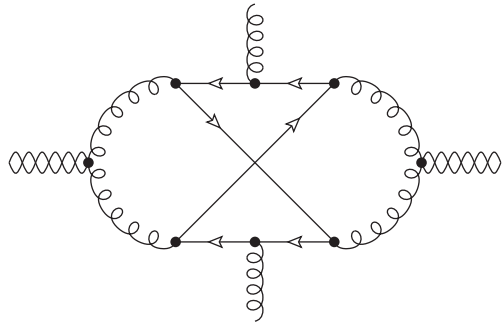
\begin{center}
\begin{axopicture}{(200,140)(0,0)}
\SetArrowStroke{0.5}
\SetArrowScale{0.8}
\Photon(7,70)(37,70){4}{3}
\Photon(7,70)(37,70){-4}{3}
\GluonArc(70,70)(30,90,270){3}{10}
\Line[arrow](100,100)(70,100)
\Line[arrow](130,100)(100,100)
\Line[arrow,arrowpos=0.25](70,100)(130,40)
\Line[arrow](100,40)(70,40)
\Line[arrow](130,40)(100,40)
\Line[arrow,arrowpos=0.75](70,40)(130,100)
\GluonArc(130,70)(30,270,450){3}{10}
\Photon(163,70)(193,70){4}{3}
\Photon(163,70)(193,70){-4}{3}
\Gluon(100,100)(100,130){3}{4}
\Gluon(100,40)(100,10){3}{4}
\Vertex(37,70){2}
\Vertex(163,70){2}
\Vertex(70,100){2}
\Vertex(70,40){2}
\Vertex(130,100){2}
\Vertex(130,40){2}
\Vertex(100,100){2}
\Vertex(100,40){2}
\end{axopicture}
\end{center}
for which the code is:
\begin{verbatim}
  \begin{center} \begin{axopicture}{(200,140)(0,0)}
  \SetArrowStroke{0.5} \SetArrowScale{0.8}
  \Photon(7,70)(37,70){4}{3}
  \Photon(7,70)(37,70){-4}{3}
  \GluonArc(70,70)(30,90,270){3}{10}
  \Line[arrow](100,100)(70,100) \Line[arrow](130,100)(100,100)
  \Line[arrow,arrowpos=0.25](70,100)(130,40)
  \Line[arrow](100,40)(70,40) \Line[arrow](130,40)(100,40)
  \Line[arrow,arrowpos=0.75](70,40)(130,100)
  \GluonArc(130,70)(30,270,450){3}{10}
  \Photon(163,70)(193,70){4}{3}
  \Photon(163,70)(193,70){-4}{3}
  \Gluon(100,100)(100,130){3}{4}
  \Gluon(100,40)(100,10){3}{4}
  \Vertex(37,70){2} \Vertex(163,70){2} \Vertex(70,100){2}
  \Vertex(70,40){2} \Vertex(130,100){2} \Vertex(130,40){2}
  \Vertex(100,100){2} \Vertex(100,40){2}
  \end{axopicture} \end{center}
\end{verbatim}
The diagrams can become a bit more complicated when more lines meet in a 
single vertex. One could compose some lines from straight lines and arcs, 
but in this case we selected some B\'ezier curves. The result is
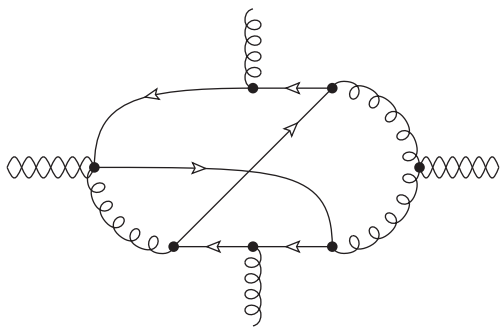
\begin{center}
\begin{axopicture}{(200,140)(0,0)}
\SetArrowStroke{0.5}
\SetArrowScale{0.8}
\Photon(7,70)(40,70){4}{3}
\Photon(7,70)(40,70){-4}{3}
\GluonArc(70,70)(30,180,270){3}{5}
\Bezier[arrow](100,100)(55,100)(40,95)(40,70)
\Line[arrow](130,100)(100,100)
\Bezier[arrow,arrowpos=0.37](40,70)(110,70)(130,70)(130,40)
\Line[arrow](100,40)(70,40)
\Line[arrow](130,40)(100,40)
\Line[arrow,arrowpos=0.75](70,40)(130,100)
\GluonArc(130,70)(30,270,450){3}{10}
\Photon(163,70)(193,70){4}{3}
\Photon(163,70)(193,70){-4}{3}
\Gluon(100,100)(100,130){3}{4}
\Gluon(100,40)(100,10){3}{4}
\Vertex(40,70){2}
\Vertex(163,70){2}
\Vertex(70,40){2}
\Vertex(130,100){2}
\Vertex(130,40){2}
\Vertex(100,100){2}
\Vertex(100,40){2}
\end{axopicture}
\end{center}
for which the code is:
\begin{verbatim}
  \begin{center}
  \begin{axopicture}{(200,140)(0,0)}
  \SetArrowStroke{0.5} \SetArrowScale{0.8}
  \Photon(7,70)(40,70){4}{3}
  \Photon(7,70)(40,70){-4}{3}
  \GluonArc(70,70)(30,180,270){3}{5}
  \Bezier[arrow](100,100)(55,100)(40,95)(40,70)
  \Line[arrow](130,100)(100,100)
  \Bezier[arrow,arrowpos=0.37](40,70)(100,70)(130,70)(130,40)
  \Line[arrow](100,40)(70,40) \Line[arrow](130,40)(100,40)
  \Line[arrow,arrowpos=0.75](70,40)(130,100)
  \GluonArc(130,70)(30,270,450){3}{10}
  \Photon(163,70)(193,70){4}{3}
  \Photon(163,70)(193,70){-4}{3}
  \Gluon(100,100)(100,130){3}{4} \Gluon(100,40)(100,10){3}{4}
  \Vertex(40,70){2} \Vertex(163,70){2} \Vertex(70,40){2}
  \Vertex(130,100){2} \Vertex(130,40){2} \Vertex(100,100){2}
  \Vertex(100,40){2}
  \end{axopicture}
  \end{center}
\end{verbatim}

%\subsection{A flowchart}

%\subsection{A histogram}
 
\subsection{A diagrammatic equation}

This example is from ref~\cite{twopap}. The equations in that paper were 
rather untransparent, because each Feynman diagram represents a complicated 
two loop integral and to solve these integrals one needed many different 
recursion relations in terms of the powers of the propagators. We defined a 
number of macro's for the diagrams, each containing one picture. Here are 
three of them:

\begin{verbatim}
  \def\TAA(#1,#2,#3,#4,#5,#6){
    \raisebox{-19.1pt}{ \hspace{-12pt}
      \begin{axopicture}{(50,39)(0,-4)}
      \SetScale{0.5}\SetColor{Blue}%
      \CArc(40,35)(25,90,270) \CArc(60,35)(25,270,90)
      \Line(40,60)(60,60) \Line(40,10)(60,10) \Line(50,10)(50,60)
      \Line(0,35)(15,35) \Line(85,35)(100,35)
      \SetColor{Black}\SetPFont{Helvetica}{14}%
      \PText(55,39)(0)[lb]{#5} \PText(55,36)(0)[lt]{#6}
      \PText(35,62)(0)[rb]{#1} \PText(65,62)(0)[lb]{#2}
      \PText(65,8)(0)[lt]{#3} \PText(35,8)(0)[rt]{#4}
      \SetColor{Red} \SetWidth{3}
      \Line(50,35)(50,60) \Line(40,60)(50,60)
      \CArc(40,35)(25,90,180) \Vertex(50,60){1.3}
      \end{axopicture}
      \hspace{-12pt}
    }
  }
\end{verbatim}
\def\TAA(#1,#2,#3,#4,#5,#6){
  \raisebox{-18.1pt}{ \hspace{-12pt}
    \begin{axopicture}{(50,39)(0,-4)}
    \SetScale{0.5}\SetColor{Blue}%
    \CArc(40,35)(25,90,270) \CArc(60,35)(25,270,90)
    \Line(40,60)(60,60) \Line(40,10)(60,10) \Line(50,10)(50,60)
    \Line(0,35)(15,35) \Line(85,35)(100,35)
    \SetColor{Black}\SetPFont{Helvetica}{14}%
    \PText(55,39)(0)[lb]{#5} \PText(55,36)(0)[lt]{#6}
    \PText(35,62)(0)[rb]{#1} \PText(65,62)(0)[lb]{#2}
    \PText(65,8)(0)[lt]{#3} \PText(35,8)(0)[rt]{#4}
    \SetColor{Red} \SetWidth{3}
    \Line(50,35)(50,60) \Line(40,60)(50,60)
    \CArc(40,35)(25,90,180) \Vertex(50,60){1.3}
    \end{axopicture}
    \hspace{-12pt}
  }
}
\begin{verbatim}
  \def\TABs(#1,#2,#3,#4,#5){
    \raisebox{-18.1pt}{ \hspace{-12pt}
      \begin{axopicture}{(50,39)(0,-4)}
      \SetScale{0.5}\SetColor{Blue}%
      \CArc(40,35)(25,90,270) \CArc(60,35)(25,270,90)
      \Line(40,60)(60,60) \Line(40,10)(60,10) \Line(50,10)(50,60)
      \Line(0,35)(15,35) \Line(85,35)(100,35)
      \SetColor{Black}\SetPFont{Helvetica}{14}%
      \PText(55,38)(0)[l]{#5}
      \PText(35,62)(0)[rb]{#1} \PText(65,62)(0)[lb]{#2}
      \PText(65,8)(0)[lt]{#3} \PText(35,8)(0)[rt]{#4}
      \SetColor{Red} \SetWidth{3}
      \Line(50,10)(50,60) \Vertex(50,60){1.3}
      \Line(40,60)(50,60) \CArc(40,35)(25,90,180)
      \end{axopicture}
      \hspace{-12pt}
    }
  }
\end{verbatim}
\def\TABs(#1,#2,#3,#4,#5){
  \raisebox{-18.1pt}{ \hspace{-12pt}
    \begin{axopicture}{(50,39)(0,-4)}
    \SetScale{0.5}\SetColor{Blue}%
    \CArc(40,35)(25,90,270) \CArc(60,35)(25,270,90)
    \Line(40,60)(60,60) \Line(40,10)(60,10) \Line(50,10)(50,60)
    \Line(0,35)(15,35) \Line(85,35)(100,35)
    \SetColor{Black}\SetPFont{Helvetica}{14}%
    \PText(55,38)(0)[l]{#5}
    \PText(35,62)(0)[rb]{#1} \PText(65,62)(0)[lb]{#2}
    \PText(65,8)(0)[lt]{#3} \PText(35,8)(0)[rt]{#4}
    \SetColor{Red} \SetWidth{3}
    \Line(50,10)(50,60) \Vertex(50,60){1.3}
    \Line(40,60)(50,60) \CArc(40,35)(25,90,180)
    \end{axopicture}
    \hspace{-12pt}
  }
}
\begin{verbatim}
  \def\TACs(#1,#2,#3,#4,#5){
    \raisebox{-19.1pt}{ \hspace{-12pt}
      \begin{axopicture}{(50,39)(0,-4)}
      \SetScale{0.5}\SetColor{Blue}%
      \CArc(40,35)(25,90,270) \CArc(60,35)(25,270,90)
      \Line(40,60)(60,60) \Line(40,10)(60,10) \Line(50,10)(50,60)
      \Line(0,35)(15,35) \Line(85,35)(100,35)
      \SetColor{Black}\SetPFont{Helvetica}{14}%
      \PText(53,38)(0)[l]{#5}
      \PText(35,62)(0)[rb]{#1} \PText(65,62)(0)[lb]{#2}
      \PText(65,8)(0)[lt]{#3} \PText(35,8)(0)[rt]{#4}
      \SetColor{Red} \SetWidth{3}
      \Line(40,60)(50,60) \CArc(40,35)(25,90,180)
      \end{axopicture}
      \hspace{-12pt}
    }
  }
\end{verbatim}
\def\TACs(#1,#2,#3,#4,#5){
  \raisebox{-19.1pt}{ \hspace{-12pt}
    \begin{axopicture}{(50,39)(0,-4)}
    \SetScale{0.5}\SetColor{Blue}%
    \CArc(40,35)(25,90,270) \CArc(60,35)(25,270,90)
    \Line(40,60)(60,60) \Line(40,10)(60,10) \Line(50,10)(50,60)
    \Line(0,35)(15,35) \Line(85,35)(100,35)
    \SetColor{Black}\SetPFont{Helvetica}{14}%
    \PText(53,38)(0)[l]{#5}
    \PText(35,62)(0)[rb]{#1} \PText(65,62)(0)[lb]{#2}
    \PText(65,8)(0)[lt]{#3} \PText(35,8)(0)[rt]{#4}
    \SetColor{Red} \SetWidth{3}
    \Line(40,60)(50,60) \CArc(40,35)(25,90,180)
    \end{axopicture}
    \hspace{-12pt}
  }
}
and together with two extra little macro's
\begin{verbatim}
\def\plus{\!+\!}
\def\minus{\!-\!}
\end{verbatim}
\def\plus{\!+\!}
\def\minus{\!-\!}
the equations became rather transparent and easy to program. This is the 
code
\begin{verbatim}
   \begin{eqnarray}
     \TAA({n,m},1,1,1,1,1) & = & \frac{1}{\tilde{N}\plus 5\plus n\minus
     m\minus D}\ (\ n\ \ \TAA({n+1,m},0,1,1,1,1)
       \ \ -n\ \ \TACs({n+1,m},1,1,1,1)  \\ & &
       +\ \ \TAA({n,m},1,0,2,1,1)
       \ \ -\ \ \TABs({n,m},1,1,2,1)
       \ \ +m\ \ \TACs({n,m-1},1,1,1,1) 
       \ \ -m\ \ \TABs({n,m-1},1,1,1,1)\ \ \ ) \, .\nonumber
   \end{eqnarray}
\end{verbatim}
and the equation becomes
\begin{eqnarray}
     \TAA({n,m},1,1,1,1,1) & = & \frac{1}{\tilde{N}\plus 5\plus n\minus
     m\minus D}\ (\ n\ \ \TAA({n+1,m},0,1,1,1,1)
       \ \ -n\ \ \TACs({n+1,m},1,1,1,1)  \\ & &
       +\ \ \TAA({n,m},1,0,2,1,1)
       \ \ -\ \ \TABs({n,m},1,1,2,1)
       \ \ +m\ \ \TACs({n,m-1},1,1,1,1) 
       \ \ -m\ \ \TABs({n,m-1},1,1,1,1)\ \ \ ) \, .\nonumber
\end{eqnarray}
The diagrams are actually four-point diagrams. A momentum $P$ flows through 
the diagram (the fat red line), but because the method of computation 
involves an expansion in terms of this momentum the remaining diagrams are 
like two-point functions. Details are in the paper.

%>>#] Some examples :
%>>#[ Acknowledgements :

\section*{Acknowledgements}

JAMV's work is part of the research program of the ``Stichting voor 
Fundamenteel Onderzoek der Materie (FOM)'', which is financially supported 
by the ``Nederlandse organisatie voor Wetenschappelijke Onderzoek (NWO)'' and 
is also supported by the ERC Advanced Grant no.~320651, HEPGAME.
JCC is supported in part by the U.S. Department of Energy under Grant
No.\ DE-SC0008745.  

We like to thank Lucas Theussl for discussions during the development of 
axodraw2.

%>>#] Acknowledgements :
%--#[ Appendix :

\appendix

\section{The axohelp program: Information for developers}
\label{sec:axohelp.devel}

This appendix provides some details on how the axohelp program works.
Most of the information is only relevant to people who wish to modify
or extend axodraw2 and therefore may need to modify axohelp as well.

The reason for axohelp's existence is that axodraw needs to perform
substantial geometric calculations.  When axodraw is used with
pdflatex to produce pdf output directly, suitable calculational
facilities are not available, neither within the PDF language nor
within \LaTeX{} itself.  Therefore when axodraw is used under
pdflatex, we use our program axohelp to perform the calculations.

The mode of operation is as follows.  Let us assume that the .tex file
being compiled by the pdflatex program is called paper.tex.  When one
issues the command
\begin{verbatim}
  pdflatex paper
\end{verbatim}
the reaction of the system is of course to translate all \TeX{}
related objects into a PDF file. Most (but not all) axodraw objects
need non-trivial calculations and hence their
specifications are placed inside a file called paper.ax1. At the end
of the processing \program{pdflatex} will place a message on the screen
that mentions that the user should run the command
\begin{verbatim}
  axohelp paper
\end{verbatim}
for the processing of this graphical information. In principle it is
possible to arrange for axohelp to be invoked automatically from
within pdflatex.  But for this to be done, the running of general
external commands from pdflatex would have to be enabled.  That is a
security risk, and is therefore normally disabled by default for
pdflatex.

When run, axohelp reads the file paper.ax1, processes the contents,
and produces a file paper.ax2. For each axodraw object, it contains
both the code to be placed in the pdf file, and a copy of the
corresponding specification that was in paper.ax1.

When pdflatex is run again, it sees that the file paper.ax2 is present
and reads it in to give essentially an array of objects, one for each
processed axodraw object.  Then during the processing of the document,
whenever axodraw runs into an axodraw object in need of external
calculation, it determines whether an exactly corresponding
specification was present in the file paper.ax2. If not, it means that
the graphical information in the file paper.tex has changed since the
last run of axohelp and the graphics information is invalidated. In
that case, at the end of the program the message to run axohelp will
be printed again. But if instead there is an exact match between an
axodraw object in the current paper.tex and its specification in
paper.ax2, then the corresponding pdf code will be placed in the PDF
file. If all axodraw commands have a proper match in the paper.ax2
file, there will be no message in the paper.log file and on the screen
about rerunning axohelp; then the PDF file should contain the correct
information for drawing the axodraw objects (at least if there are no
\TeX{} errors).

In a sense the situation with axohelp is no different from the use of
makeindex when one prepares a document that contains an index. In that
case one also has to run \LaTeX{} once to prepare a file for the
makeindex program, then run this program which prepares another file
and finally run \LaTeX{} again. Note that if you submit a paper to
arXiv.org, it is likely that their automated system for processing the
file will not run axohelp. So together with paper.tex, you one should
also submit the .ax2 file.

The complete source of the axohelp program can be found in the file 
axohelp.c. This file contains a bit less than 4000 lines of C code but
should translate
without problems with any C compiler --- see Sec.\ \ref{sec:axohelp}
for an appropriate command line on typical Unix-like systems.

The axohelp program functions as follows:
\begin{enumerate}
\item The .ax1 file is located, space is allocated for it and the complete 
file is read and closed again.
\item The input is analysed and split in individual object
  specifications, of which a list is made.
\item The list of object specifications is processed one by
  one. Before the processing of each object specification, the system
  is brought to a default state to avoid that there is a memory of the
  previous object. 
\item In the .ax2 file, for each object is written both the
  corresponding pdf code and a copy of the specification of the object
  as was earlier read from the .ax1 file.  Before the output for an
  object is written to the .ax2 file it is optimized a bit to avoid
  superfluous spaces and linefeeds.
\end{enumerate}

Processing an object from the input involves finding the proper routine for 
it and testing that the number of parameters is correct. Some objects have 
a special input (like the Curve, DashCurve, Polygon and FilledPolygon 
commands). All relevant information is stored in an array of double 
precision numbers. Then some generic action is taken (like setting the 
linewidth and the color) and the right routine is 
called. The output is written to an array of fixed (rather large) length. 
Finally the array is optimized and written to file.

A user who would like to extend the system with new objects should
take the above structure into account. There is an array that gives
the correspondence between axodraw object names and the corresponding
routine in axohelp.  For each object, this array also gives the number
of parameters and whether the stroking or non-stroking color space
should be used. 

Naturally, when adding new kinds of object, it is necessary to add new
items to the just-mentioned array, and to add a corresponding
subroutine.  One should also try to do all the writing of PDF code by
means of some routines like the ones sitting in the file in the
section named ``PDF utilities''. This is important from the viewpoint
of future action. When new graphical languages will be introduced and
it will be needed to modify axodraw2 such that it can produce code for
those languages, it should be much easier if code in the supporting
axohelp program needs to be changed in as few places as possible.
They form a set of graphics primitives used by other subroutines.
Some of these subroutines in the ``PDF utilities'' section of
axohelp.c have names similar to operators in the postscript language
that perform the same function.

%--#] Appendix :
%>>#[ bibliography :

%>>#] bibliography :

\begin{thebibliography}{9}

\bibitem{axodraw1} J.A.M. Vermaseren,
  Comput.\ Phys.\ Commun.\ {\bf 83} (1994) 45--58

\bibitem{jaxodraw1} D. Binosi and L. Theussl,
  Comput.\ Phys.\ Commun.\ {\bf 161} (2004) 76--86.

\bibitem{jaxodraw2}
D. Binosi, J. Collins, C. Kaufhold, L. Theussl, 
  Comput.\ Phys.\ Commun.\ {\bf 180} (2009) 1709--1715

\bibitem{GPL} GNU General Public
  License. \url{http://www.gnu.org/copyleft/gpl.html}. 
  
\bibitem{qcdbook}
J.C. Collins, ``Foundations of Perturbative QCD'' (Cambridge
  University Press, 2011).

\bibitem{twopap} S. Moch and J.A.M. Vermaseren,
  Nucl.\ Phys.\ {\bf B573} (2000) 853.
  %%CITATION = NUPHA,B573,853;%%.

\end{thebibliography}
\end{document}